\documentclass[acmsmall]{acmart}
\usepackage{tikz}
\usepackage{amsmath}
\usepackage{xspace}
\usepackage{hyphenat}
\usepackage{hyperref}
\usepackage{tabularx}
\usepackage{enumitem}
\usepackage{caption,subcaption}
\usepackage[noend]{algpseudocode}
\usepackage[export]{adjustbox}

\usepackage{caption,subcaption}
\microtypecontext{spacing=nonfrench}
\newcommand{\parahead}[1]{\vspace*{0.4ex plus 0.15ex minus 0.15ex}\noindent %
  {\bfseries #1.}}
\newcommand{\parabreak}{\vspace*{1ex}\noindent}
\renewcommand{\paragraph}[1]{\vspace{2pt plus 0pt minus 2pt}\noindent{\bfseries #1}}

\usepackage[utf8]{inputenc}
\newcommand{\ie}{{\itshape i.e.}\xspace}
\newcommand{\eg}{\emph{e.g.}\xspace}

\newcommand{\systemname}{NG-Scope}
\newcommand{\systemnames}{NG-Scope's}

\setitemize{itemsep=1pt,topsep=2pt,parsep=1pt,partopsep=0pt,leftmargin=1em}
\setenumerate{itemsep=1pt,topsep=2pt,parsep=1pt,partopsep=0pt,leftmargin=1.5em}
\setlist{itemsep=1pt,parsep=1pt}

\setcopyright{acmlicensed}
\acmJournal{POMACS}
\acmYear{2022} \acmVolume{6} \acmNumber{1} \acmArticle{12} \acmMonth{3} \acmPrice{}\acmDOI{10.1145/3508032}

\received{January 2021}
\received[revised]{October 2021}
\received[accepted]{December 2021}

\title{\systemname: Fine-Grained Telemetry for NextG Cellular Networks}
\author{Yaxiong Xie}
\email{yaxiongx@cs.princeton.edu}
\author{Kyle Jamieson}
\email{kylej@cs.princeton.edu}
\affiliation{
 \institution{Princeton University}
 \city{Princeton}
 \state{New Jersey}
 \country{USA}
}

\newcolumntype{P}[1]{>{\centering\arraybackslash}p{#1}}
\newcolumntype{M}[1]{>{\centering\arraybackslash}m{#1}}

\begin{document}

\begin{abstract}
Accurate and highly-granular channel capacity telemetry of the 
cellular last hop is crucial for 
the effective operation of transport layer protocols and
cutting edge applications, such as video on demand and video telephony.  
This paper presents the design, implementation,
and experimental performance evaluation of \systemname{}, the first such
telemetry tool able to
fuse physical-layer channel occupancy readings from the cellular
control channel with higher-layer packet arrival statistics and
make accurate capacity estimates.
\systemname{} handles the latest cellular innovations, such as when 
multiple base stations aggregate their signals together to serve mobile users. 
End-to-end experiments in a commercial cellular network 
demonstrate that wireless capacity varies significantly 
with channel quality, mobility, competing traffic within each cell, 
and the number of aggregated cells.
Our experiments demonstrate significantly improved cell load estimation accuracy, 
missing the detection of less than 1\% of data capacity overall,
a reduction of 82\% compared to OWL~\cite{OWL}, 
the state-of-the-art in cellular monitoring.
Further experiments show that 
MobileInsight\hyp{}based \cite{MobileInsight}
CLAW~\cite{CLAW} has a root\hyp{}mean\hyp{}squared capacity error
of 30.5~Mbit/s, which is 3.3$\times$ larger than \systemname{} (9.2~Mbit/s).
\end{abstract}

\setcopyright{acmlicensed}
\acmJournal{POMACS}
\acmYear{2022} \acmVolume{6} \acmNumber{1} \acmArticle{12} \acmMonth{3} \acmPrice{}\acmDOI{10.1145/3508032}

\maketitle
\section{Introduction}
\label{s:intro}

Cellular traffic has accounted for 72\% of global mobile 
traffic in the past few years, 
with this trend expected to continue through 2022~\cite{CiscoReport}.  
Motivated by this, network designers are devoting significant 
efforts to improve the congestion control design of end\hyp{}to\hyp{}end transport protocols, 
so key applications such as video streaming, video 
telephony~\cite{CellurVideo,MuVi,Pensieve,Sprout,Salsify}, 
and bulk data transfer~\cite{PCC-v,PCC,Copa,CUBIC,BBR,Verus}, 
can work seamlessly with cellular networks, 
whose channel capacity varies significantly within a very short period of time. 

\begin{figure}[tb]
    \centering
    \begin{subfigure}[b]{0.31\linewidth}
        \centering
        \includegraphics[width=0.98\textwidth]{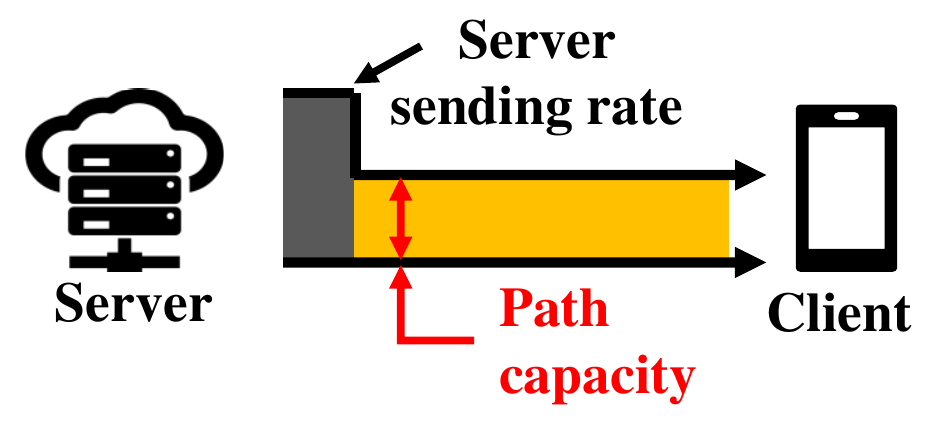}
        \caption{Congestion control matches the sending rate to path capacity.}
        \label{fig:3app_transport}
    \end{subfigure}
    \hfill
    \begin{subfigure}[b]{0.31\linewidth}
        \centering
        \includegraphics[width=0.98\textwidth]{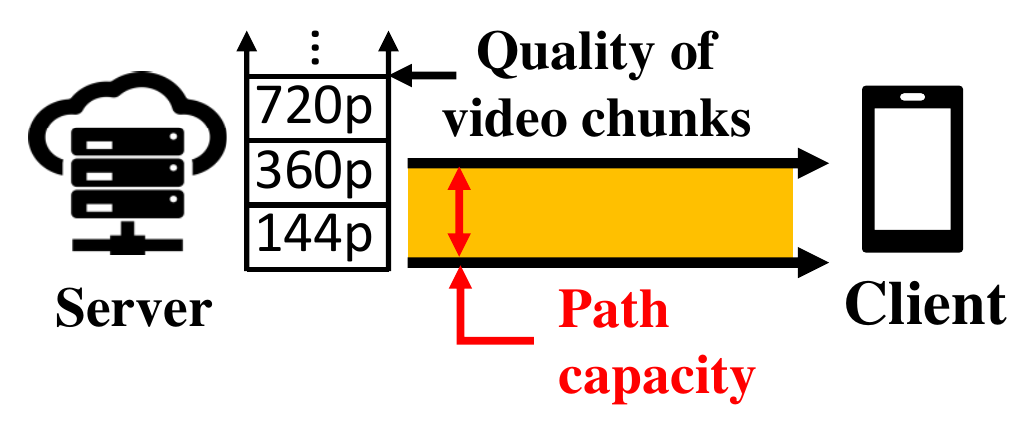}
        \caption{The server adapts the streaming video quality to path capacity.}
        \label{fig:3app_vod}
    \end{subfigure}
    \hfill
    \begin{subfigure}[b]{0.32\linewidth}
        \centering
        \includegraphics[width=0.98\textwidth]{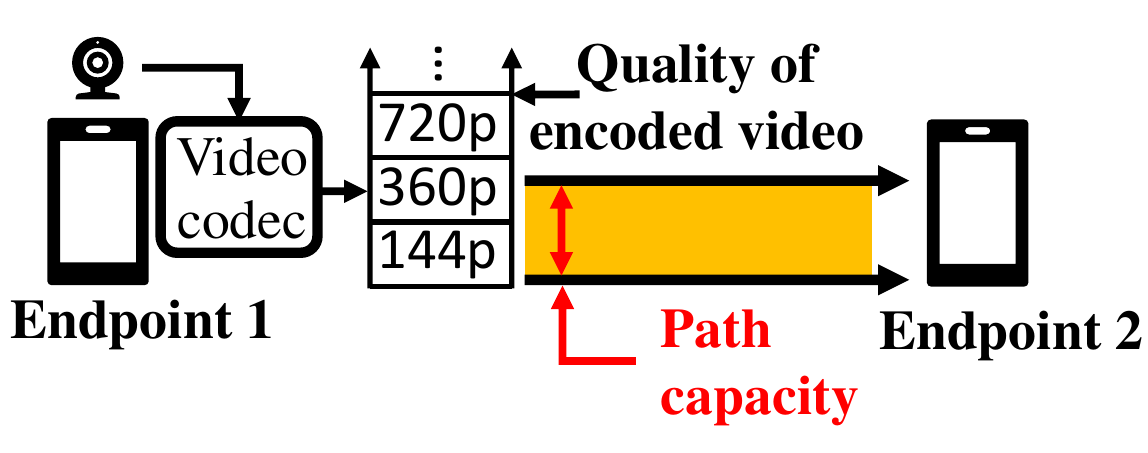}
        \caption{A mobile user adjusts video encoding quality to path capacity.}
        \label{fig:3app_conference}
    \end{subfigure}
    
   \caption{ 
    Congestion control algorithms of the transport layer 
    protocols (\textbf{a}); video streaming (\textbf{b}); and video telephony (\textbf{c}) applications,
    require accurate and fine-grained channel capacity to achieve a maximized system performance.}
    \label{fig:intro_3app}
\end{figure}

Path capacity, \textit{i.e.,} the capacity of a TCP connection,  
is a fundamental system parameter 
that many end-to-end protocols and applications require as an input. 
For example, to maximize throughput and minimize delay, 
modern congestion control algorithms, 
like BBR~\cite{BBR}, PCC~\cite{PCC,PCC-v}, Copa~\cite{Copa}, and PBE~\cite{PBE},
try to match senders' rates to path capacity, 
as shown in Figure~\ref{fig:3app_transport}, 
so high performance requires accurate and fine-grained capacity estimation~\cite{PBE}.
Another example is a video streaming application, 
where the server fragments whole videos into small chunks, 
encodes each chunk into multiple bit rates, 
and adaptively selects the bit rate (and thus resolution) of each chunk, 
according to its path capacity estimate
and buffer occupancy at the mobile client~\cite{bufferABR,BOLA,robustMPC,Pensieve,Oboe}, 
as shown in Figure~\ref{fig:3app_vod}.
Unlike video streaming, which pre-encodes the video into all available bit rates,
video telephony endpoints encode the video in real time~\cite{Concerto,Vantage}.
Consequently, endpoints use available path capacity to 
adjust their encoded video quality,
minimizing video delivery latency and maximizing user quality of experience (QoE)
as shown in Figure~\ref{fig:3app_conference}.


One determinant of the end-to-end path capacity is the capacity of the last-(or first-)hop:
wireless cellular Radio Access Network (RAN). 
Estimating and tracking the capacity of a wireless RAN, 
is challenging,
since its capacity varies over millisecond\hyp{}level 
time scales for the following three reasons.
First, cellular networks adopt OFDMA to share bandwidth among all
mobile users at a subframe (millisecond) granularity. 
Therefore, the start and termination of any user's data flow can cause abrupt capacity fluctuations. 
Second, the wireless channel quality between the cell tower and the user varies
due to user mobility, multipath propagation, and interference.
The highest data rate the cellular wireless channel can support changes accordingly.
Third, both 4G LTE~\cite{LTE-Rel-10} and the 5G~\cite{5G} 
implement a technique called \textit{carrier aggregation} (CA),
via which, the cellular network aggregates two or more cell towers 
(called \textit{component carriers})
to boost maximum per\hyp{}user data rates.
Because of carrier aggregation, one user is not 
only affected by the dynamics from one cell but from all of its aggregated cells.


The dynamic nature of a cellular network motivates a telemetry tool  
that is able to provide accurate and millisecond-granular capacity estimation updates for the cellular network.
On the one hand, with the provided accurate capacity estimation, the congestion controller and the upper layer applications
can adjust its system parameters, such as send rate or video resolution, with the underlying network condition.
On the other hand, with the millisecond-granular capacity updates, 
the controller and applications can detect network variations with up to millisecond-level delay and 
thus take quick actions to mitigate the impact of such network dynamics.
We note that even though there may exist a delay between the observation of the network variations 
and the time the actions of the controller or end-to-end applications take effect 
(the maximum delay is one propagation delay depending on where the observation is taken),
(the delay is minimized when the action and the observation are taken at the same place and 
the delay is maximized to one RTT if the observation has to be delivered back to the sender),
our experimental results in Section~\S\ref{s:congestion} and Section~\S\ref{s:video} demonstrate that
such a timely observation still improves the performance of congestion control and video streaming applications significantly. 

\begin{figure}[tb]
    \centering
    \begin{subfigure}[b]{0.49\linewidth}
        \centering
        \includegraphics[width=0.95\textwidth]{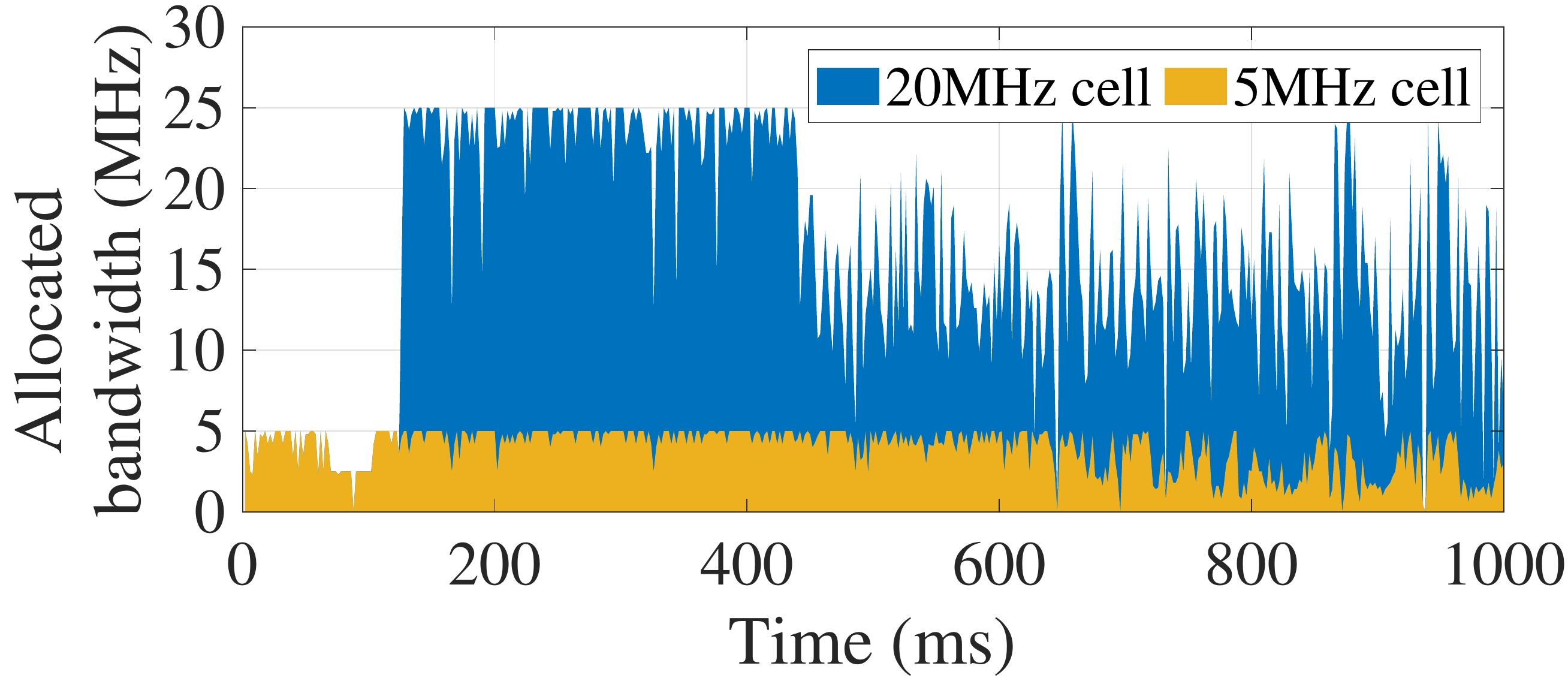}
        \caption{Allocated RAN bandwidth.}
        \label{fig:CA_allocatedB}
    \end{subfigure}
    \hfill
    \begin{subfigure}[b]{0.49\linewidth}
        \centering
        \includegraphics[width=0.95\textwidth]{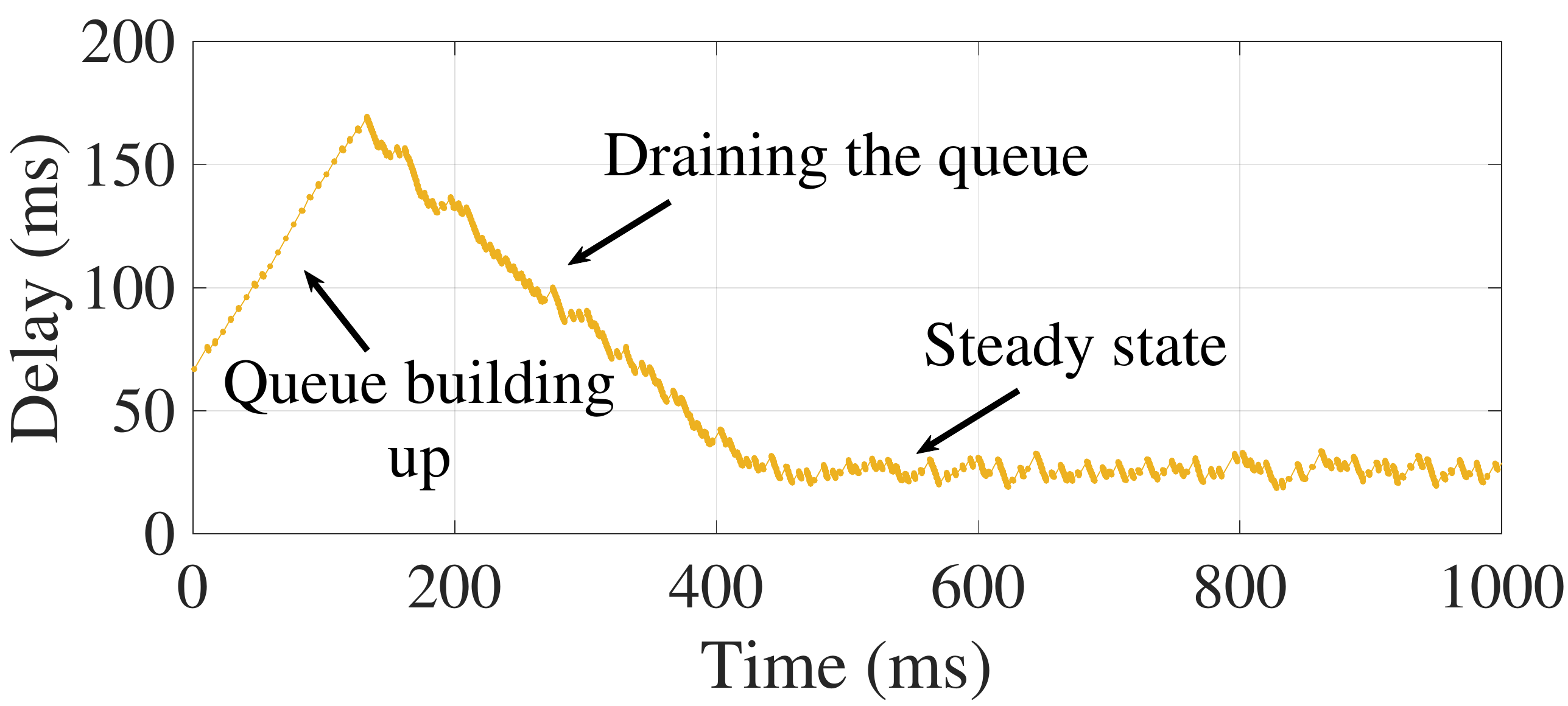}
        \caption{One way packet delay.}
        \label{fig:CA_oneway}
    \end{subfigure}
    \caption{An example of carrier aggregation. The offered load of the server (25~Mbit/s) exceeds the 
    capacity of the 5~MHz cell, resulting in packet queuing. A cellular network aggregates a 20~MHz cell to boost the maximum capacity available to such a user.  }
    \label{fig:CA_example}
\end{figure}

The opportunity to design such an accurate and fine-grained telemetry tool arises from one observation we make:
the cell tower broadcasts all the wireless parameters that affect the wireless cellular channel capacity, 
including the \textit{modulation and coding rate} (MCS), 
the number of MIMO spatial streams, 
the allocated frequency bandwidth, and the aggregated cell towers, 
to the mobile users via a physical control channel, every one millisecond.
By carefully decoding this physical control channel, we expose the internal state of the wireless cellular network to transport layer and applications at the endpoints,
opening new possibilities for designing agile transport layer protocols and implementing video applications with maximized QoE.
Passive cellular sniffers such as LTEye~\cite{LTEye} and OWL~\cite{OWL} already decode the control channel to infer some of the above wireless attributes, 
but cannot work with the most cutting edge cellular networks~\cite{LTE-Rel-10,5G} that uses techniques such as carrier aggregation, 
and thus are not able to track the abrupt capacity changes the carrier aggregation causes, as the example in Figure~\ref{fig:CA_example} shows.
MobileInsight~\cite{MobileInsight} can only analyze the radio resource allocation of a single user, 
\ie, the mobile device the MobileInsight is implemented on, 
rather than cell-wide information, which is mandatory in order to
estimate the capacity available to the mobile user.

\parabreak
This paper presents the design and implementation of \textit{\systemname{}},
a tool that exposes the fine-grained capacity-related information
applications and transport protocols require for superior performance. 
\systemname{} simultaneously decodes the physical layer control channels of
multiple cell towers, extracting highly granular, per\hyp{}user link and physical 
layer transmission status information for all users associated with those cells.  
By combining these data across users within one cell, and fusing data across cells, 
\systemname{} observes and accounts for the effects of carrier aggregation, 
estimating wireless cellular channel capacity more accurately than previous cellular monitors.
By reconciling control messages from the cellular physical layer with the packet arrival time series from the transport layer of the User Equipment (UE), 
we accurately monitor the downlink delivery process of every packet the user receives across layers and down to the Physical Layer (PHY), 
enabling the better functionality of congestion control algorithms and video applications, 
which heretofore lacked insight into the PHY.

We have implemented a proof of concept prototype of \systemname{} with multiple USRP 
software-defined-radios, one host PC, and one mobile phone, in a design that synchronizes 
and fuses information from these sources~(\S\ref{s:imp})\footnote{The source code of \systemname{} is available at https://github.com/YaxiongXiePrinceton/NG-Scope}.  
We implement using USRP instead of directly deploying on mobile devices 
because our design requires customization to the firmware of the cellular module, 
which is closed-source.
However, MobileInsight-based systems like CLAW and BurstTracker that we compare in our evaluation section 
are directly implementable on mobile devices. 
Our performance evaluation in a commercial cellular network
validates the accuracy and measurement resolution of \systemname{}. 
\systemnames{} control channel decoder achieves a reduction in missed control messages to 0.8\% (a 82\% reduction compared to OWL). 
\systemname{} reduces the rate of false positives (detection of nonexistent control messages) by 92\% compared to OWL.
Further results show that even though wireless capacity varies significantly with mobility, competing traffic within each cell and the
number of aggregated cells,
\systemname{} is still capable of tracking the resulting capacity variations~(\S\ref{s:eval_capEst}).

We also make new experimental observations 
on cellular wireless physical\hyp{} and link\hyp{}layer operations.
Firstly, we find that there exist dedicated cell towers that only function 
as a secondary cell to deliver carrier\hyp{}aggregated downlink traffic. 
Without competition from primary users, 
such a cell ensures a stable capacity boost for UEs~(\S\ref{s:eval_capEst}), 
motivating a congestion control that can exploit such an opportunity. 
Secondly, we observe that with carrier aggregation enabled, 
the cellular network saturates the primary cell 
before the potentially higher speed secondary 
and tertiary cells~(\S\ref{s:eval_CA_cells}),
resulting in inefficient bandwidth usage. 
Lastly, we observe cross-cell traffic correlation (\S\ref{s:eval_compete}), the effect of competing for traffic propagating to other cells via carrier aggregation.

To further demonstrate the advantage of \systemnames{} complete cell-wide information over the partial information MobileInsight~\cite{MobileInsight} observes, 
we compare \systemname{} with two systems built atop of MobileInsight: CLAW~\cite{CLAW} for capacity tracking~(\S\ref{s:CLAW}) and BurstTracker~\cite{BurstTracker} 
for bottleneck determination~(\S\ref{s:burstTracker}).
Experimental results show that \systemname{} achieves superior performance compared to these two systems.  
\section{LTE Primer}
\label{s:lte_primer}
\label{s:OFDMA}

Here we introduce relevant parts of LTE's Layer~1 and~2 data\hyp{}plane design, 
focusing on frequency division duplexing (FDD), the mode cellular operators use most widely.

\begin{figure}[t]
    \centering
    \begin{subfigure}[b]{0.49\linewidth}
        \centering
        \includegraphics[width=0.85\textwidth]{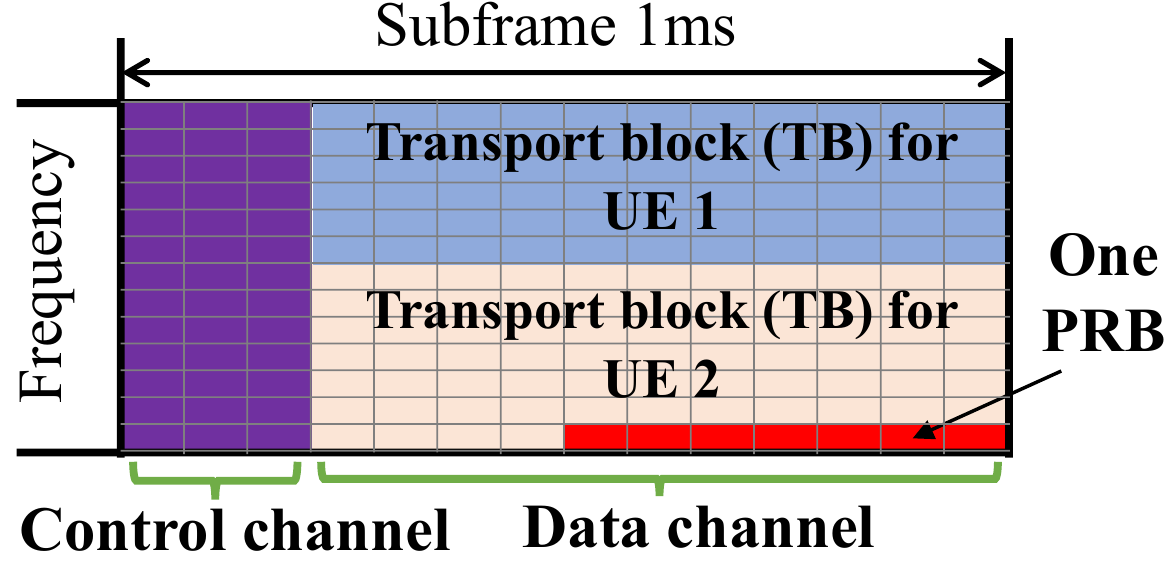}
        \caption{LTE subframe structure.}
        \label{fig:pdsch}
    \end{subfigure}
    \hfill
    \begin{subfigure}[b]{0.49\linewidth}
        \centering
        \includegraphics[width=0.85\textwidth]{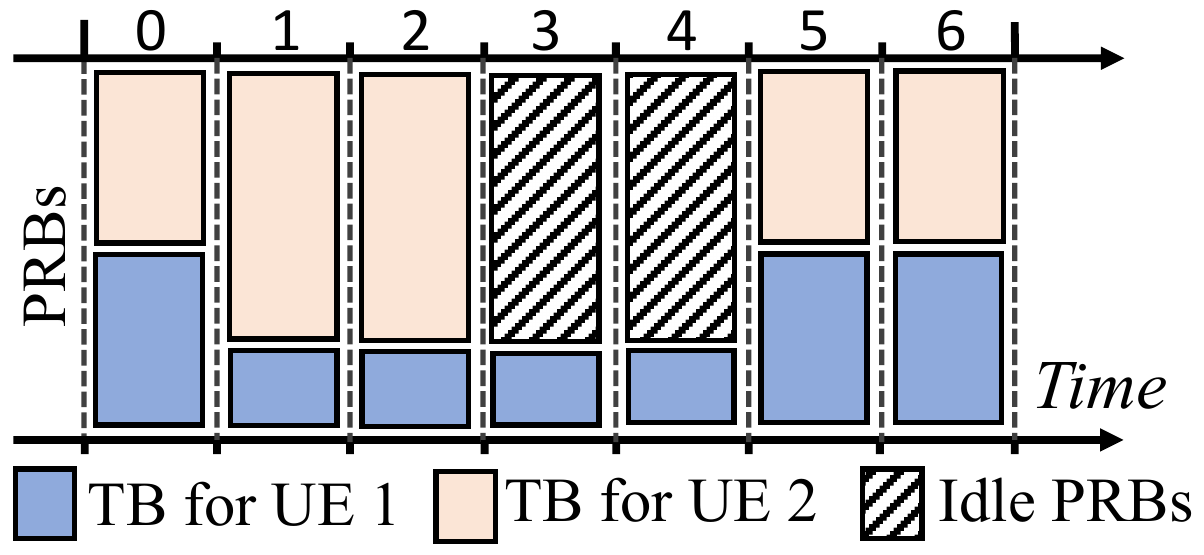}
        \caption{Resource allocation.}
        \label{fig:OFDMA}
    \end{subfigure}
    \caption{LTE divides time into millisecond\hyp{}length subframes. 
    The OFDM symbols inside a subframe are allocated to a control channel carrying control information and a data channel carrying transport blocks (TBs) for specific users. 
    }
    \label{fig:t_f_structure}
\end{figure}

\parahead{Resource allocation}
LTE uses OFDMA in the physical layer \cite{TS211}, dividing the whole channel into 
smaller time\hyp{}frequency blocks called \textit{physical resource blocks} 
(PRB), each spanning 12 OFDM subcarriers in frequency and 500~$\mu$s in time, 
as shown in Figure~\ref{fig:pdsch}.
LTE also divides the time into millisecond\hyp{}length \textit{subframes}, 
consisting of 14 OFDM symbols in time. 
Roughly speaking, within the subframe, 
LTE allocates one to three OFDM symbols to 
a \textit{control channel} (containing resource allocation information), 
and the remainder to a \textit{data channel}.
Depending on traffic load, LTE allocates PRBs inside the data channel to one or more UEs, 
with each user's allocation termed a \textit{transport block} (TB).
TB size varies across time, 
depending on the user's traffic load and competition across users, 
as Figure~\ref{fig:OFDMA} shows.

\parahead{Retransmission}
Wireless transmissions are subject to errors.
LTE relies on a Hybrid ARQ (HARQ) mechanism at the link layer to retransmit errored data bits~\cite{LTE-MAC}. 
LTE's HARQ adopts a \textit{stop-and-wait} scheme, sending one transport block in each subframe 
and then waiting for an acknowledgment from the receiver. 
If the decoding of a transport block fails, 
the cell tower resends the erroneous transport block, 
after eight milliseconds of the original transmission.
An explicit \textit{new-data indicator} is included in the control message for every transport block to 
differentiate the original transmission from the retransmission. 
One stop-and-wait HARQ process cannot send any data while awaiting an acknowledgment. 
For continuous data delivery, LTE starts eight HARQ processes that run in parallel, 
for every associated mobile device.

\section{System Design}
\label{s:design}

We first introduce \systemnames{} architecture (\S\ref{s:sys_arch}), and then the control channel decoder, a crucial component in our 
system (\S\ref{s:decoder}). Finally, we detail how we fuse
the cross-layer information in our \textit{fusion layer} (\S\ref{s:fusion}), providing a 
complete view of wireless cellular communication to higher layers.
\begin{figure}[thb]
    \centering
    \includegraphics[width=0.6\linewidth]{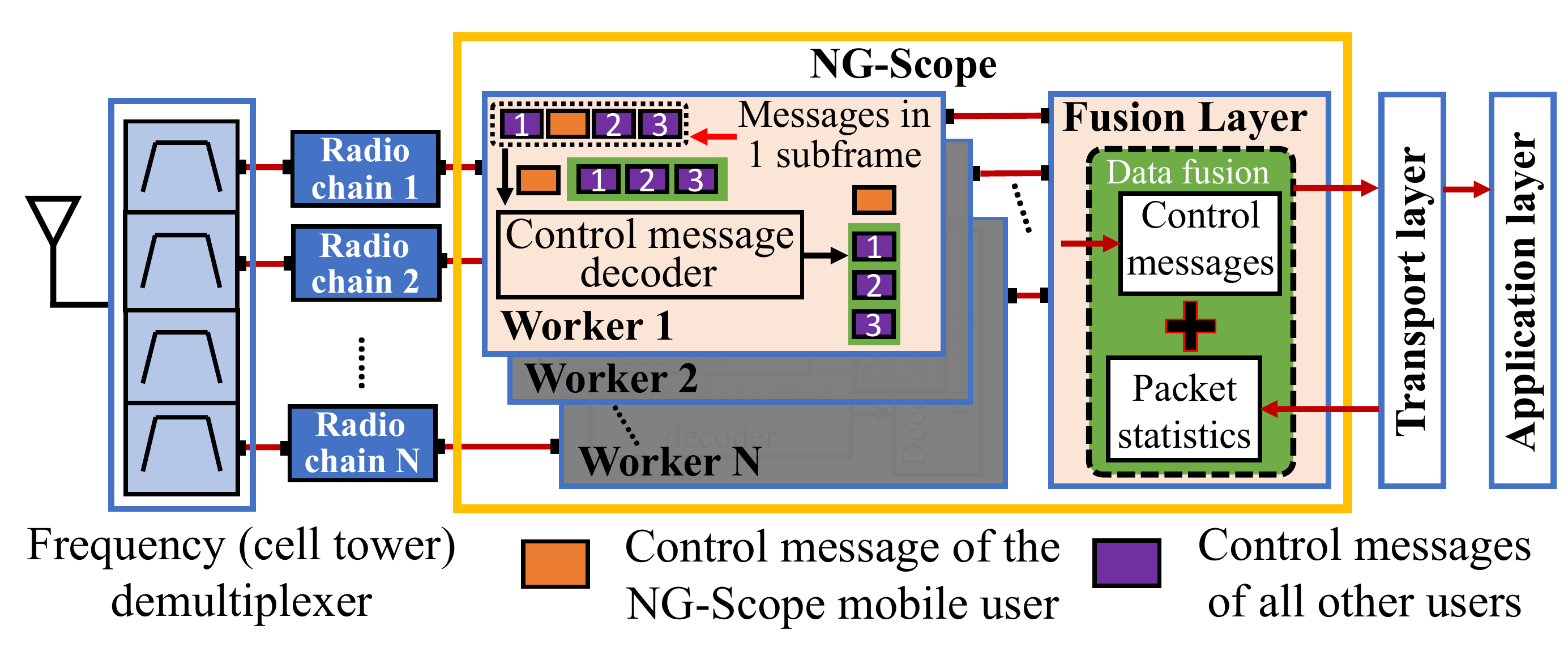}
    \caption{\textit{\systemname{} architecture.} A customized control message
    decoder decodes all control messages inside one subframe. A fusion layer fuses control 
    messages from the physical layer and packet statistics from the transport layer.}
    \label{fig:system}
\end{figure}

\subsection{System Architecture}
\label{s:sys_arch}
A mobile user with carrier aggregation triggered communicates simultaneously 
with multiple cell towers, as shown in Figure~\ref{fig:system}.
The mobile user starts one worker to decode the control messages from one cell.
\systemname{} designs a control message decoder that is capable of
decoding all users' control messages in the control channel, 
as shown in Figure~\ref{fig:system}.
\systemname{} fuses the decoded control messages from the physical layer with packet statistics such as packet arrival time, 
one\hyp{}way delay and packet size, from the transport layer, 
which are reported back to the transport and application layer 
to facilitate congestion control or video quality selection.

\subsection{Control Message Decoder}
\label{s:decoder}
In this section, we introduce \systemnames{} control message decoder, which is capable of decoding 
every control message inside one subframe, thus providing a full picture of the bandwidth usage 
of the cell tower at millisecond time granularity.

\subsubsection{Message encoding and transmission inside control channel.}\label{s:dci_encode}
In this section, we introduce the background of control message encoding and its transmission via the physical control channel.
\begin{figure*}[t]
    \begin{minipage}[htb]{0.49\linewidth}
        \centering
        \includegraphics[width=0.9\linewidth]{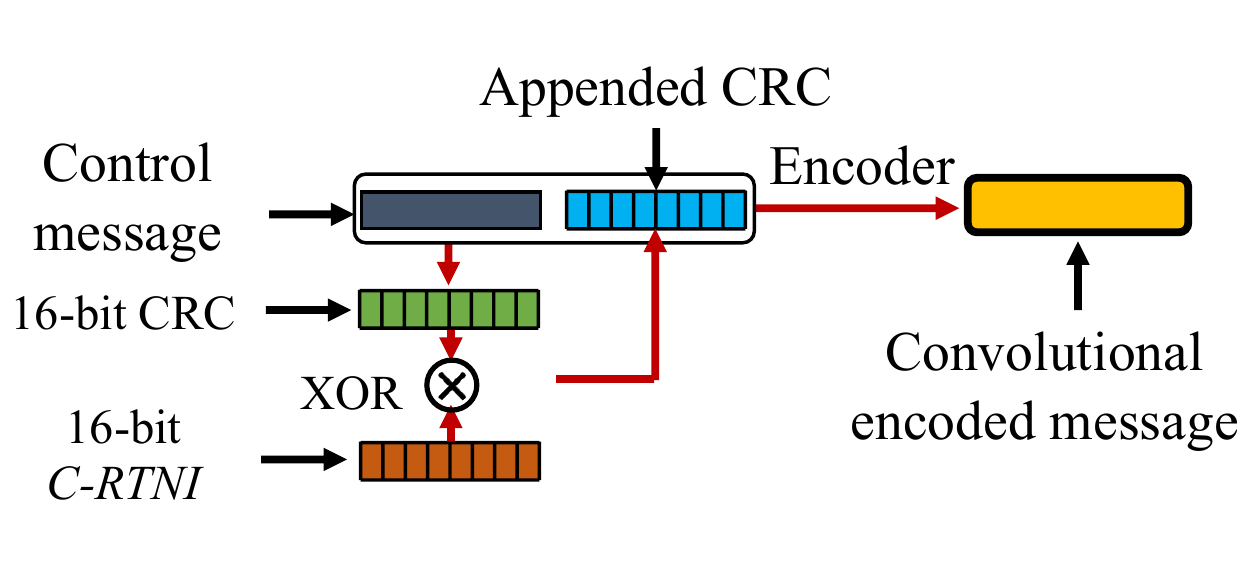}
        \caption{Control message's encoding process.
        A 16\hyp{}bit CRC is calculated from the message, 
        which is XOR\hyp{}ed with the receiver's ID, \ie, another 16\hyp{}bit \textsf{\small C-RNTI}.
        The XOR\hyp{}ed value is appended at the end of the packet.}
        \label{fig:dci_encode}
    \end{minipage}
    \hfill
    \begin{minipage}[htb]{0.49\linewidth}
        \centering
        \includegraphics[width=0.9\linewidth]{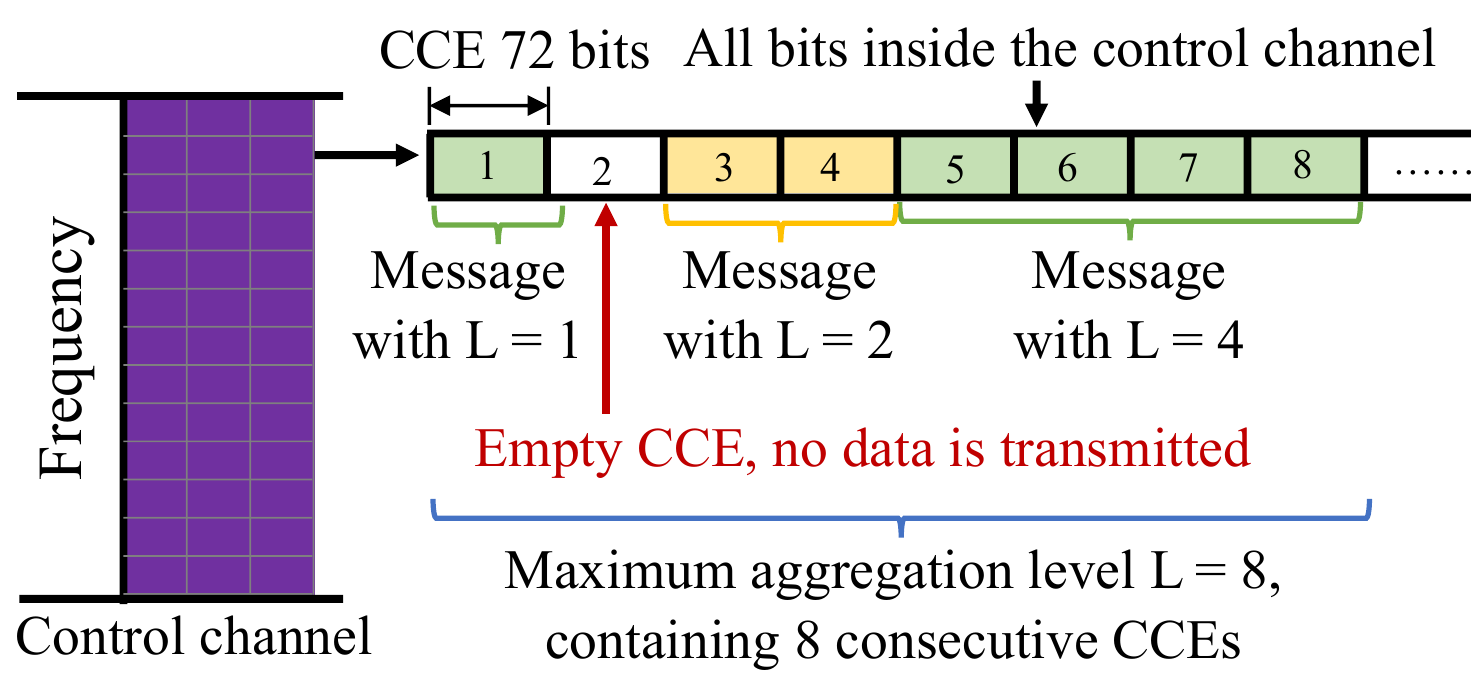}
        \caption{The distribution of control messages inside the control channel. 
        The bits inside the control channel are aligned in order and then grouped into 72\hyp{}bit CCEs.
        Each control message occupies $L$ (aggregation level) consecutive CCEs. }
        \label{fig:dci_cce}
    \end{minipage}
\end{figure*} 

\parahead{Message encoding and decoding}
A control message is a bit\hyp{}string, 
with every single bit or group of bits inside the message representing various control information, 
such as the PRB allocation or the MCS index.
The length of the bit\hyp{}string and the exact information each bit conveys 
depends on the \textit{message format} the base station selects for each message. 
Base station calculates a 16\hyp{}bit CRC based on the control message, 
and then XORs the calculated CRC with another 16\hyp{}bit physical layer ID of a mobile user (which is the receiver of this message), 
\ie, the \textit{cell radio network temporary identifier} (\textsf{\small C-RNTI}),
as shown in Figure~\ref{fig:dci_encode}.
Base station appends the XOR\hyp{}ed value at the end of the control message 
and encodes the message and the appended CRC using convolutional code.


During the decoding process, the appended CRC serves two purposes: message verification and receiver identification.
After convolutional decoding, a mobile device separates the 16\hyp{}bit appended CRC with the control message. 
To verify the correctness of the decoding process, 
the mobile device calculates the CRC using the decoded message 
and then XOR the calculated CRC with its own \textsf{\small C-RNTI}.
When and only when the XOR\hyp{}ed value matches with the appended CRC, 
the decoding is successful and 
and such a mobile device is indeed the intended receiver of the message. 

\parahead{The distribution of control messages inside control channel}
The base station transmits the encoded control messages via the physical control channel, as shown in Figure~\ref{fig:pdsch}.
Even though the number of bits that each control channel contains varies with the bandwidth of the base station,
such a number is much larger than the size of one encoded control message, 
\eg, the control channel of a 20~MHz base station is capable of transmitting a maximum of 84 control messages.
Therefore, to organize the transmission of control messages, 
the base station aligns the bits of the control channel and groups every 72 consecutive bits into a 
\textit{control channel element} (CCE), as shown in Figure~\ref{fig:dci_cce}. 
Each encoded control message occupies $L$ consecutive CCEs, where $L = [1, 2, 4, 8]$ is referred to as the \textit{aggregation level}.
We note that the size of the convolutional encoded control message is smaller than one CCE so a message is repeated multiple times if it occupies $L>1$ CCEs, 
which provides extra redundancy over the convolutional code and thus extra protection over bit errors. 
The base station transmits nothing in empty CCEs that contain no control messages. 
\begin{figure*}[htb]
    \begin{minipage}[htb]{0.59\linewidth}
        \begin{subfigure}[b]{0.47\linewidth}
            \centering
             \includegraphics[width=0.99\linewidth]{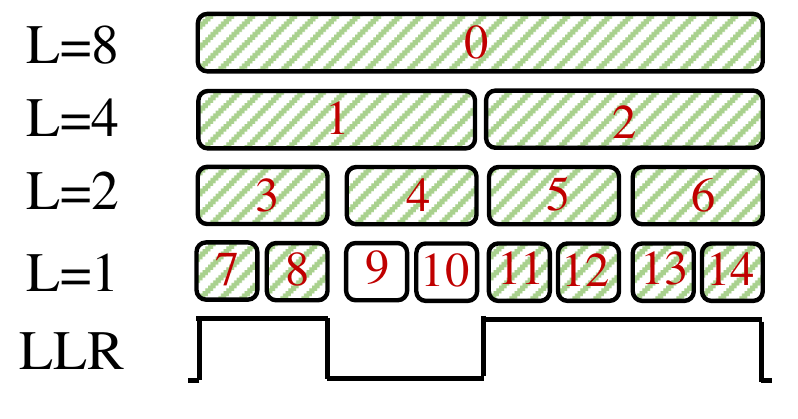}
            \caption{All the candidate message locations with various aggregation level.}
            \label{fig:tree_a}
        \end{subfigure}
        \hfill
        \begin{subfigure}[b]{0.52\linewidth}
            \centering
            \includegraphics[width=0.99\linewidth]{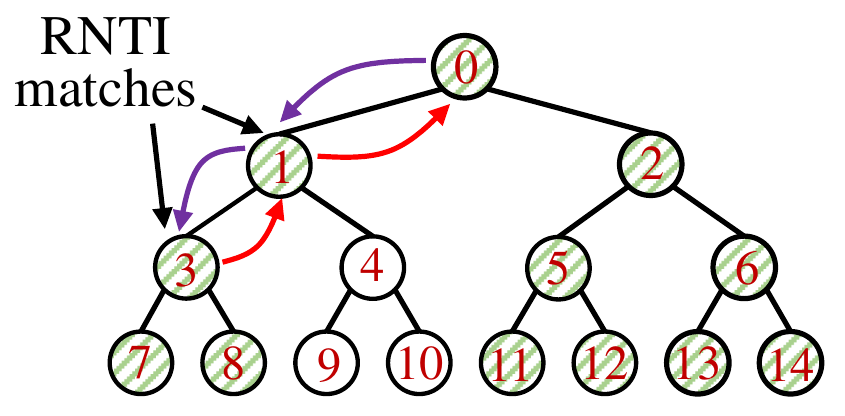}
            \caption{Organizing the search space into the hierarchy of a binary tree.}
            \label{fig:tree_b}
        \end{subfigure}
        \caption{\textbf{(a)} The candidate message locations within each 8\hyp{}CCE segment.
            \textbf{(b)} The search space can be organized into the hierarchy of a binary tree. }
        \label{fig:tree_search}
    \end{minipage}
    \hfill
    \begin{minipage}[htb]{0.38\linewidth}
        \centering
        \includegraphics[width=0.85\linewidth]{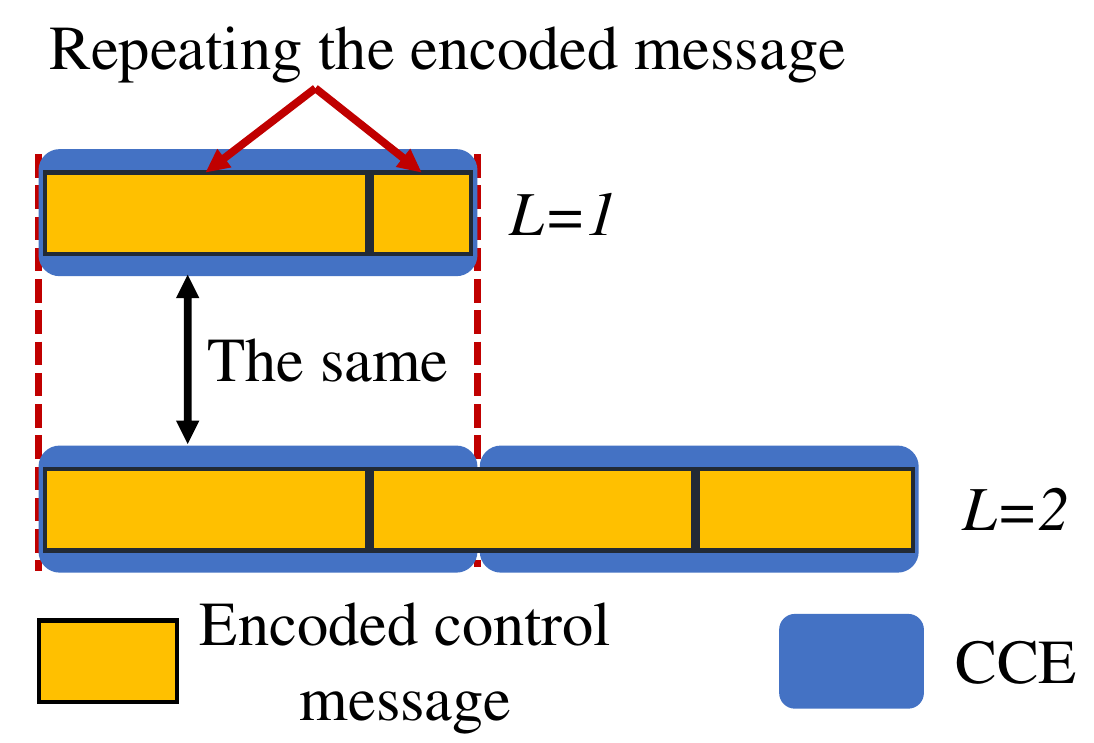}
        \caption{The first half of a message with $L$ is the same as the whole message with $L-1$.} 
        \label{fig:rate_match}
    \end{minipage}
\end{figure*}

\subsubsection{Decoding the whole control channel}
Typically, one mobile user only decodes its own control message, while \systemname{} has to decode the whole control channel
to extract every control message contained inside the control channel.
In this subsection, we introduce the challenges we met when trying to achieve that goal and the techniques we propose to address each challenge.

\parahead{Challenge: the unknowns} The key challenge of designing the decoder is that many important parameters of both the mobile users and the control messages are unknown. 
Firstly, the physical layer ID of mobile users, \ie, the \textsf{\small C-RNTI}, are unknown.
which is required for both message verification and receiver identification, as we have introduced in Section~\S\ref{s:dci_encode}. 
Secondly, the total number of control messages contained and their distribution inside the control channel is unknown. 
At last, for each possible control message, its format and aggregation level are also unknown. 
Exhaustively searching all possible combinations of location, 
aggregation level, and format is a possible solution but results in significant computational overhead.
To address the above challenges, we propose a tree-based search algorithm 
and two message validation and user identification algorithms.

\parahead{Tree-based searching}
The locations of the control message are well-organized. 
Figure~\ref{fig:tree_a} lists all the candidate locations for eight consecutive CCEs segments, 
from which we see that a message with aggregation level $L$ has $2^{4-L}$ possible locations.
The whole control channel is divided into multiple such 8\hyp{}CCE segments. 
We re-organize the search space of every 8\hyp{}CCE segment into a binary tree hierarchy, 
where eight consecutive CCEs represent the eight leaves 
and the root represents the message that aggregates the eight leaves, as shown in Figure~\ref{fig:tree_b}. 
The advantage of using a tree-based search is that once any parent node is decoded successfully, 
we skip searching all of its children nodes since one bit cannot be decoded twice. 

We propose to further prune the tree by identifying empty CCEs.
The received empty CCE contains no data, only Gaussian noise,
which results in significant uncertainty during the demodulation.
Therefore, \systemname{} inspects the confidence of demodulating the bits inside a CCE, 
\ie, the log-likelihood ratios (LLRs), as shown in Figure~\ref{fig:tree_a}, 
and identifies a CCE as empty if the average LLR of the demodulated bits is below a threshold.
\systemname{} identifies empty leaf nodes using LLR and marks a parent node as empty if all of their children are empty.
\systemname{} skips searching all identified empty node.
We note that the majority of CCEs inside the control channel is actually empty in a commercial LTE network
since we have demonstrated in section~\S\ref{s:comp} 
that the base station transmits less than four control messages in more than 99\% subframes. 
Consequently, \systemname{} significantly reduces the computational overhead by identifying and then skipping the empty CCEs.

\parahead{Deriving the \textsf{\small C-RNTI}}
\textsf{\small C-RNTI} is necessary for the message validation and receiver identification.
As we mentioned in Section~\S\ref{s:dci_encode} that 
we are able to separate the message and the appended CRC after decoding,
and then calculate the CRC based on the decoded message, as shown in Figure~\ref{fig:dci_encode}.
We, therefore, derive the \textsf{\small C-RNTI} by XOR\hyp{}ing the calculated CRC and the appended CRC. 



\parahead{Message validation via child-ancestor matching}
We introduce our first technique to verify the decoded message and the derived \textsf{\small C-RNTI}.
Our first observation is that the first half of the message with $L$ is exactly the same as 
the whole message with $L-1$, just as shown in Figure~\ref{fig:rate_match}. 
The fundamental reason behind the observation is that the base station needs to repeat the short encoded message to fill in multiple CCEs, 
as Figure~\ref{fig:rate_match} shows.
According to the observation, we know that we will get exactly the same results when decoding the whole message with $L$ and decoding only its first half. 
We, therefore, traverse the tree shown in Figure~\ref{fig:tree_b} according to an in-order tree traversal.
After a search of one node in the tree, we compare the decoded messages from this node with the messages from its ancestors.
If these two messages include the same decoded payload and derived \textsf{\small C-RNTI}, 
the decoded message is identified as \textit{validated} and the associated \textsf{\small C-RNTI} is the ID of a real mobile user. 
Accordingly, the children of the identified ancestor are removed from the tree.

\begin{figure}[t]
       \centering
        \includegraphics[width=0.7\linewidth]{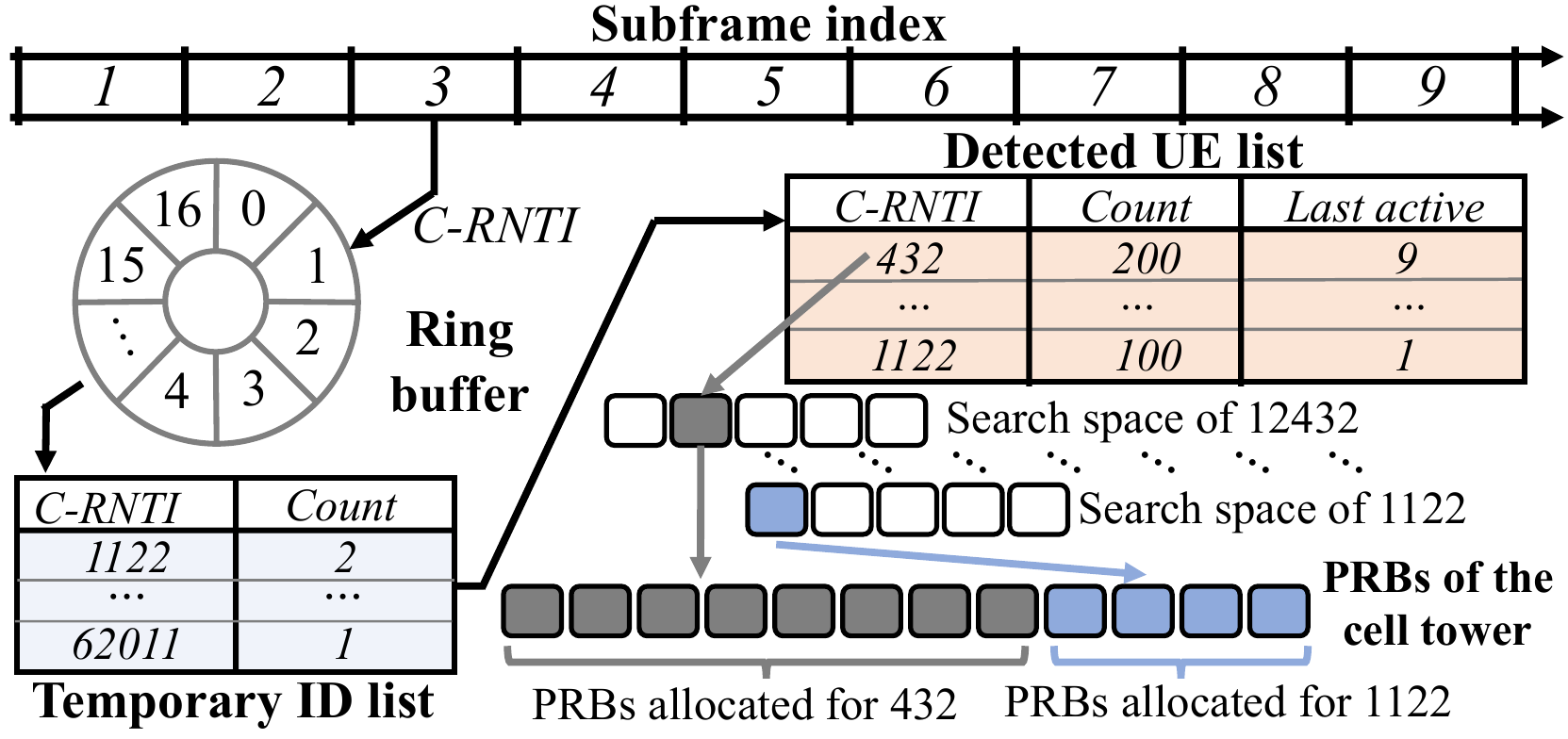}
        \caption{The \systemname{} \textit{UE Tracker}. 
        \systemname{} buffers recent decoding results, 
        maintaining a list tracking
        each \textsf{\small C-RNTI}'s recent prevalence.} 
        \label{fig:ue_tracker}
\end{figure}
\parahead{Message validation via temporal UE tracking}
The above children-ancestor matching algorithm cannot handle messages with low aggregation levels.
We, therefore, design a validation scheme by leveraging the temporal user pattern.
Our key observation is that \textsf{\small C-RNTI}s that reappear within a short 
period of time (16~$ms$ in our implementation) are likely 
real, as \textsf{\small C-RNTI}s calculated using incorrect control message parameters (location, aggregation, and format) are random
and evenly distributed. The possibility of re-hitting the same 
\textsf{\small C-RNTI} generated using wrong parameters can be calculated according to the birthday paradox~\cite{Birthday}.
 As the space of possible \textsf{\small C-RNTI}s is large in size ($2^{16}$), the re-hitting possibility is nonetheless extremely small.  
For example, the possibility that two \textsf{\small C-RNTI}s reappear within 1,000 messages are $1\times10^{-4}$.


We design a \textit{UE Tracker} to first store the derived 
\textsf{\small C-RNTI}s together with the corresponding control messages and track the ID of real UE. 
The tracker uses a ring buffer to buffer the messages from the 16 most recent subframes, and
maintains a \textit{temporary ID list} that stores the count of appearances of 
each \textsf{\small C-RNTI} in the ring buffer, 
as shown in Figure~\ref{fig:ue_tracker}.
If the \textsf{\small C-RNTI} count is larger than 
two, we identify it as the ID of a real UE and move it to the 
\textit{detected UE} list.
The tracker searches the ring buffer for messages associated with
\textsf{\small C-RNTI}s in the UE list, which  identified as 
valid control messages. 
We record the \textit{last active time} of each \textsf{\small C-RNTI} 
(the index of the most recent subframe that this \textsf{\small C-RNTI} is observed). 
A \textsf{\small C-RNTI} that is inactive for 10 seconds is removed from the list.

To reduce the number of buffered messages, we filter out messages with 
a large number of coded bit errors by re-encoding each decoded control message
into coded bits, then comparing the result with the original coded bits 
inside the received CCEs to calculate the ratio of 
bits that are erroneous. We drop messages with more than 25\% code bits flipped,
which is a very high threshold that filters out extremely noisy messages and misses almost no true positive messages, while allowing
false positives through, which however are not identified as the ID of real UEs by our UE Tracker.



\parahead{UE tracking with carrier aggregation} 
When carrier aggregation is active, each aggregated cell tower transmits its 
own control messages to a UE through its own respective control channel.
A UE has only one physical layer ID (\textsf{\small C-RNTI}), and so
the \textsf{\small C-RNTI} of a UE with carrier aggregation enabled
appears in the control messages transmitted by all aggregated cells. 
\systemname{} extracts the intersected \textsf{\small C-RNTI}s across the UE lists of all cells, 
which are identified as having carrier aggregation enabled.
\systemname{} identifies the primary and secondary cells of the UE according to the time the ID appears in each aggregated cell, 
\ie, the earliest cell that the \textsf{\small C-RNTI} appears in is identified as the primary cell, the second earliest cell is identified as a secondary cell, and so on so forth. 

 \begin{figure}[htb]
    \begin{subfigure}[b]{\linewidth}
    \centering
    \includegraphics[width=0.78\linewidth]{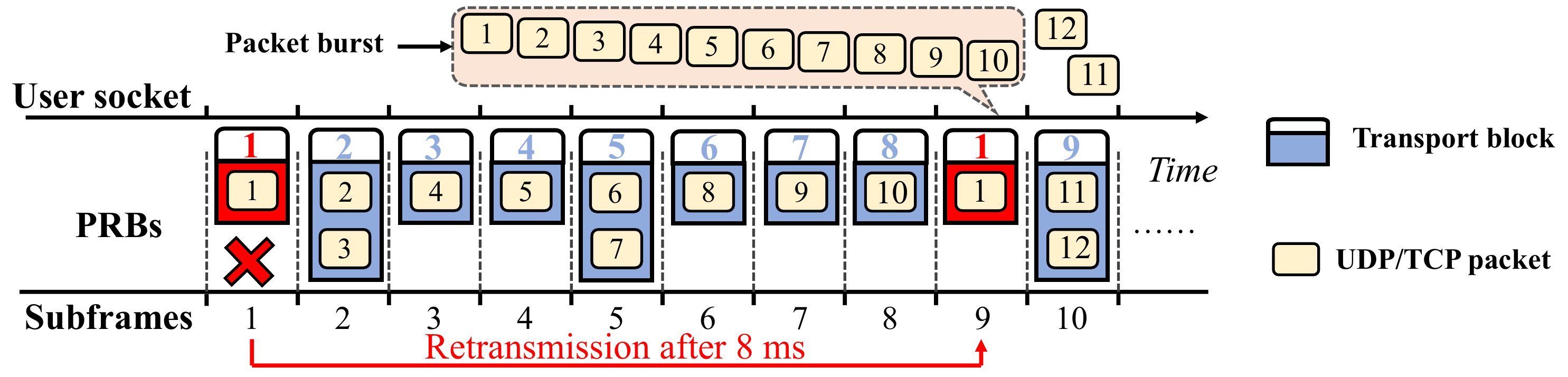}
    \caption{TB loss and reception; packet delivery at the socket.}
    \label{fig:tb_err_burst}
    \end{subfigure}
    
    \begin{subfigure}[b]{\linewidth}
    \centering
    \includegraphics[width=0.78\linewidth]{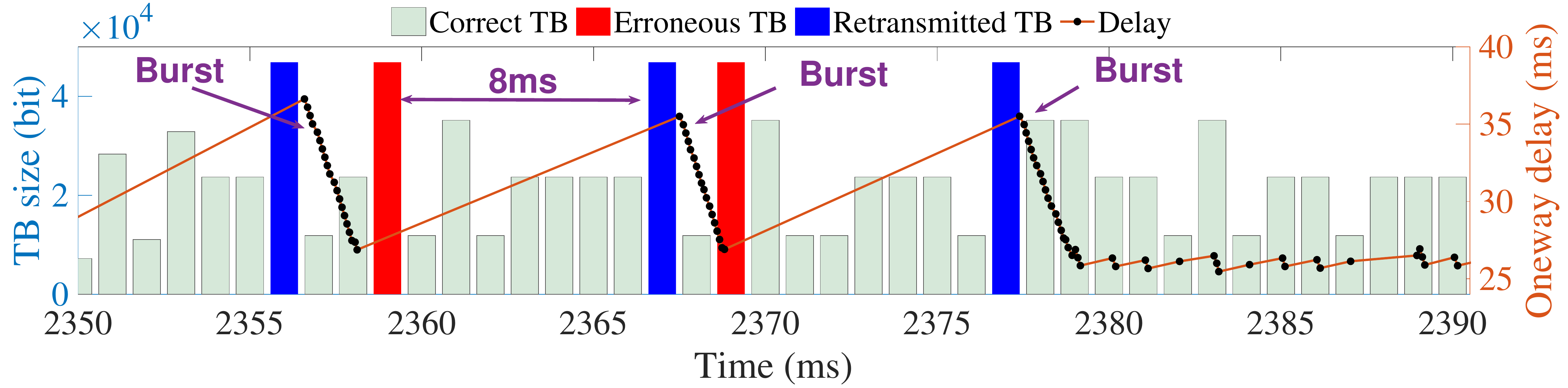}
    \caption{Received TB size (bits) and one\hyp{}way delay.}
    \label{fig:tb_reTx_delay}
    \end{subfigure}
    \caption{Packet loss and retransmission interacts with
    LTE's HARQ design to cause a packet arrival delay of eight milliseconds
    followed by a burst of packets.}
    \label{fig:tb_delay_combined}
\end{figure}

\subsection{Data Fusion at Fusion Layer}
\label{s:fusion}

The transport layer timestamps each received packet.
Subframe index naturally provides a one-millisecond resolution timestamp for the control messages contained inside it.
The timestamps of control messages and packet arrivals, however, are unsynchronized related to each other.
Therefore, 
\systemname{} must synchronize and align these two cross-layer information before fusing them.

We leverage the packet arrival pattern of 
retransmissions to eliminate the shift between timestamps.
Figure~\ref{fig:tb_err_burst} shows an example of TB failure and retransmission,
where the UE fails to decode a TB in Subframe~1 that contains one transport layer packet. 
Eight milliseconds after the original transmission, 
the cell tower retransmits the lost TB, in Subframe~9.  
However, as introduced in Section~\S\ref{s:lte_primer}, 
the base station continues transmitting data to the UE 
between the original transmission and the retransmission.  
Assuming the TBs sent over subframes two to eight are decoded correctly, 
the UE receives the second to tenth packets before 
the reception of the first packet, causing many out-of-order packet receptions.
To prevent out-of-order packet reception from happening at a higher layer, 
the LTE link-layer ensures in-order TB delivery by 
buffering all out-of-order TBs in a \textit{reorder buffer} 
until the retransmission recovers the missing TB(s). 
The link-layer then extracts packets from the TBs and delivers them to the 
receiving socket, resulting in a burst of packets received at that layer. 
Furthermore, Figure~\ref{fig:tb_reTx_delay} shows that the
burst also reshapes the inter-packet-arrival time. 
The packets inside the burst in fact arrive at the UE uniformly 
but are instead reported together, creating a large interval between the 
burst and the packets received before the burst.

From the decoded message, 
we identify the starting and ending subframe for each 
retransmission, as shown in Figure~\ref{fig:tb_err_burst}.
From the packet log, we detect retransmission based on 
the eight-millisecond interval between consecutive packets and the packet burst.
We, therefore, shift the timestamp of the packet log to match the locations of 
the retransmissions inside the packet log and the control messages.
By doing so, we synchronize the packet log with control messages to 
millisecond precision, the highest resolution possible since
the subframe index is at the same scale.

\section{Implementation}
\label{s:imp}

Our design is a pure software solution, involving no hardware modifications, 
so we believe that it can be implemented on a commercial mobile phone 
by customizing the cellular firmware. 
The source code of the cellular firmware, however, 
is proprietary to cellular equipment manufacturers, and is not accessible. 
We, therefore, implement an open\hyp{}source proof\hyp{}of\hyp{}concept prototype. 
To emulate multiple radio chains of the commercial phones, 
we use multiple off\hyp{}the\hyp{}shelf software-defined radios, (the USRP~\cite{USRP}) as a front\hyp{}end to collect cellular signals.
The signals the USRP collects are sent to the connected PC for control message decoding.
We use a mobile phone that is tethered to the PC for cellular data transmission and reception. 
Packet statistics and the decoded control messages are fused by the middle layer at the PC. 


\begin{figure}[htb]
    \begin{minipage}[htb]{0.49\linewidth}
        \centering
        \includegraphics[width=0.99\linewidth]{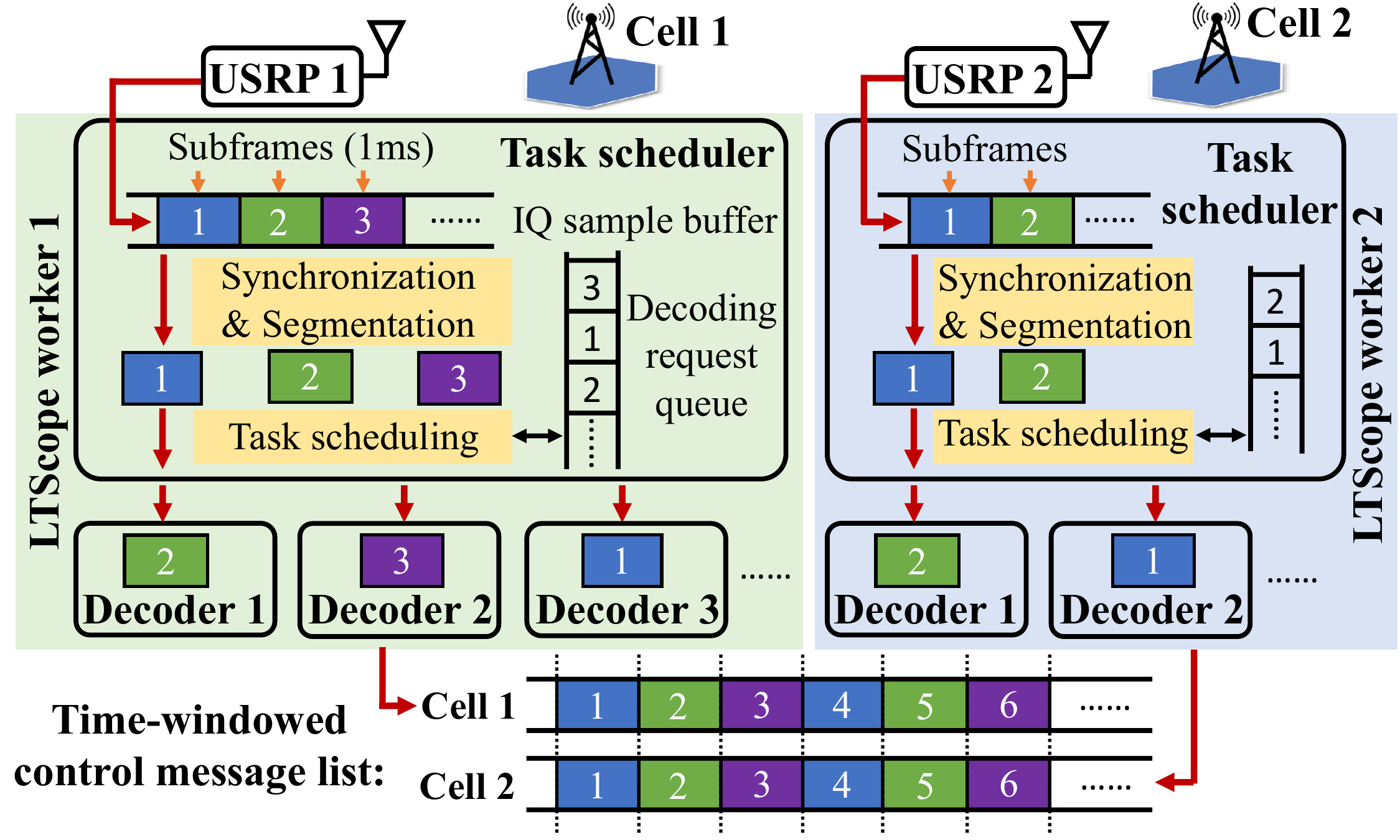}
        \caption{\textit{\systemnames{} parallel decoding framework.} \systemname{} starts multiple
        workers in parallel, each of which decodes the control channel of one cell. 
        }
        \label{fig:multi_thread}
    \end{minipage} 
    \begin{minipage}[htb]{0.49\linewidth}
        \begin{subfigure}[htb]{0.49\linewidth}
            \centering
            \includegraphics[width=0.99\linewidth]{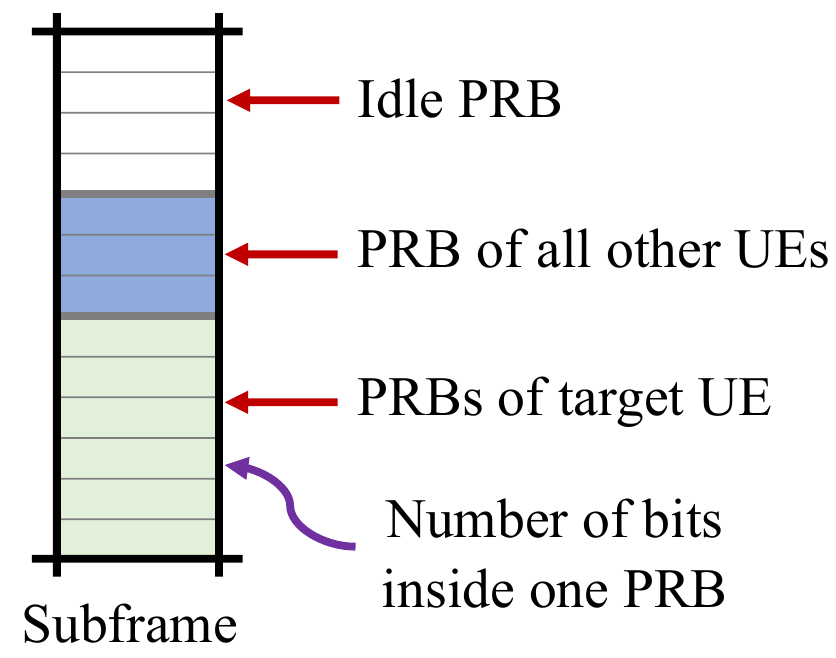}
            \caption{Calculating capacity at a single subframe.}
            \label{fig:cap_single_sub}
        \end{subfigure}
       \hfill
        \begin{subfigure}[htb]{0.49\linewidth}
            \centering
            \includegraphics[width=0.99\linewidth]{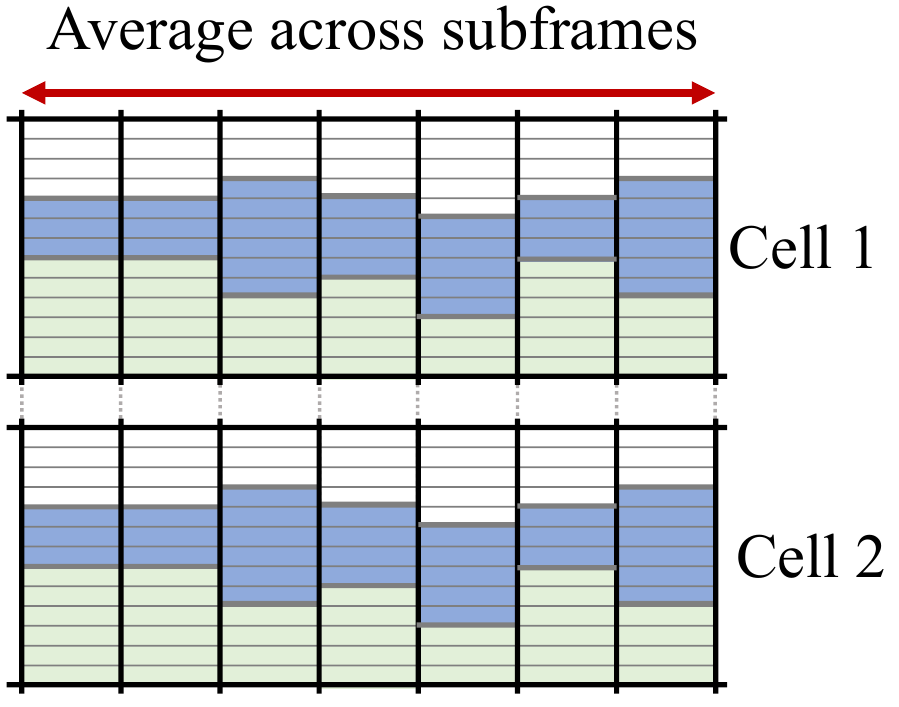}
            \caption{Smoothing capacity estimation over subframes.}
            \label{fig:cap_across_sub}
        \end{subfigure}
        \caption{Capacity calculation based on information of single subframe~\textbf{(a)};
            Smoothing the capacity estimation over subframes and aggregates the capacity across cells if CA is enabled~\textbf{(b)}.}
        \label{fig:cap_update}
    \end{minipage} 
\end{figure}


We implement a parallel decoding framework as shown in Figure~\ref{fig:multi_thread}. 
We start one \textit{\systemname{} worker} to handle the signal one USRP streams.
Each worker starts one \textit{task scheduler} thread to synchronize with the cell tower and separate the received signal into subframes. 
The worker also creates multiple \textit{control message decoder} threads, each of which takes the signal of one subframe as input and decodes all control messages inside the subframe. 
\systemname{} starts more than one decoder threads to guarantee that no matter 
when a subframe has been separated by the task scheduler, 
there is a decoder ready to decode it, so the worker buffers a minimum number of subframes,
avoiding overflowing the limited socket buffer shared by multiple USRPs. 
When one decoder thread finishes its decoding task and becomes idle, it sends a decoding request to the task scheduler.
The task scheduler allocates decoding tasks according to the order of the decoding requests it receives.
The decoded control messages from one worker are ordered according to the index of its subframe and stored in a list.
The control message lists of multiple workers are aligned using the subframe index, as shown in Figure~\ref{fig:multi_thread}. 
We only store the control messages decoded from the most recent 320 subframes.

\parahead{Capacity estimation}
\systemname{} estimates and updates the capacity at the frequency of every one millisecond. 
After decoding all the control messages of one subframe, 
\systemname{} knows the PRBs the base station allocations for the target UE and 
the total PRBs allocated for all other UEs, just as shown in Figure~\ref{fig:cap_single_sub}.
\systemname{} calculates the available PRB for target UE 
as the allocated PRBs for the UE plus the idle PRBs that are not allocated. 
The base station also tells the UE the number of bits that each PRB can carry, via the control message.  
Therefore, \systemname{} calculates the capacity of the target UE 
as the available PRB multiplies the number of bits inside each PRB. 
To smooth the estimation, \systemname{} averages the available PRB and the number of bits each PRB carries 
across multiple decoded subframes in the past, as shown in Figure~\ref{fig:cap_across_sub}. 
If CA is enabled, \systemname{} obtains the overall capacity of the UE by summarizing the capacities of all aggregated cells. 

\section{Evaluation}
\label{s:eval}

Our performance evaluation quantifies \systemnames{} accuracy and responsiveness in measuring cell load 
in a head\hyp{}to\hyp{}head comparison with OWL (\S\ref{s:eval:accuracy}).  
We then demonstrate \systemnames{} ability to provide extremely high-granular 
and accurate cell load estimation in the presence of bit rate adaptation and carrier aggregation, 
for both stationary and mobile user scenarios (\S\ref{s:eval_capEst}).  
We then look at carrier aggregation in-depth, demonstrating unique
insights into its operation that \systemname{} enables
for the first time (\S\ref{s:eval_CA}).
At last, we compare with CLAW~(\S\ref{s:CLAW}) and BurstTracker~(\S\ref{s:burstTracker}), 
two systems built atop of MobileInsight, to demonstrate the advantage of complete cell-wide information \systemname{} decodes 
from control channel.

\begin{table}[htb]
	\centering
	\begin{tabularx}{\linewidth}{@{}l@{\hspace{5pt}}X@{}|@{}l@{\hspace{5pt}}X@{}|@{}l@{\hspace{5pt}}X@{}}
		\toprule
		\textbf{\#} & \textbf{Configuration} & \textbf{\#} & \textbf{Configuration} & \textbf{\#} & \textbf{Configuration}\\
		\hline
		1  & 20~MHz, 1.94~GHz, 2 ant.& 2 &  10~MHz, 739~MHz, 2 ant. & 3  & 10~MHz, 723~MHz, 2 ant. \\
		
		4  &  10~MHz, 2.36~GHz, 4 ant. & 5 &  5~MHz, 872~MHz, 4 ant. & 6  &5~MHz, 1.95~GHz, 4 ant.\\ 
		\bottomrule
	\end{tabularx}
	\caption{Evaluation \textit{cell tower configurations}: 
	frequency bandwidth, center frequency and antenna count.}
	\label{t:cell_config}
	\vspace{-0.4cm}
\end{table}

\parahead{Experimental configuration} 
We experiment with
the six AT\&T cell towers that provide LTE service
for the testing area, a university area near a busy
street. The detailed configurations of all six cell towers are listed in Table~\ref{t:cell_config}.
The index of each cell is used to refer to that cell in later sections.

\subsection{Decoding Accuracy}
\label{s:eval:accuracy}\label{s:micro}
In this section, we investigate \systemnames{} decoding accuracy. We compare head\hyp{}to\hyp{}head with OWL~\cite{OWL}, which has demonstrated superior performance over LTEye~\cite{LTEye}. 

\parahead{Methodology}
Without hacking the cell tower, we cannot get the exact ground truth of control messages 
the cell tower sends in the control channel of each subframe.
To infer the ground truth, we set up four USRPs to listen to the same cell tower 
at four locations that are one meter apart from each other 
and apply \systemname{} to decode the received signal.
Since the signal received by these four USRPs are uncorrelated, 
\systemnames{} decoding results using signals from different USRPs are independent of each other.
Therefore, messages that appear in the decoding results 
of all four USRPs are highly likely to be correct
and thus are treated as the ground truth of control messages that the cell tower sends.
At each location, we repeat the decoding using OWL. 
We also repeat the entire experiment in 20 combinations of URRP location.
\begin{figure}[htb]
   \centering
    \begin{subfigure}[b]{0.49\linewidth}
        \centering
        \includegraphics[width=0.85\textwidth]{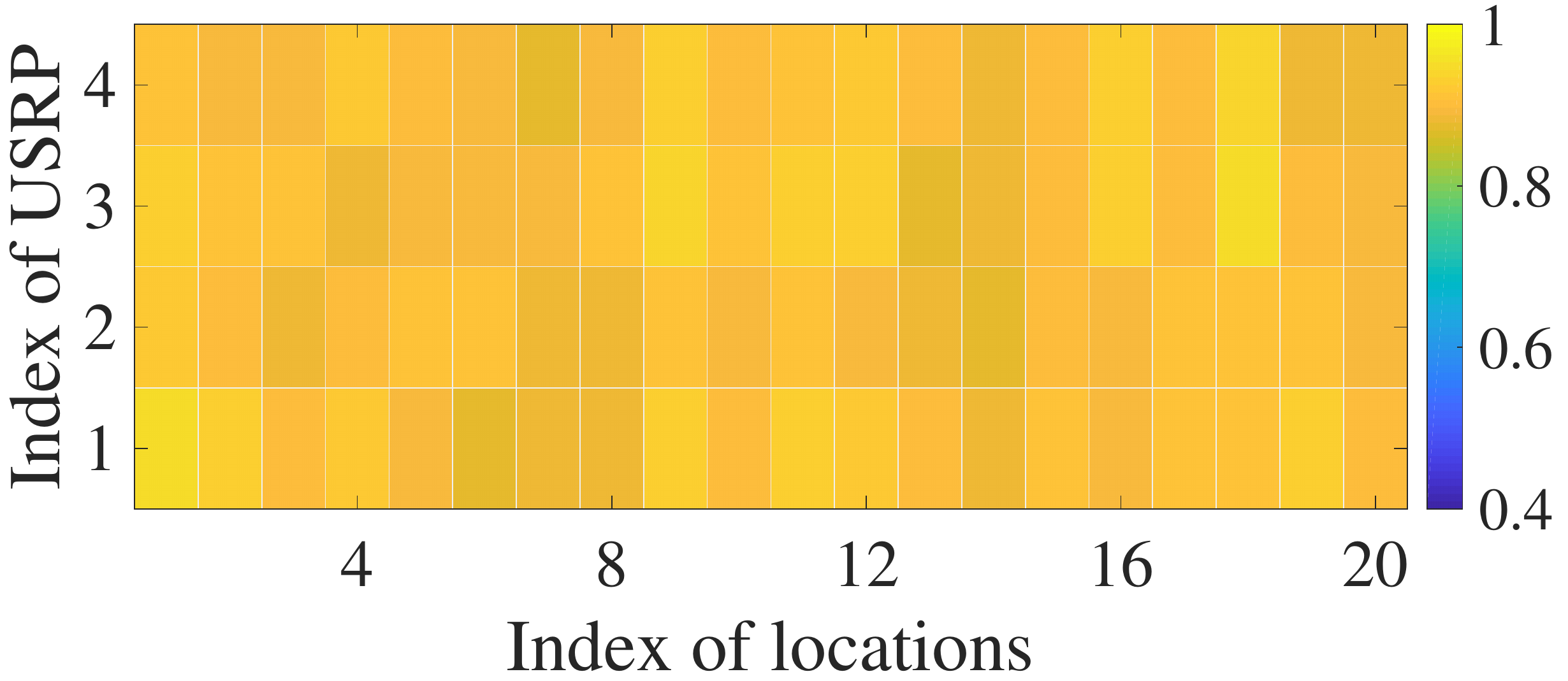}
        \caption{Messages decoded by \systemname{}}
        \label{fig:acc_ltscope}
    \end{subfigure}
    \hfill
    \begin{subfigure}[b]{0.49\linewidth}
        \centering
        \includegraphics[width=0.85\textwidth]{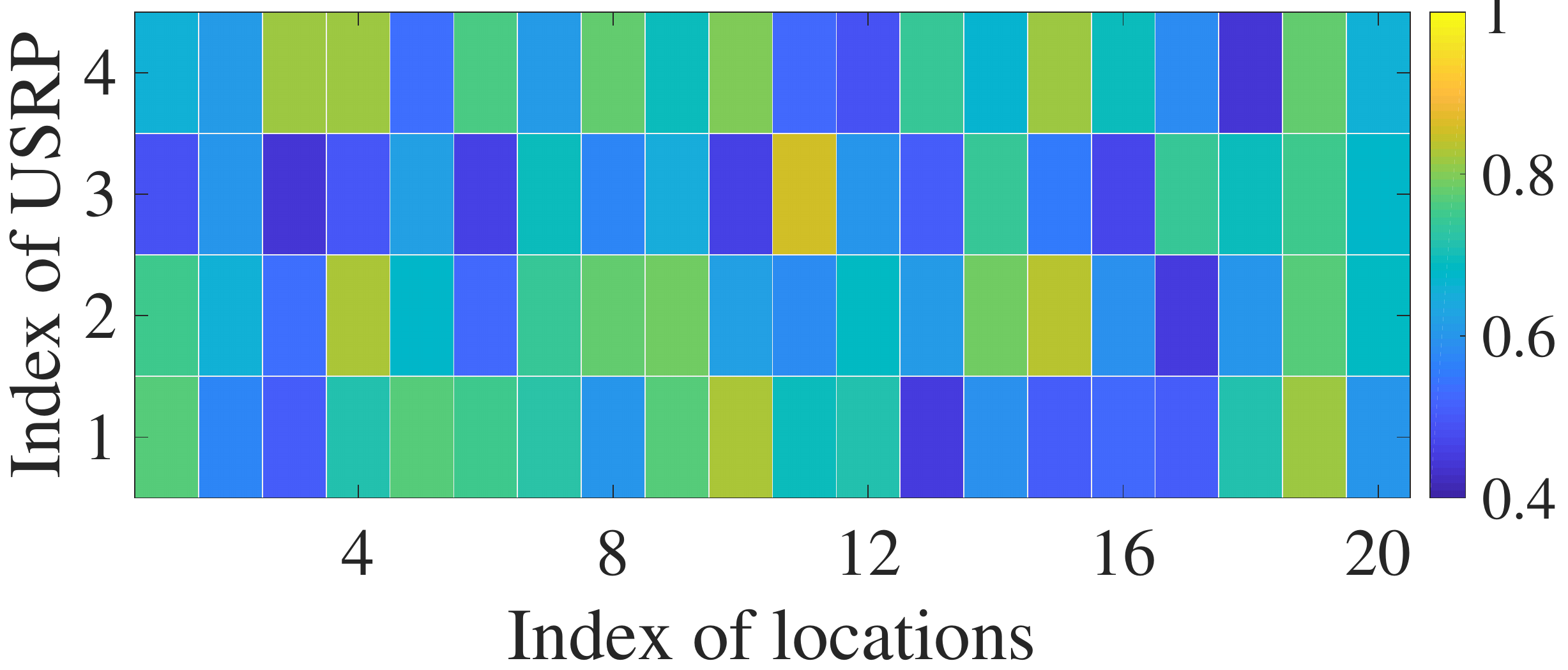}
        \caption{Messages decoded by OWL.}
        \label{fig:acc_owl}
    \end{subfigure}
    \caption{The percentage of control messages that are correctly decoded by \systemname{}~(\textbf{a}) and OWL~(\textbf{b}). We repeat the experiments using four USRPs, across 20 combinations of location.}
    \label{fig:accuracy}
\end{figure}

We calculate the percentage of correctly decoded messages in the decoding results of each USRP 
and plot the results in Figure~\ref{fig:accuracy}.
\systemnames{} decoding performance is much more stable and accurate than OWL across locations.
On average, 90.4\% of messages decoded by \systemname{} are correct, while the percentage is 65.3\% for OWL.
The \textit{true negative}--messages that are missed on any USRP,
and \textit{false positive}--messages that are decoded with error and not filtered out, 
are two main sources of the incorrectly decoded messages. 
Since we do not have the exact ground truth of every message the base station sends, 
we cannot get the exact ratio of the true negatives and false positives. 
We, therefore, conduct the following two experiments to infer these two ratios of \systemname{} and OWL. 

\begin{figure}[htb]
    \centering
    \begin{subfigure}[b]{0.48\linewidth}
        \centering
        \includegraphics[width=0.97\textwidth]{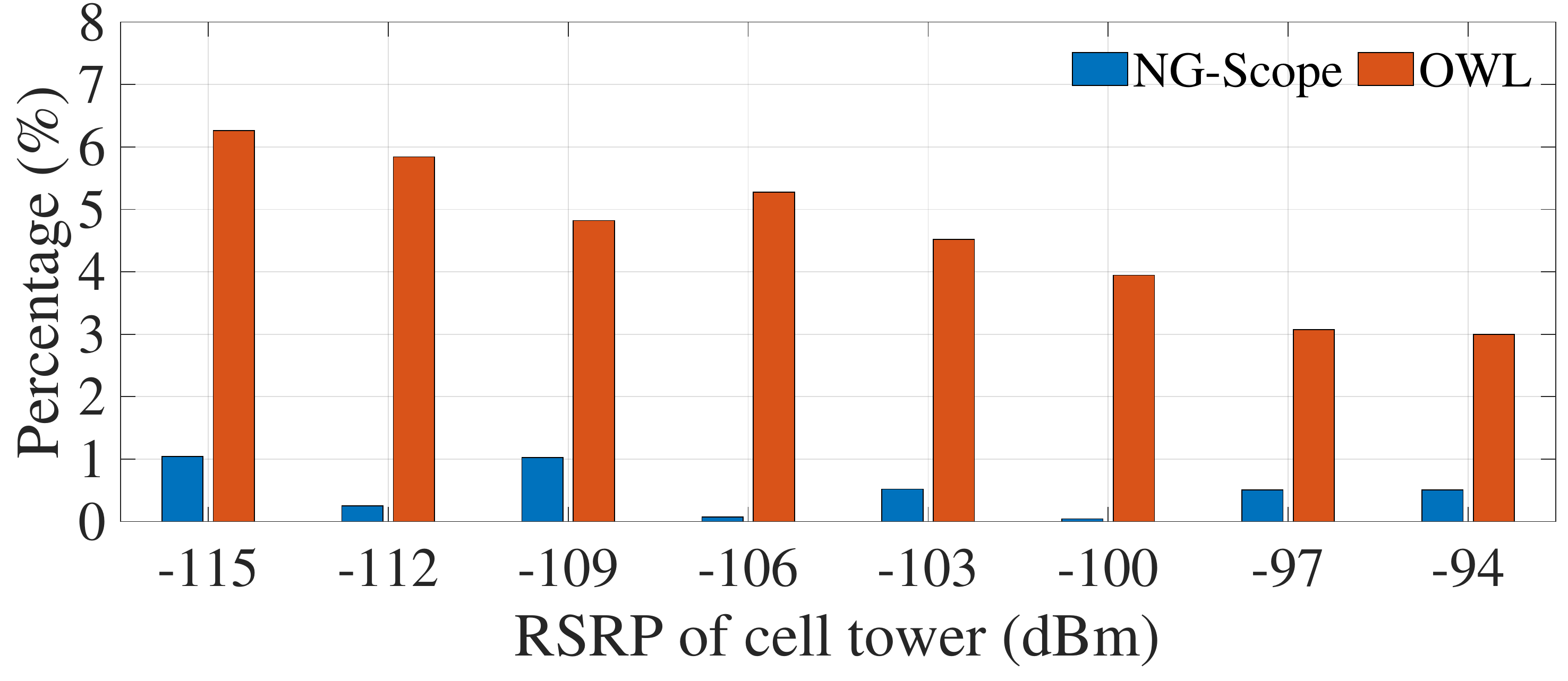}
        \caption{Missing data.}
        \label{fig:micro_data}
    \end{subfigure}
    \hfill
    \begin{subfigure}[b]{0.49\linewidth}
        \centering
        \includegraphics[width=0.97\textwidth]{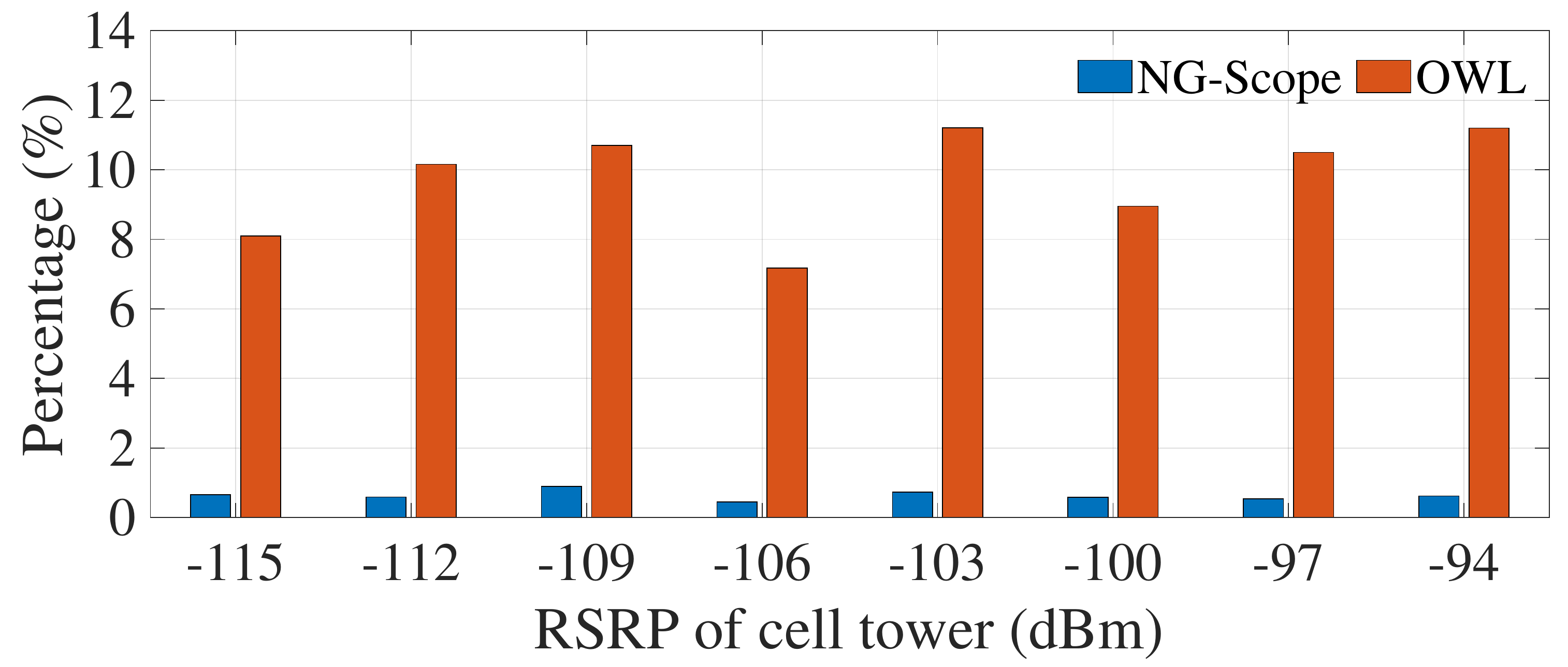}
        \caption{False positives.}
        \label{fig:micro_TB_rate}
    \end{subfigure}
    \caption{Precision and recall of collected packet data~\textbf{(a)} and generated false positives~\textbf{(b)}.}
    \label{fig:Micro}
\end{figure}
We send 300~Mbit data from the server to a Galaxy S8. 
We record the amount of data received from the socket of the phone 
and count the amount of data delivered from cell towers to this phone 
from the control messages decoded using \systemname{} and OWL. 
We repeat the experiment at eight locations varying \textit{reference signal received power} (RSRP).
We compare the size of data we derive from the control channel with the data received from the socket 
to get the percentage of data that is missing from the control channel
and plot the result in Figure~\ref{fig:Micro}(a), 
We see that \systemname{} captures almost all the UDP traffic sent from the server, 
with an average missing percentage of 0.83\%, a reduction of 83\% compare to OWL's 4.5\%.
Missing control messages results in missing data,
so \systemname{} has a smaller percentage of true negative in its decoding results.

We also quantify the generated false positives, 
a false control message, and thus a nonexistent resource allocation. 
The total allocated PRBs of one subframe calculated by summing up the allocated PRBs of every control message 
may exceed the maximum number of PRBs the cell supports, due to false positives. 
We calculate the percentage of such subframes to infer the generated false positives and plot the results in Figure~\ref{fig:Micro}(b). 
We see that OWL generates a large number of false positives (8\%-12\%), 
so that it may overestimate the overall usage of the whole cell. 
\systemname{} reduces the false positives to the greatest extent possible via its message validation schemes.

\begin{figure}[htb]
    \begin{minipage}[htb]{0.49\linewidth}
       \begin{subfigure}[htb]{0.49\linewidth}
            \centering
            \includegraphics[width=0.99\linewidth]{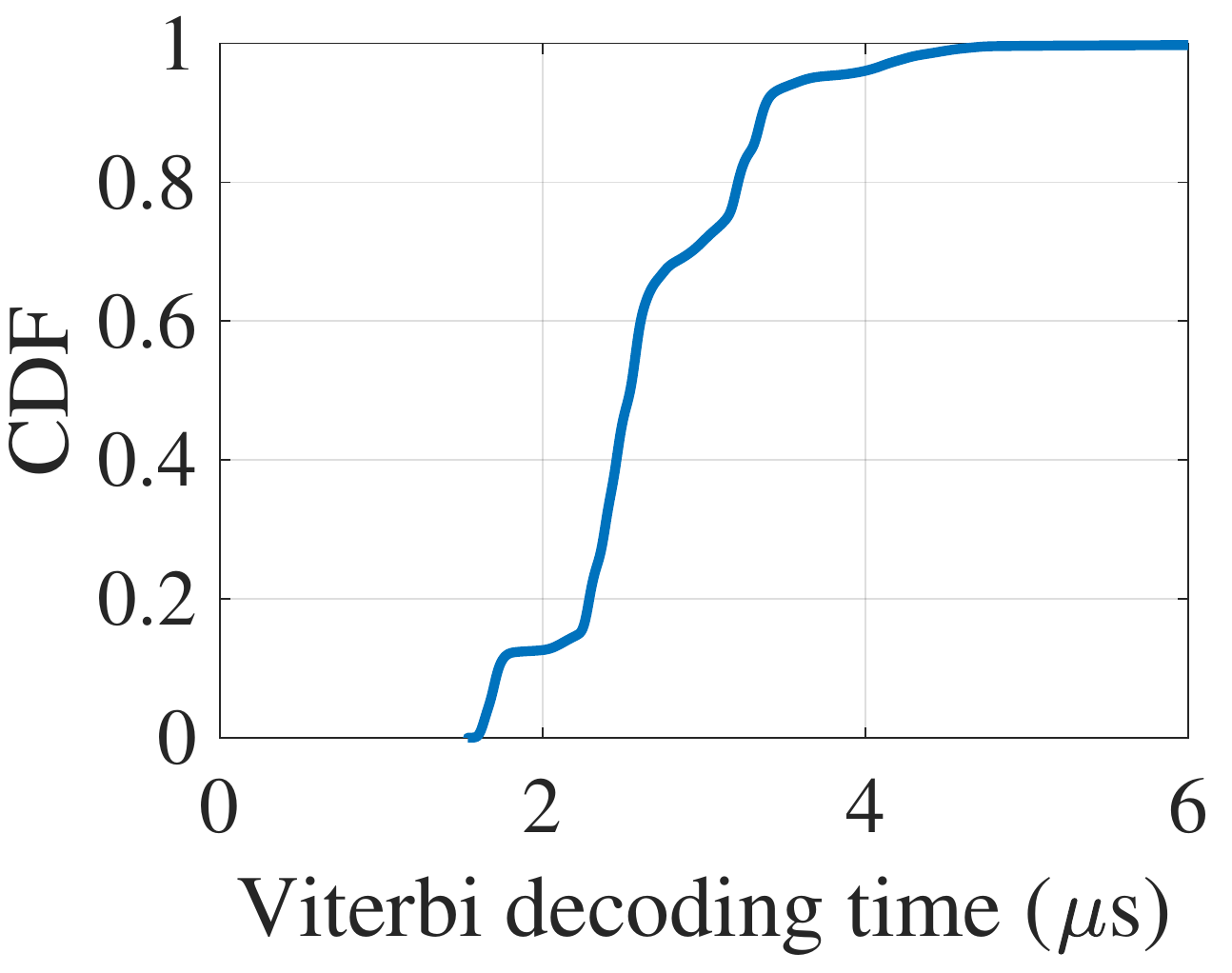}
            \caption{The CDF of the time each Viterbi decoding consumes.}
            \label{fig:decode_time}
        \end{subfigure}
        \begin{subfigure}[htb]{0.48\linewidth}
            \centering
            \includegraphics[width=0.99\linewidth]{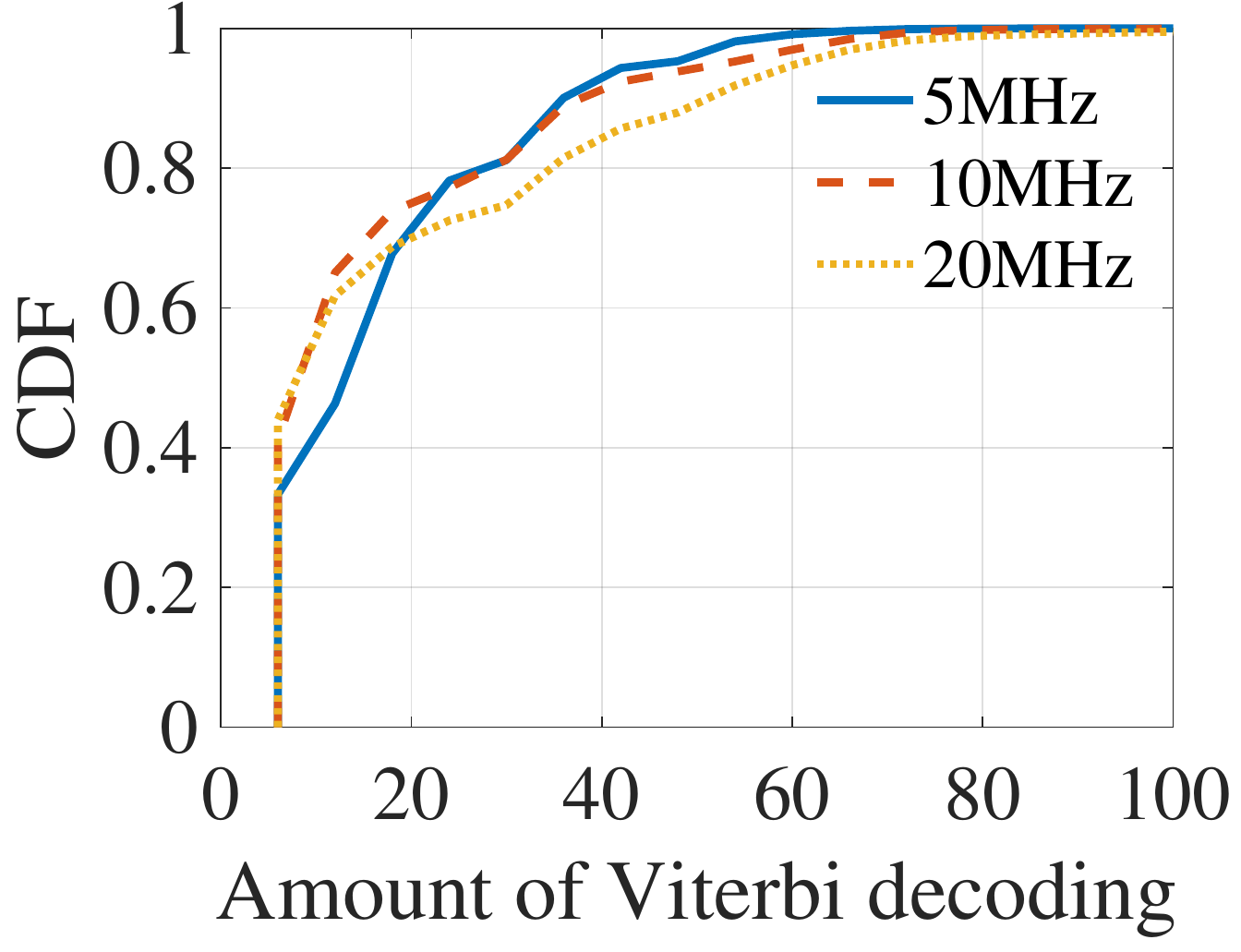}
            \caption{The CDF of the number of Viterbi decoding performed.}
            \label{fig:nof_decode}
        \end{subfigure}
        \caption{The time the Viterbi decoder takes to decode one control message \textbf{(a).}
            The number of Viterbi decoding \systemname{} performs to decode all control message inside each subframe \textbf{(b)}.}
    \end{minipage}
    \hfill
    \begin{minipage}[htb]{0.49\linewidth}
       \begin{subfigure}[htb]{0.49\linewidth}
            \centering
            \includegraphics[width=0.99\linewidth]{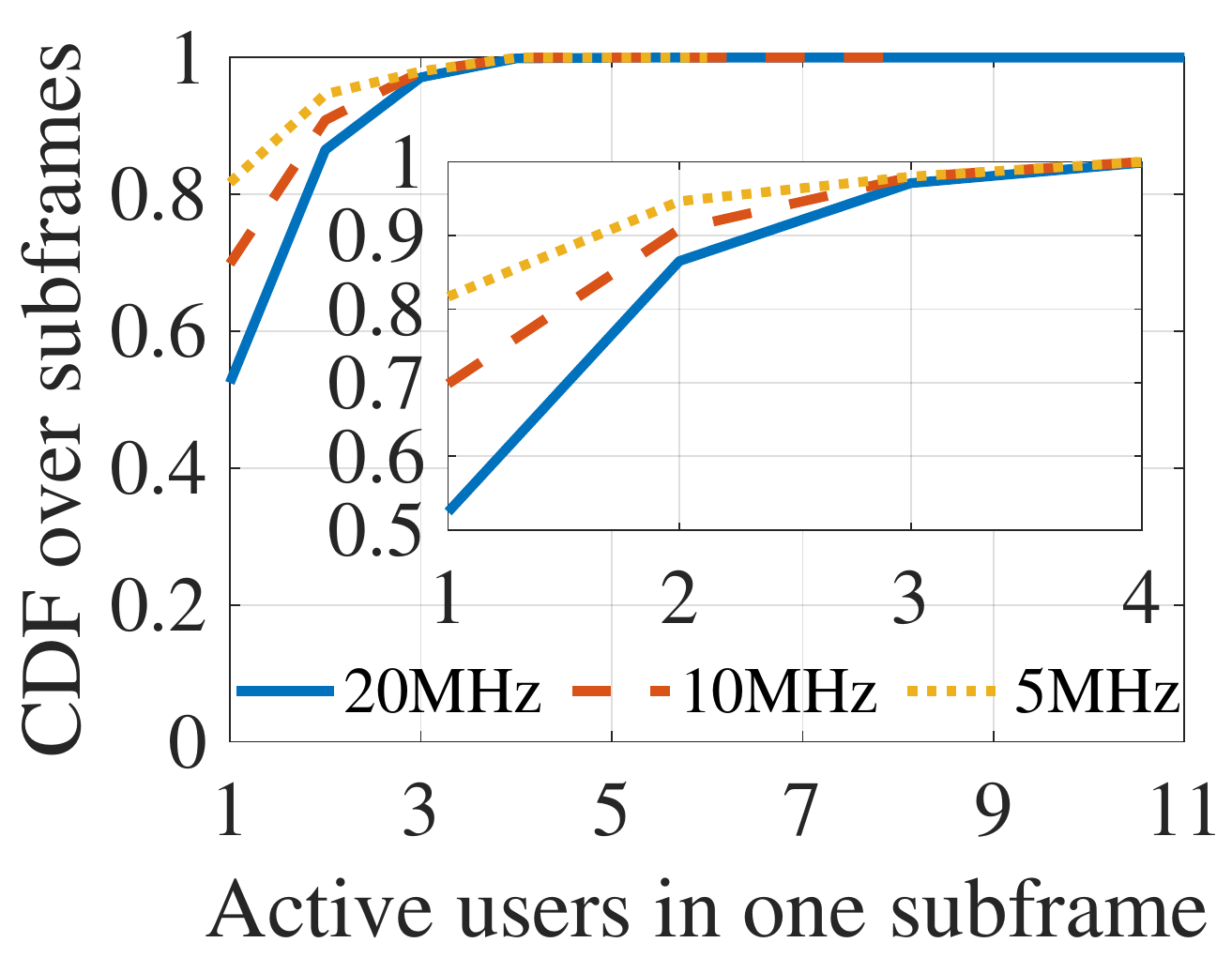}
            \caption{The CDF of the number of active users in one subframe.}
            \label{fig:comp_ueFreq}
        \end{subfigure}
       \hfill
        \begin{subfigure}[htb]{0.49\linewidth}
            \centering
            \includegraphics[width=0.99\linewidth]{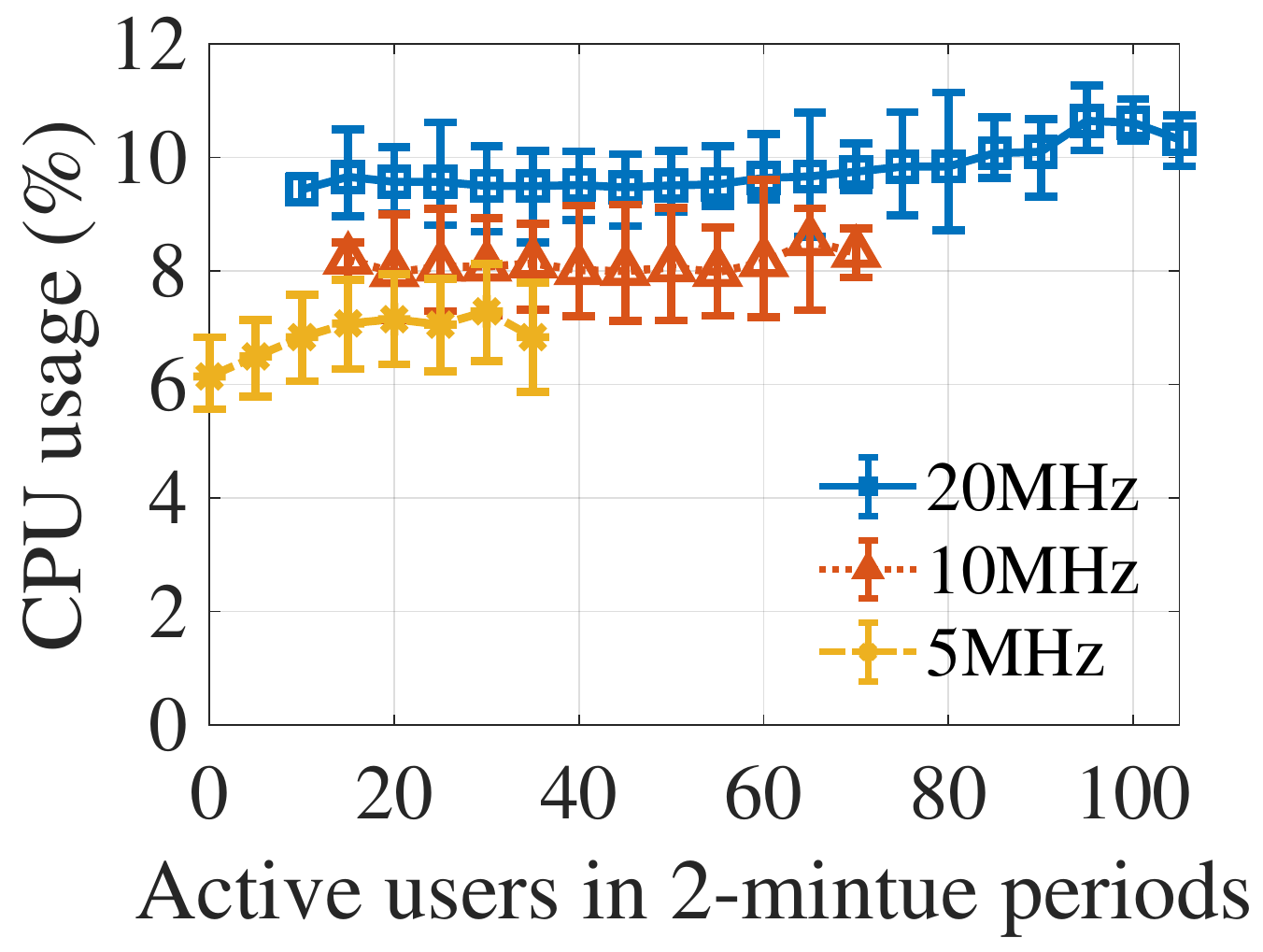}
            \caption{The number of active users in two-minute time window.}
            \label{fig:comp_cpu}
        \end{subfigure}
        \caption{The number of active users inside the subframes~(\textbf{a});
        the number of user the base station serves within a 2\hyp{}minute interval
        and the CPU usage of \systemname{} during each interval~(\textbf{b}).}
        \label{fig:comp_cost}
    \end{minipage}
\end{figure}

\subsection{Computational Cost}\label{s:comp}
In this section, 
we investigate the computational cost of \systemname{}. 
The basic operation that \systemname{} performs is the convolutional decoding of one control message. 
Therefore, the computation of each convolutional decoding and 
the total number of decoding \systemname{} conducts in each subframe determines its computational costs.
\systemname{} use the software\hyp{}implemented Viterbi decoder. 
Figure~\ref{fig:decode_time} plots the distribution of the time such a Viterbi decoder takes 
to decode 3.8 million control messages transmitted by a 20~MHz base station, 
using an Intel i7-8700 CPU, from which we see that the time is smaller than 4.51~$\mu$s for 99\% of cases.
We note that the hardware Viterbi decoder the mobile phone leverages for convolutional decoding is much faster than our software version.

We plot the number of convolutional decoding attempts \systemname{} performs inside each subframe, 
when decoding the control channel of a 5, 10, and 20~MHz base station,
in Figure~\ref{fig:nof_decode}.
We could observe that \systemname{} performs less than 80 convolutional decoding attempts
inside 99\% subframes of all three base stations.
We note that even though a mobile phone only needs to decode its own control message, 
the phone still needs to blindly perform multiple rounds of convolutional decoding attempts
as the mobile phone does not know all the parameters that are required to decode its control message. 
The mobile phone even does not know whether the base station transmits a control message for it or not, 
before the blind decoding. 
According to the standard~\cite{TS213}, the maximum decoding attempts each mobile phone needs to perform is 44.
Therefore, \systemname{} merely introduces reasonable extra computational cost, 
\ie, around 1$\times$ more convolutional decoding attempts compared with a legacy mobile phone. 

\parahead{The number of active users inside each subframe}
The fundamental reason that \systemname{} only introduces limited extra computational overhead 
is that the number of active users inside each subframe is limited.
By active we mean that the base station allocates bandwidth for one user in one subframe 
and thus transmits one corresponding control message to that user.
We derive the number of active users in each subframe by counting the number of decoded control messages and 
plot the distribution of the number of active users inside every subframe of a base station in 72 hours in Figure~\ref{fig:comp_ueFreq}.
We clearly see that there are less than four active users and thus less than four control messages in 99.9\% of the subframes for all three base stations.
Due to the limited amount of control messages to be transmitted,
there exist a large number of empty CCEs inside the control channel and thus most of the nodes of the tree in Figure~\ref{fig:tree_b} are empty, 
significantly reducing the search space.

\parahead{The mobile users the base station serves}
Even though the base station transmits data to a limited number of mobile devices inside each subframe, 
the base station serves tens or even hundreds of mobile devices on a longer time scale.
To demonstrate that, we divide the 72 hours into 4,320 two\hyp{}minute periods 
and plot the number of uses that the base station serves in Figure~\ref{fig:comp_cpu}, 
from which we see that a 20~MHz base stations talks to 100 different mobile devices within 2 minutes. 
We also plot the CPU usage of \systemname{} during each interval. 
We could see that \systemnames{} computational costs do not increase proportionally 
with the number of active users the base station serves in a longer period.

\subsection{Highly-granular Capacity Tracking using \systemname{}}\label{s:eval_capEst}
In this section, we evaluate \systemnames{} highly-granular capacity tracking. We begin with the evaluation of a single cell followed by the aggregated capacity with CA triggered. 
\begin{figure}[tb]
    \includegraphics[width=0.65\linewidth]{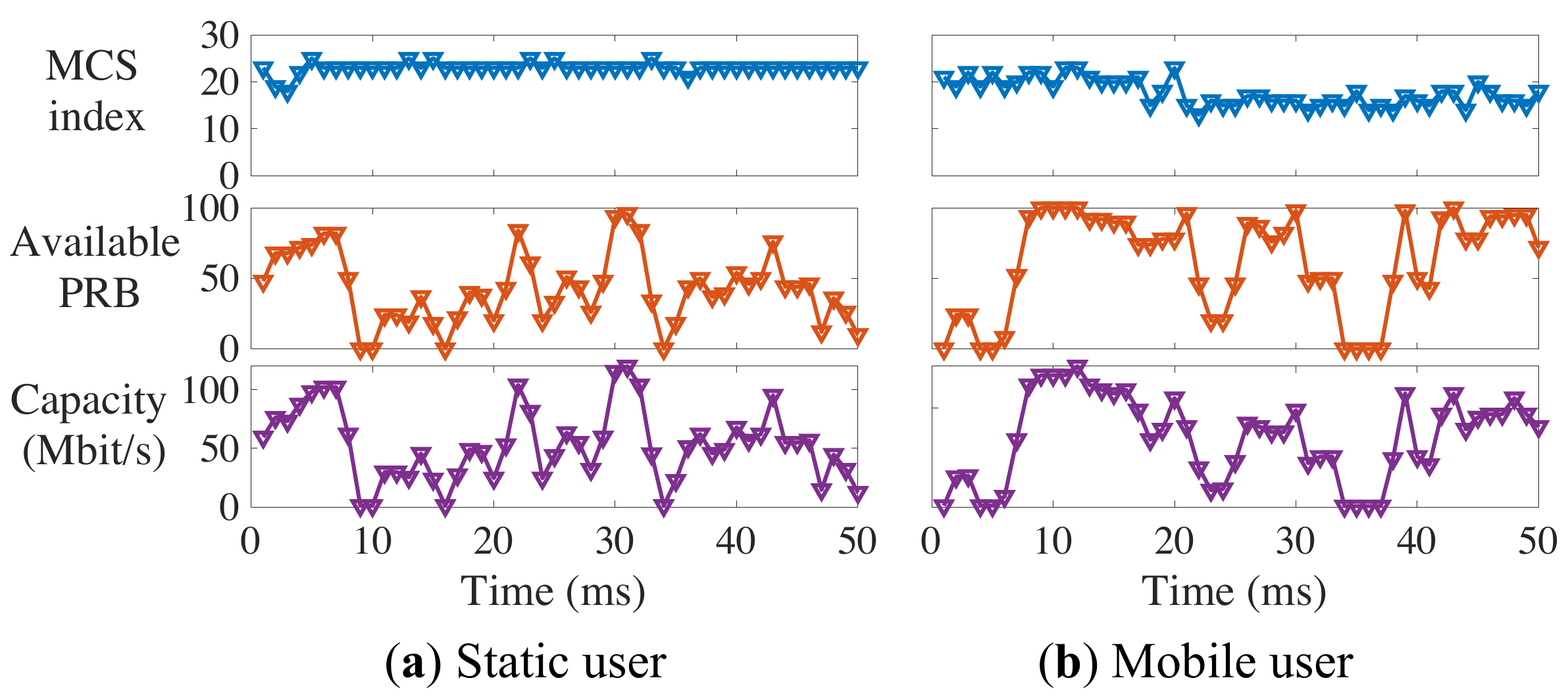}
    \caption{The estimated wireless link capacity, the decoded modulation 
        and coding rate, and the calculated available PRBs for a static user (\textbf{a}) 
        and mobile user (\textbf{b}).}
    \label{fig:CapEst_all}
\end{figure}

\parahead{Single cell capacity} 
\systemname{} tracks the LTE wireless link capacity based on the available PRB and the bits each PRB carries for a certain user.
The bits each PRB carriers for one use is calculated based on the MCS index and the number spatial stream inside the control message. 
We plot the MCS index that the one cell selects for the 
UE (only one spatial stream), the available PRBs,
and the calculated LTE link capacity for the static UE at a location with average RSRP $-98$~dBm, in Figure~\ref{fig:CapEst_all}(a). 
We see that the selected MCS is stable, while the available PRBs of the cell changes dramatically---the capacity varies accordingly.  
As a comparison, we also plot the same statistics for a mobile UE, in Figure~\ref{fig:CapEst_all}(b). 
We see that, for the mobile user, both the MCS index and available PRBs fluctuate, but at different time scales. 



\begin{figure}[htb]
    \centering
    \begin{minipage}[h]{0.7\linewidth}
        \centering
        \includegraphics[width=0.99\linewidth]{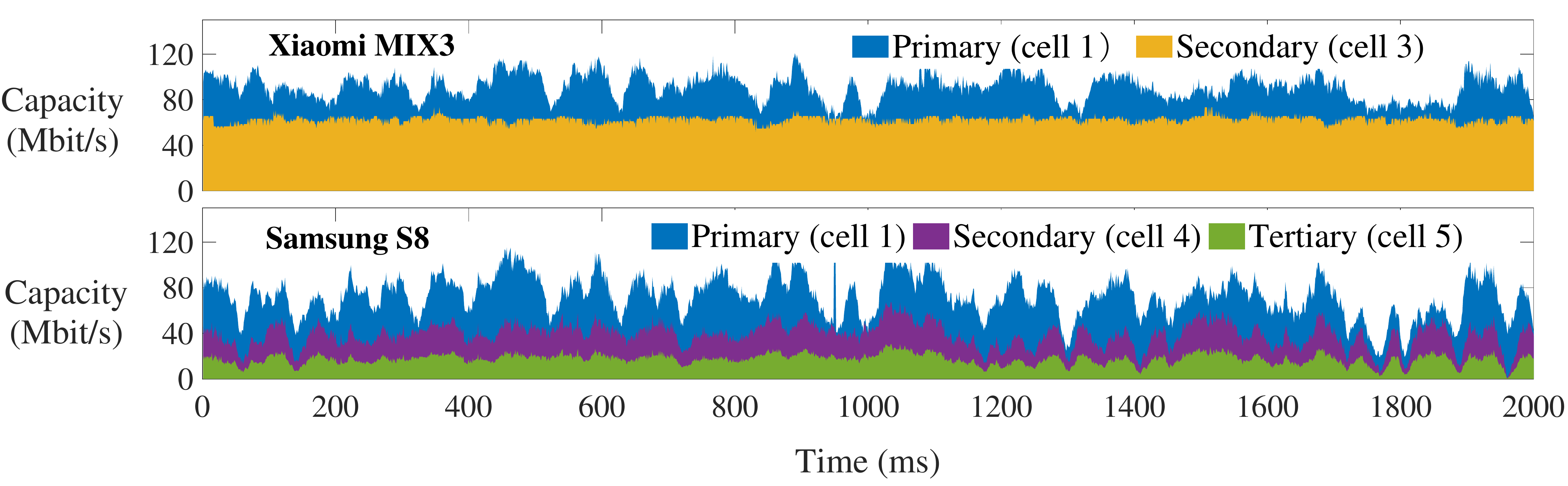}
        \caption{Aggregated capacities of Xiaomi MIX3 (two cells) and Samsung S8 (three cells). Curves are stacked top 
        to bottom in the order they appear in the legend.}
        \label{fig:CapEst_CA}
    \end{minipage}
    \hfill
    \begin{minipage}[h]{0.29\linewidth}
        \centering
        \includegraphics[width=0.99\linewidth]{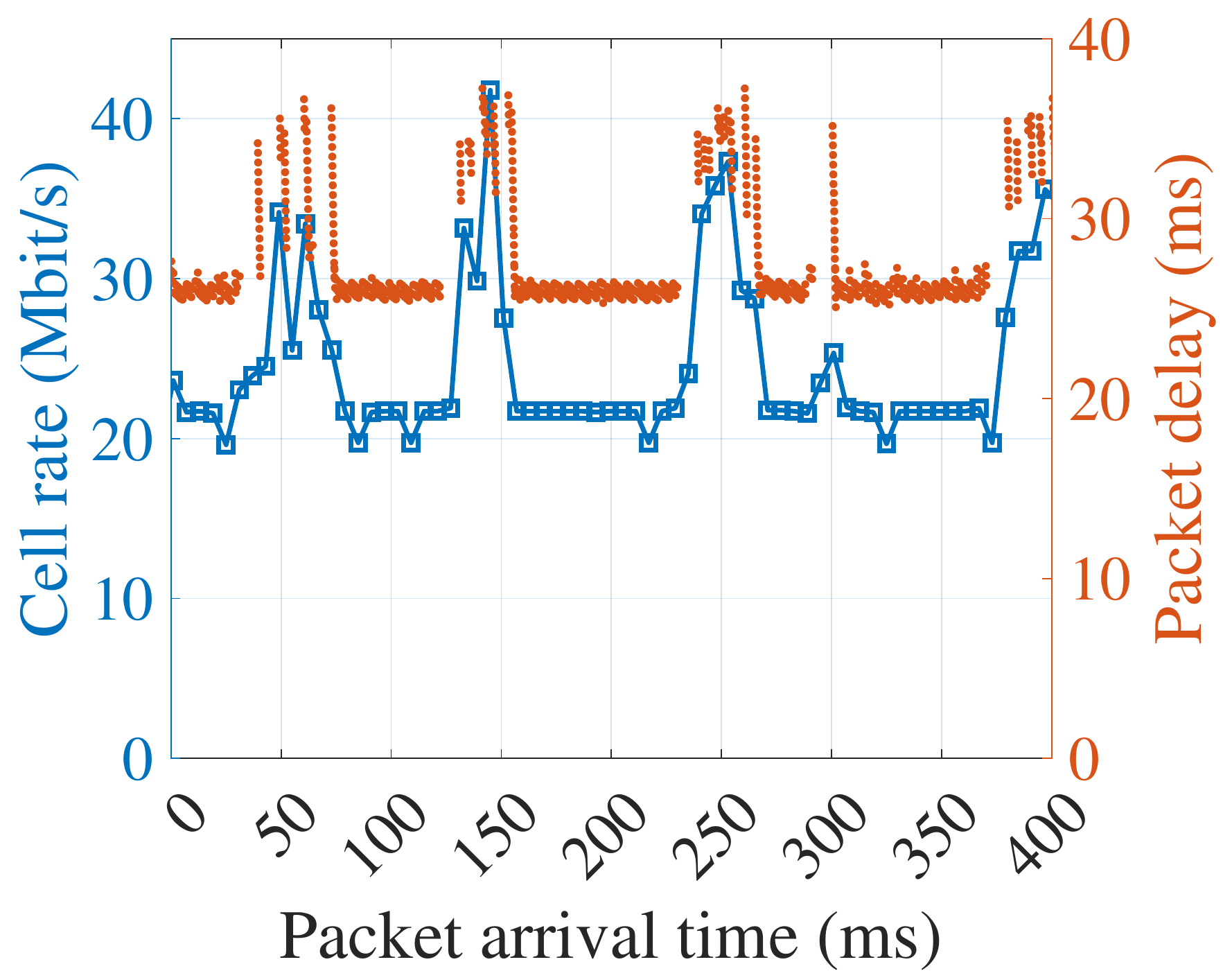}
        \caption{The LTE physical layer data rate increases with retransmission.}
        \label{fig:cell_rate_inc}
    \end{minipage}
\end{figure}

\parahead{Aggregated capacity}
We investigate the aggregated capacity for a certain user with CA triggered. 
We test with two mobile phones---a Samsung Galaxy S8 and a Xiaomi MIX3.
The MIX3 is aggregated with two cells.
We plot the aggregated capacity for MIX3 in Figure~\ref{fig:CapEst_CA} (\textit{upper}), from which, 
we see that the capacity provided by the secondary cell is large and stable, 
while the capacity provided by the primary cell varies significantly. 
The reason is that the secondary cell operates on LTE band 29, 
which has only a downlink frequency 
(generally, one FDD band is divided into uplink and downlink frequency blocks), 
and so no UE associates with it. 
This cell is installed purely for carrier aggregation, \ie to work as a secondary cell for downloading data to the UE, so it is idle for most of the time. 
The S8 is aggregated with three cells. 
We plot the aggregated capacity for S8 in Figure~\ref{fig:CapEst_CA} (\textit{lower}), 
from which, we see that the capacities provided by three aggregated have large variations. 
The aggregated capacity varies accordingly.


\parahead{Impact of TB errors and its retransmissions}
We note that not all physical layer capacity is used for transmitting data. 
When TB error happens, the cell tower needs to allocate bandwidth for both retransmissions of the erroneous TB and ongoing data transmission, resulting in an increase of the instantaneous data rate at the LTE physical layer. To demonstrate this, we plot the LTE physical layer data calculated from the control message and the one-way delay recorded at UE, with a server offered load of 20~Mbit/s. The packet bursts and eight-millisecond interval in one way delay tell us where the retransmission happens, as we have discussed in Section~\ref{s:fusion}. We observe that even without retransmissions, the LTE physical layer rate is higher than the offered load (20~Mbit/s) due to protocol overhead (LTE protocol headers). When retransmissions happen, the instantaneous PHY rate increases significantly. The maximum rate can be larger than 40~Mbit/s in this example, which is double the offered load.
Since the total PHY capacity of each cell is bounded by the bandwidth, the increasing allocation of capacity for retransmission affects the available capacity for original data transmission, resulting in dynamics of the final available capacity at the transport layer.

\subsection{Congestion control using \systemname{}}\label{s:congestion}
We implemented a \systemname{} based congestion control algorithm 
that fully leverages the highly-granular capacity reported by \systemname{},  
achieving high throughput and at the same time low latency. 
By default, our congestion control algorithm sets its sending rate to the capacity reported by \systemname{}.
Such a rate causes no congestion when the bottleneck of the TCP connection is at the cellular link. 
A congestion is detected when the bottleneck shifts from cellular link to the Internet and 
accordingly our algorithm falls back to a CUBIC algorithm to match its sending rate to the capacity of the Internet link.  
Our algorithm identifies the bottleneck has shifted back to the cellular link 
if the rate selected by CUBIC is the same or larger than the capacity reported by \systemname{}.

\begin{figure}[!t]
    \centering
    \begin{minipage}[b]{0.49\linewidth}
        \centering
        \includegraphics[width=0.86\textwidth]{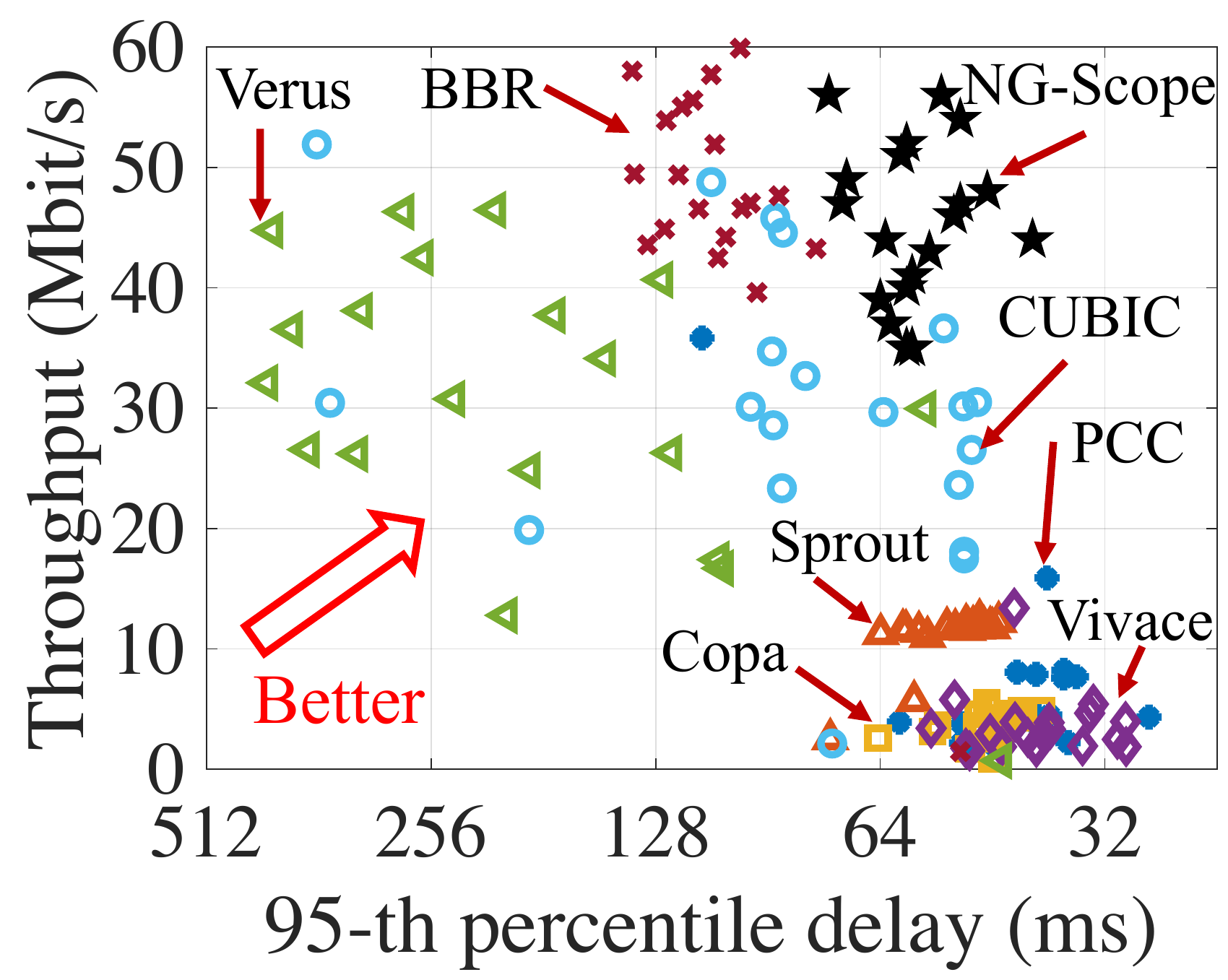}
        \caption{The average achieved throughput and the $95^{th}$ percentile oneway delay of eight congestion control algorithms.
        }
        \label{fig:thput_delay}
    \end{minipage}
    \hfill
    \begin{minipage}[b]{0.49\linewidth}
        \centering
        \includegraphics[width=0.97\textwidth]{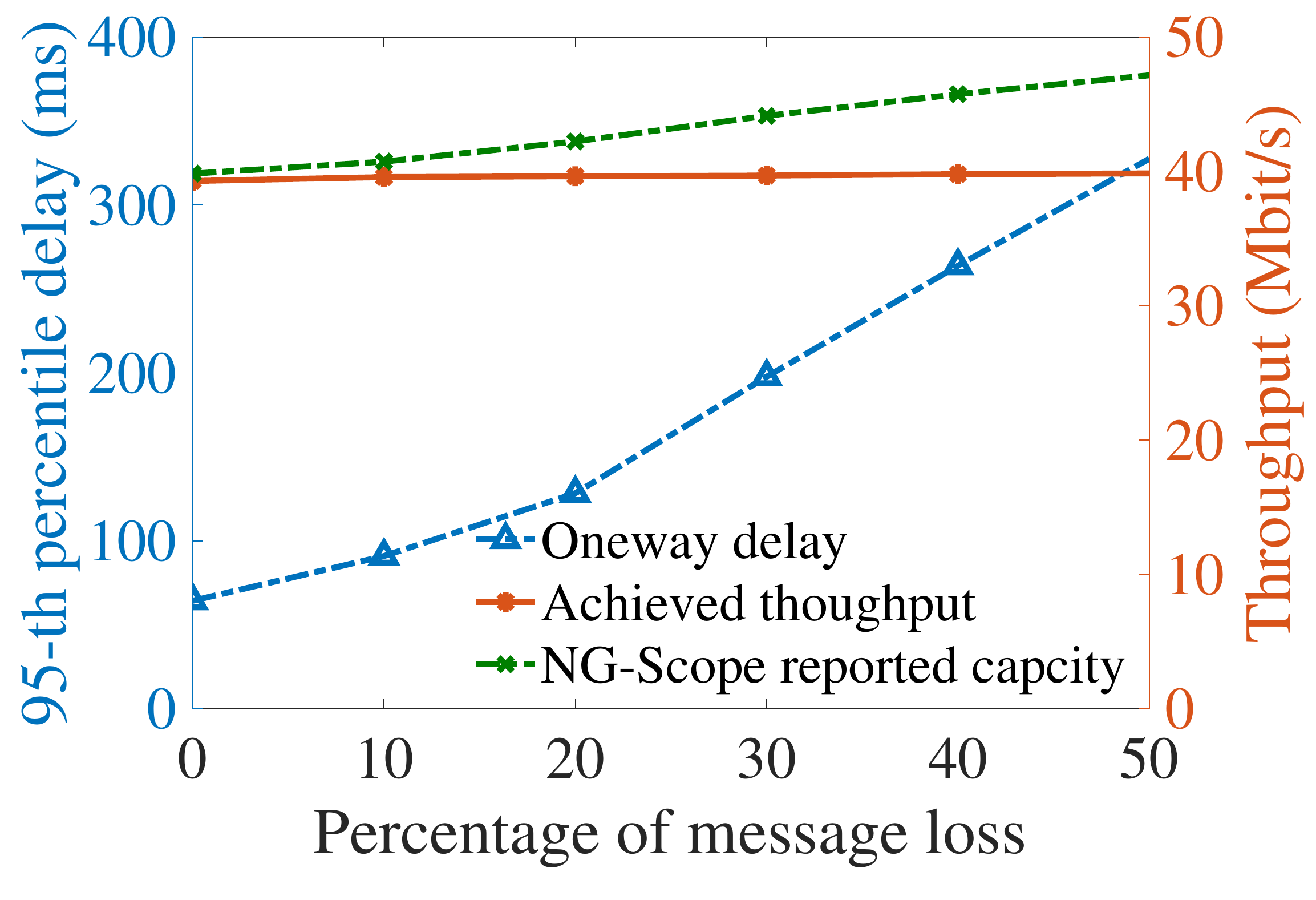}
        \caption{The average achieved throughput and $95^{th}$ percentile oneway delay with a varying number of missing control messages.}
        \label{fig:message_loss}
    \end{minipage}
\end{figure}

We measure the performance of our congestion control algorithm in a commercial LTE network (AT\&T)
and compare its performance with seven other congestion control algorithms, 
including BBR~\cite{BBR}, CUBIC~\cite{CUBIC}, COPA~\cite{Copa}, Sprout~\cite{Sprout}, PCC~\cite{PCC}, and PCC-Vivace~\cite{PCC-v}.
We test each algorithm for 30 seconds and repeat the test 20 times for each algorithm. 
All the performance tests are conducted on workdays when the base station is busy.
We measure the performance of the algorithms using Pantheon~\cite{Pantheon},
and plot the averaged achieved throughput and $95^{th}$ percentile of oneway delay of eight algorithms in Figure~\ref{fig:thput_delay}. 
We see from the performance results that, with the accurate per-millisecond capacity updates, 
\systemname{}based algorithm is able to fully utilize the capacity provided by the cellular network, maximizing its throughput,
and at the same time avoid congesting the network, minimizing its latency.  


\parahead{Impact of missing control messages}
We investigate the impact of missing control messages on the end\hyp{}to\hyp{}end 
performance of the \systemname{} based congestion control algorithm. 
In this evaluation, we perform trace-driven emulation using a link emulator: Mahimahi~\cite{Mahimahi}.
We decode and record the control messages the base station transmits in a 30 second period,
base on which we calculate the available capacity for one mobile user 
and then feed such a capacity trace to Mahimahi.
We build a TCP connection over the link emulated by Mahimahi and 
feed the recorded control messages to the TCP sender for capacity calculation and congestion control. 
To emulate the missing messages, we randomly dropped 0\% to 50\% of the decoded control messages before feeding to the TCP sender. 

We measure the achieved throughput and delay of the congestion control algorithm over the emulated link
and plot the results in Figure~\ref{fig:message_loss}.
We see that, with the percentage of missing messages increases from 0\% to 50\%, 
the achieved throughput increases slightly, \ie, from 39.2 to 39.8~Mbit/s, 
while the delay increases significantly, \ie, from 64 to 327~$ms$.  
With more missing control messages, the sender observes more idle bandwidth from the base station 
and thus over-estimate the available capacity, as shown in Figure~\ref{fig:message_loss}, 
causing frequent congestions inside the network.

\subsection{Video streaming using \systemname{}}\label{s:video}
To further demonstrate the value of the telemetry data provided by \systemname{}, we implemented a \systemname{} based video streaming system.
MPC~\cite{robustMPC} is the state-of-the-art video bitrate adaptation system, 
which estimates the capacity as the harmonic mean of the video downloading speed in the near past.
We integrate \systemname{} with MPC~\cite{robustMPC} by replacing the capacity with 
the telemetry data measured by \systemname{}, which we refer to as \textit{NG-MPC}.
We modify the dash.js to implement our algorithm. 
We compare our algorithm with MPC~\cite{robustMPC}, Buffer based ABR~\cite{bufferABR}, BOLA~\cite{BOLA}, and deep learning based ABR -- Pensieve~\cite{Pensieve}.

We evaluate the video QoE provided by all five ABR algorithms. 
There exist a wide range of QoE metrics that characterizes the user-perceived video quality, 
but three important factors are included in most of the QoE metrics:
the average video quality, the quality variations, and the rebuffering.
We use the following equation to summarize their impact on the QoE:
\begin{equation}
    QoE = \sum^N_{n=1} q(R_n) - \mu \sum^N_{n=1}T_n - \sum^N_{n=1}T_n |q(R_{n+1} - q(R_n) | 
    \label{eqn:qoe}
\end{equation}
where $N$ represents the total number of chunks inside a video. 
The $R_n$ describes the bitrate of $n-th$ chunk and the function $q(R_n)$ translate the bitrate to the user-perceived video quality;
the $T_n$ is the rebuffering time when downloading $n$ chunk at bit rate $R_n$;
and the final term characterizes the video bitrate changes which penalizes the overall QoE. 

We consider three types of QoE metrics. 
We start with the first QoE metric $QoE_{lin}$ that has been considered in both MPC and Pensieve,
where the quality mapping function is linear $q(R_n) = R_n$.
The second metric $QoE_{log}$ is used in BOLA~\cite{BOLA} where the quality mapping function $q(R_n) = log(R/R_{min})$,
which captures the phenomenon that the improvement of user-perceived quality decreases at higher video bitrates. 
The third metric $QoE_{HD}$ is used in Pensieve~\cite{Pensieve} which favors the high definition video 
by assigning a low score to a low-quality video and a much higher score to a high-quality video. 
Table 1 in Pensieve~\cite{Pensieve} gives the detailed score values.  

\begin{figure}[htb]
    \centering
    \begin{subfigure}[b]{0.32\linewidth}
        \centering
        \includegraphics[width=0.99\textwidth]{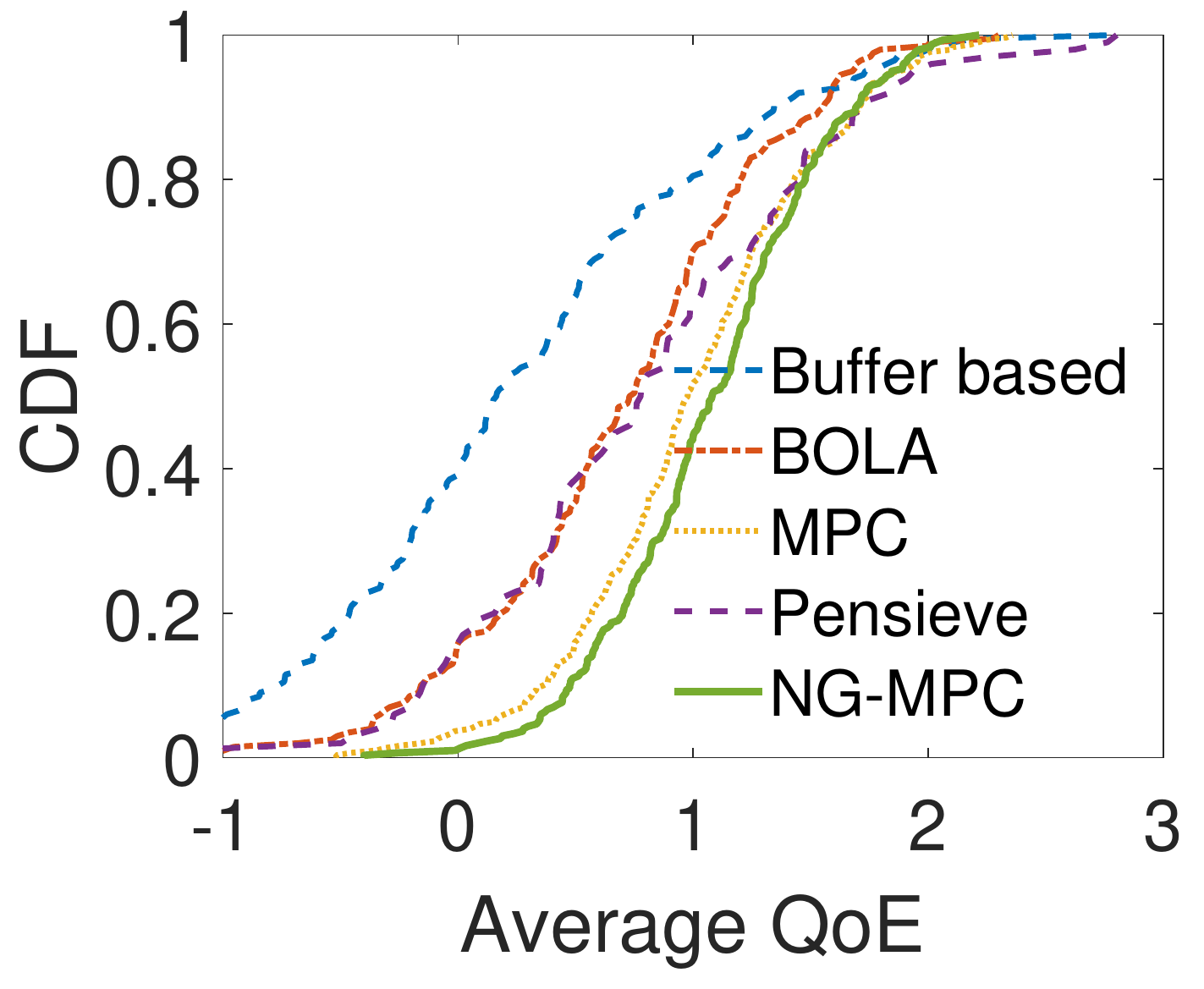}
        \caption{$QoE_{lin}$.}
        \label{fig:QoE_lin}
    \end{subfigure}
    \hfill
    \begin{subfigure}[b]{0.32\linewidth}
        \centering
        \includegraphics[width=0.99\textwidth]{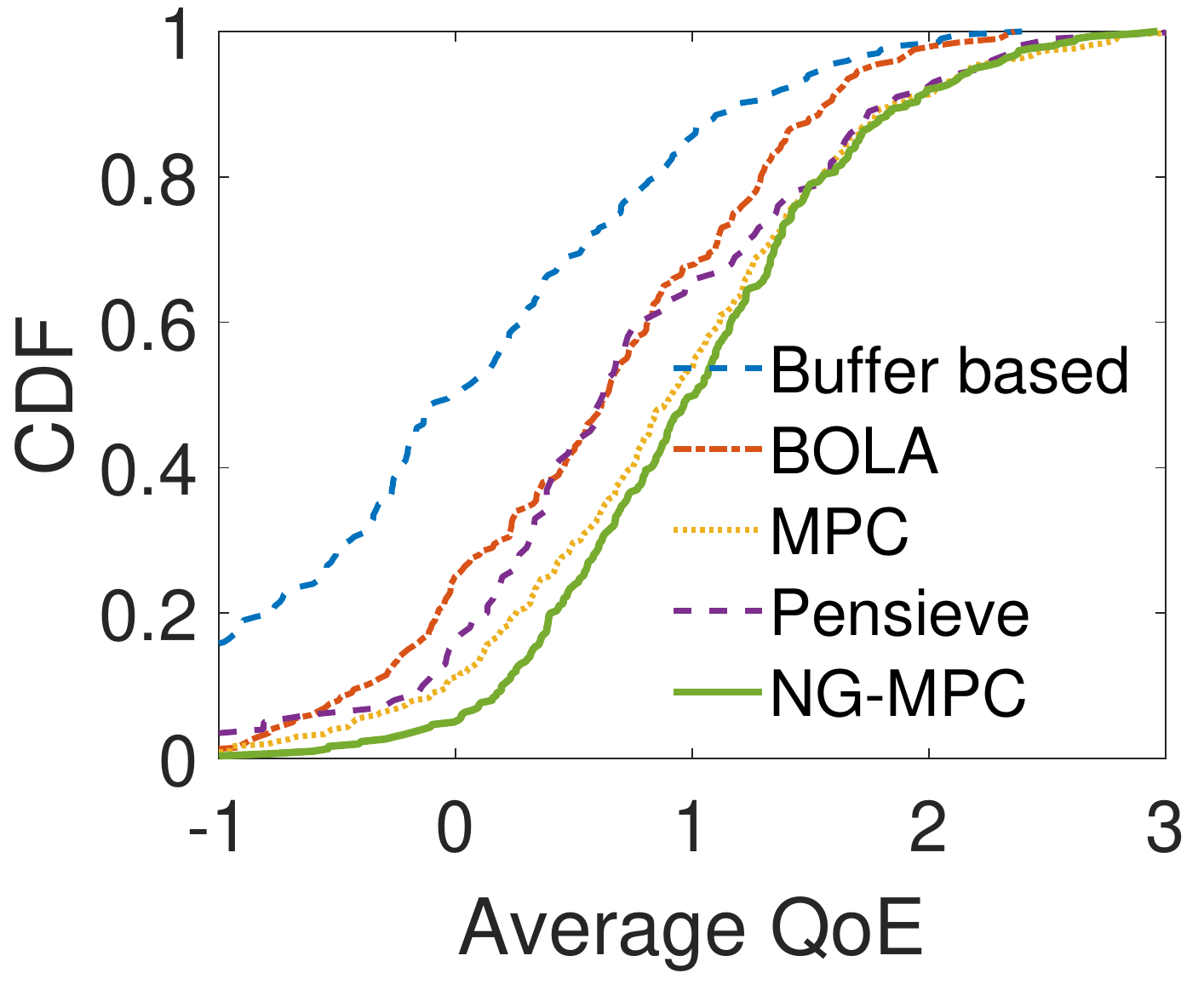}
        \caption{$QoE_{log}$.}
        \label{fig:QoE_log}
    \end{subfigure}
     \hfill
    \begin{subfigure}[b]{0.32\linewidth}
        \centering
        \includegraphics[width=0.99\textwidth]{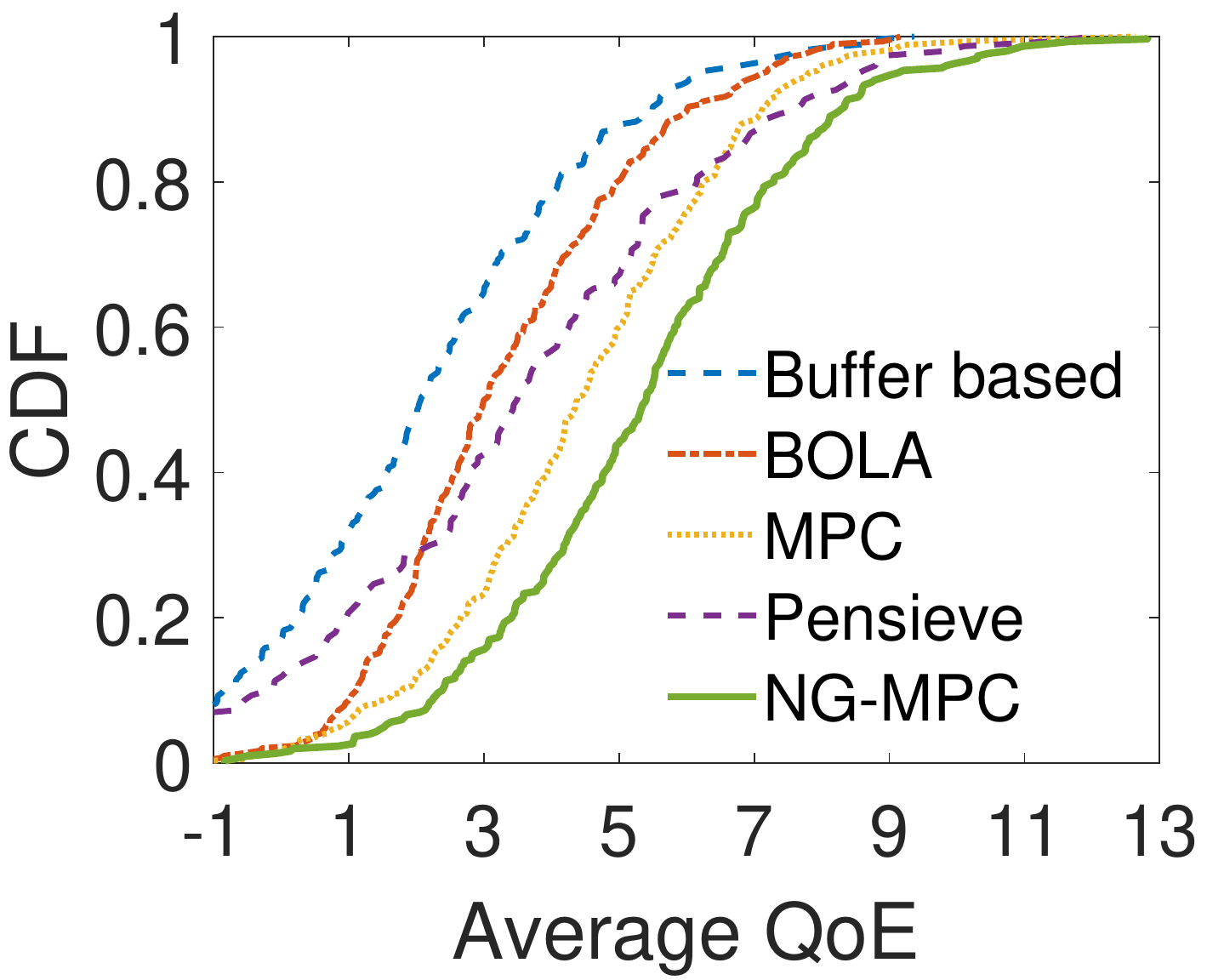}
        \caption{$QoE_{HD}$.}
        \label{fig:QoE_HD}
    \end{subfigure}
    \caption{Comparing NG-MPC with existing ABR algorithms on the three QoE metrics. }
    \label{fig:VideoQoE}
\end{figure}

We set up a DASH~\cite{DASH} server and use Google Chrome as the client video player. 
We stream 10 videos using all aforementioned ABR algorithms. 
We use Mahimahi~\cite{Mahimahi} to emulate the network conditions from our cellular traces described in Section~\S\ref{s:congestion}.
We measure the final achieve video QoE of different ABR algorithms and plot the results in Figure~\ref{fig:VideoQoE},
from which we could observe obvious QoE improvement achieved by NG-MPC over the original MPC.
Since we directly use the trained model from Pensieve without retraining it using our cellular traces, 
we could see that Pensieve cannot provide high quality video streaming as its model does not generalize well, 
which has also been reported by prior works~\cite{fugu, Oboe}.
With the accurate capacity provided by \systemname{}, the NG-MPC improves the QoE by 7.6\%, 10.1\% and 11.4\% over 
MPC for $QoE_{lin}$, $QoE_{log}$, and $QoE_{HD}$, over MPC, respectively.

\subsection{Tracking Frame Loss and Size using \systemname{}}\label{s:eval_reTx}
\systemname{} is able to identify retransmitted transport block using the new-data indicator of the control message.
In this section, we investigate how frequently retransmissions happen in commercial cellular networks and their impact on cellular packet transmissions. 

\parahead{Experimental methodology} We use the same setup of the remote server and mobile phone as in \S\ref{s:micro}. In our static experiment, we place the mobile phone at one location and let the remote server send UDP packets with a payload length of 1,400~bytes for 10 seconds. We vary the speed of the remote server from 10~Mbit/s to 55~Mbit/s.
The phone is connected with the same cell tower (20~MHz bandwidth at 1.94~GHz) during both the static and mobile experiments.

\subsubsection{Static user} 
\begin{figure}[htb]
    \centering
    \begin{minipage}[b]{0.25\linewidth}
        \centering
        \includegraphics[width=\textwidth]{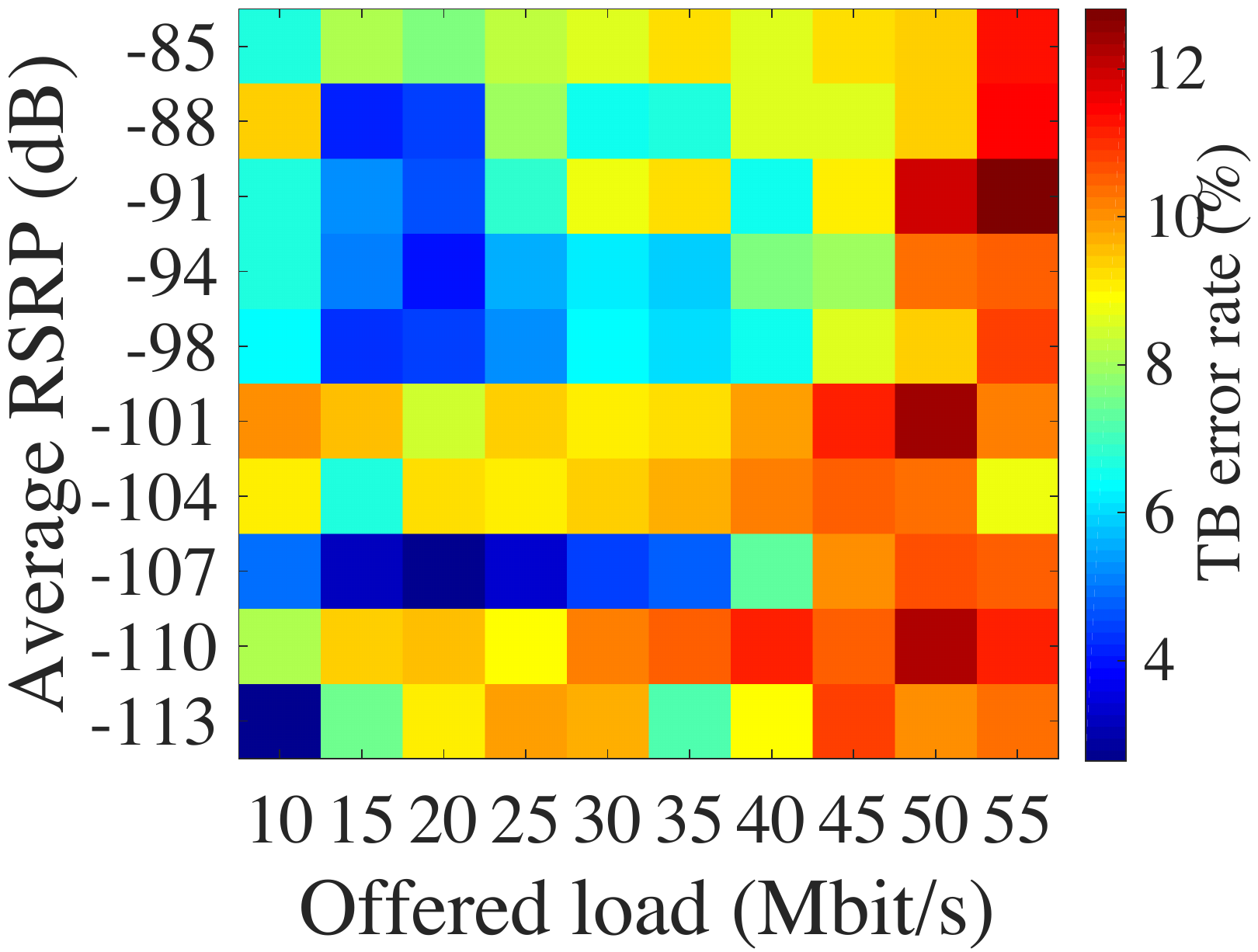}
        \caption{TB error rate versus offered load and the locations of the user.}
        \label{fig:reTx_ratio}
    \end{minipage}
    \hfill
    \begin{minipage}[b]{0.23\linewidth}
        \centering
        \includegraphics[width=\textwidth]{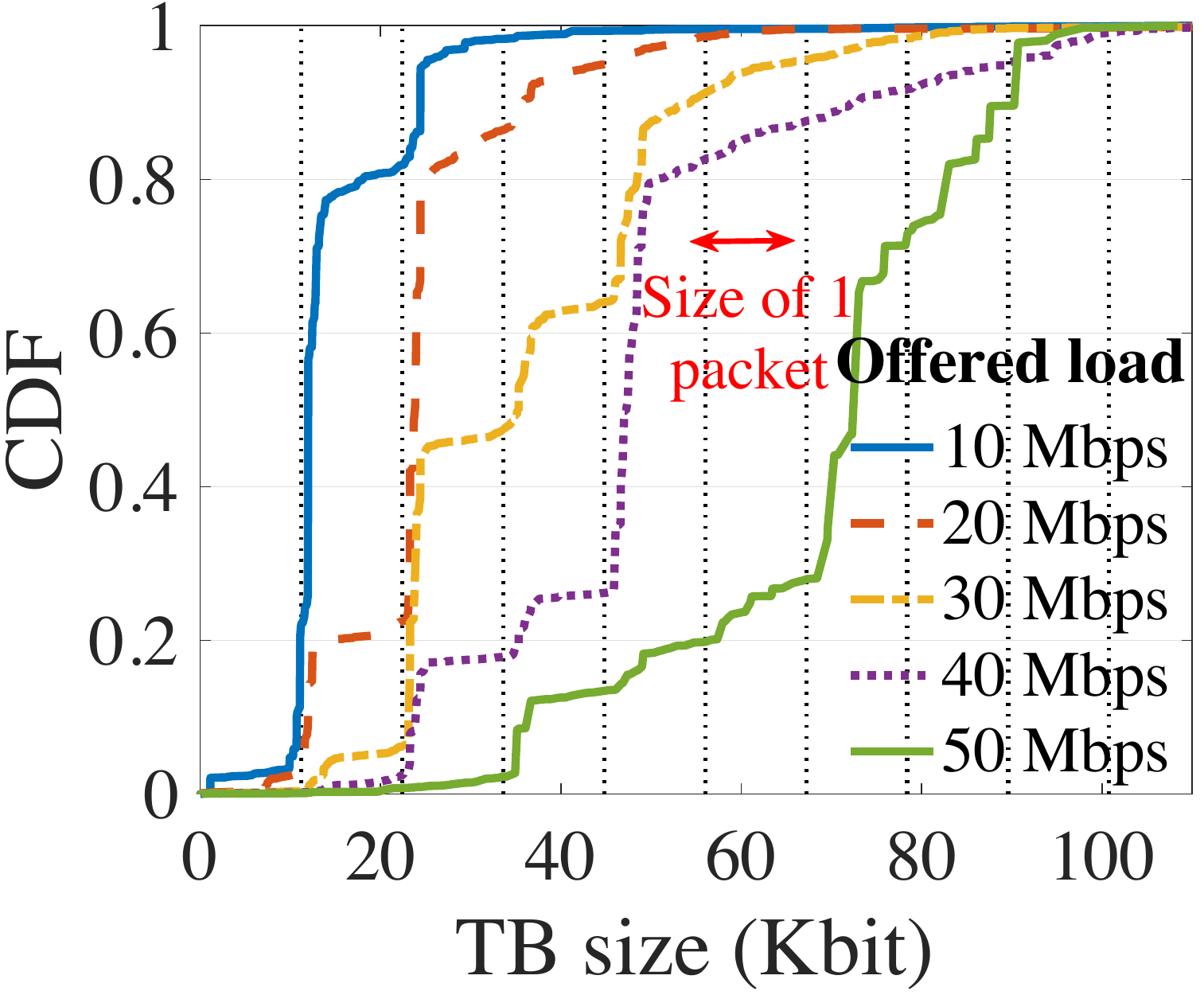}
        \caption{TB size of TBs sent to the UE with varying offered load.}
        \label{fig:tbs_cdf}
    \end{minipage}
    \hfill
    \begin{minipage}[b]{0.23\linewidth}
        \centering
        \includegraphics[width=\textwidth]{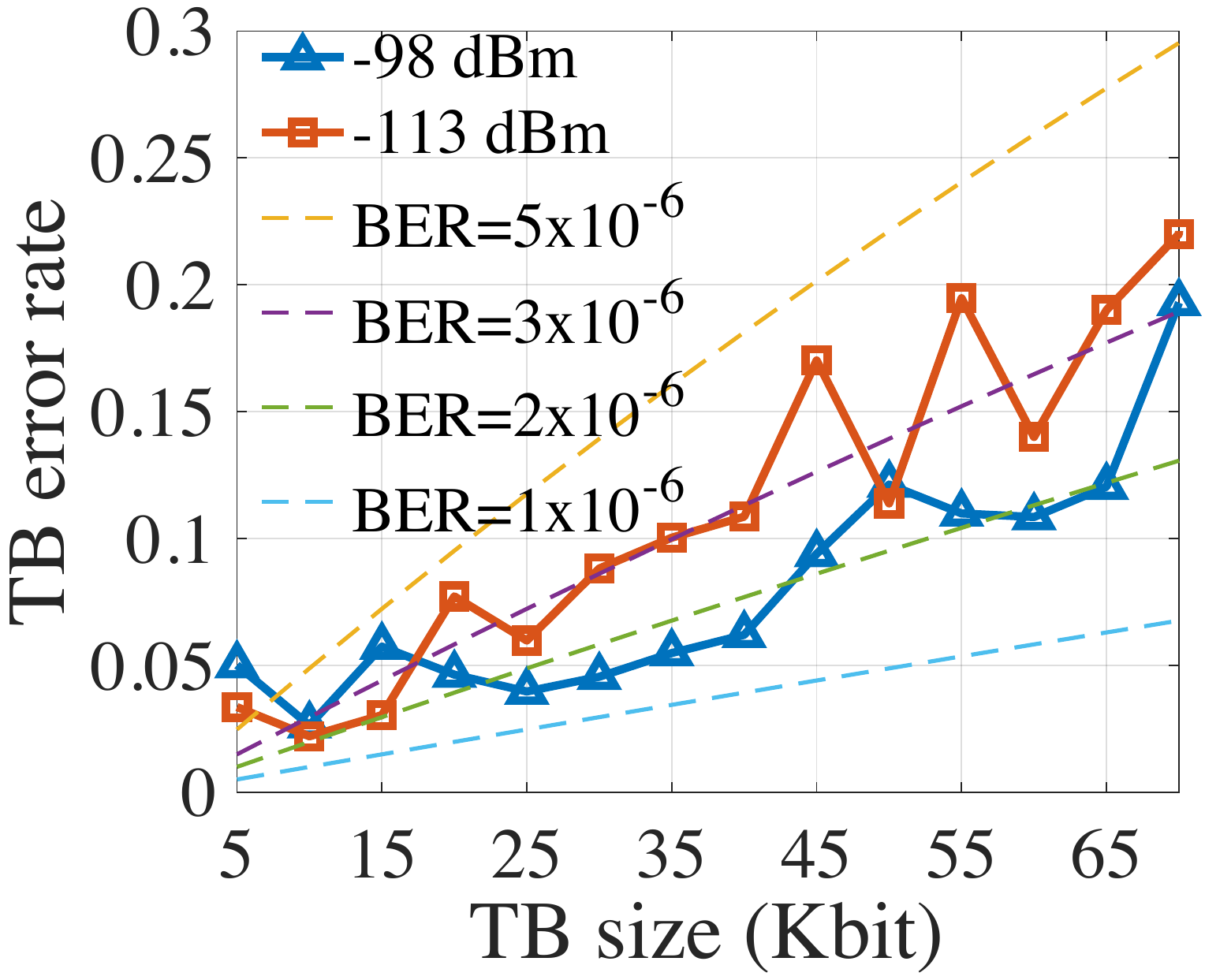}
        \caption{The TB error rate varies with the size of the TB.}
        \label{fig:FER_size}
    \end{minipage}
    \hfill
    \begin{minipage}[b]{0.23\linewidth}
        \centering
        \includegraphics[width=\textwidth]{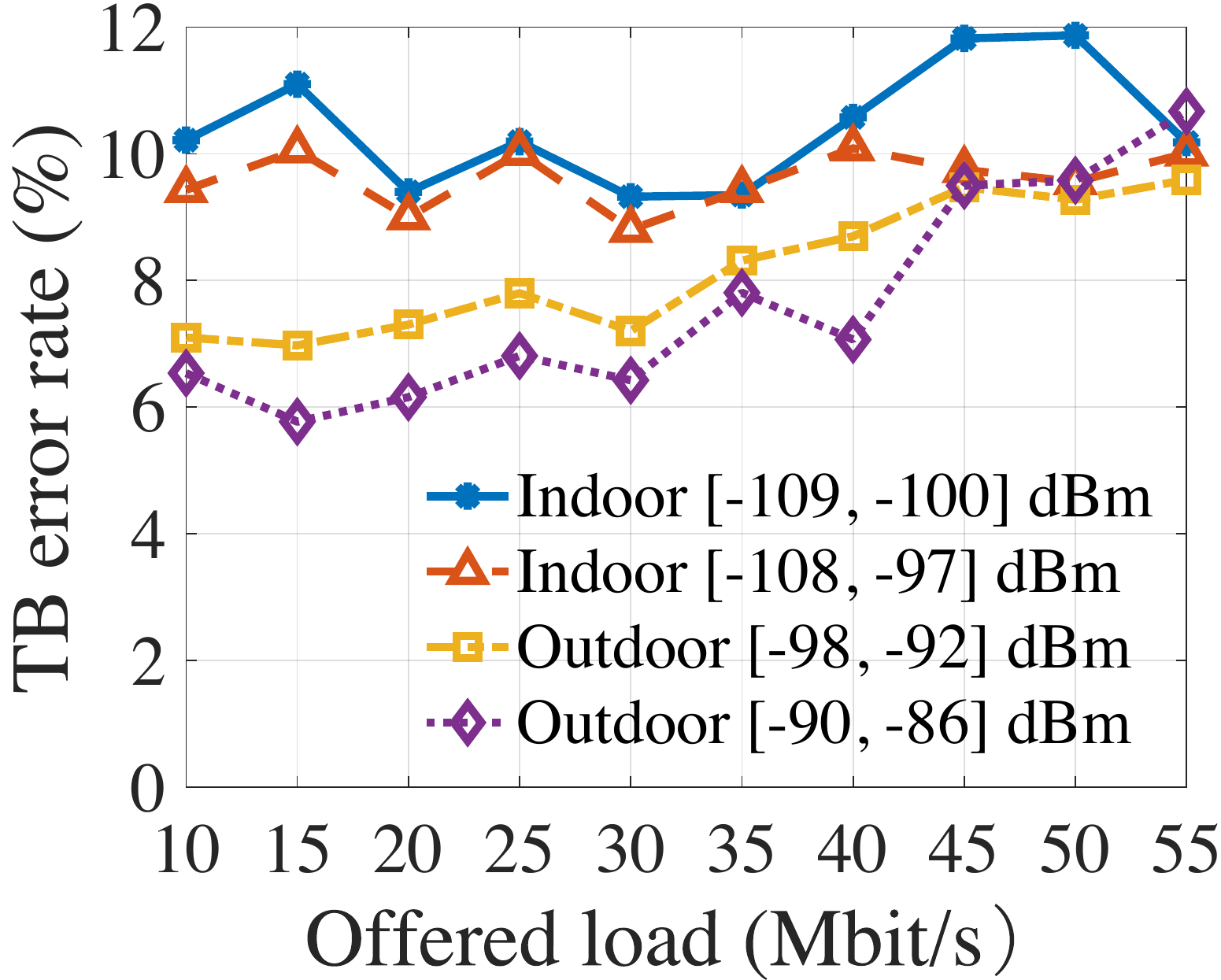}
        \caption{TB error rate versus offered load for a mobile user.}
        \label{fig:mobile_reTx_tb}
    \end{minipage}
    \hfill
\end{figure}

We move the phone to 10 different locations with varying signal strength in a building.
From the decoded control message, we count the number of \textit{original} (not retransmitted) transport blocks that the cell tower sends to the UE. 
Among all those original transport blocks, we also count the number of transport blocks that have been decoded with bit errors and thus require retransmissions, according to the new data indicator (\textsf{ndi}).
We calculate the \textit{TB error rate}, \ie, the ratio of erroneous transport blocks (requiring retransmissions), to the total number of original transport blocks sent. 
In Figure~\ref{fig:reTx_ratio}, each row of data comes from a different location and is indexed by the average RSRP at that location. 
Color represents the value of TB error rate (the lighter colors represent higher TB error rates).
We see that the TB error rate varies from 2\% to 12\%, but there is no obvious pattern of transport block errors across locations. 
The likelihood of the transport block error at a certain location is determined by the channel at that specific location and how the rate adaptation algorithm works on that channel. 
We observe however that at each location, the transport block error rate significantly increases with the offered load. 

\parahead{Relationship with TB size} We hypothesize that the increased TB error rate is due to the increased size of the TBs themselves. 
We plot the distribution of the size of the TBs sent from the cell tower to the UE at the location 
in Figure~\ref{fig:tbs_cdf}. We see that with a higher data rate, TB size increases accordingly, at a multiple of the packet size, \ie, 1,400~byte. 
For example, at an offered load of 10~Mbit/s, only one packet arrives at the receiver in each millisecond interval. 
Accordingly, the cell tower embeds only one packet inside the TB of each subframe, so more than 75\% of the TB have a size of only around 1,400~bytes. 
With increased offered load, more packets are grouped into one TB, increasing its size.

To further validate our hypothesis, we group all received TBs into 14 bins, \ie, zero to 70,000 bits with a step size of 5,000 bits, according to their size. 
We calculate the TB error rate for blocks in each bin and plot the calculated error rate for locations with RSRP $-98$~dBm and $-113$~dBm, in Figure~\ref{fig:FER_size}. 
We can see that the TB error rate increases with TB size. 
We also compare error rates with theory: supposing the error rate of each data bit inside one TB is $p$ and that bit errors are \textit{i.i.d.}, then the TB error rate can be calculated as $1-(1-p)^N$, where $N$ is the TB size. We plot the calculated TB error rate for bit error rate $p$ of $5\times10^{-6}$, $3\times10^{-6}$, and $1\times10^{-6}$, in Figure~\ref{fig:FER_size}. Firstly, we see that the experimental data fit the theoretical predictions, including the \textit{i.i.d.} bit error probability assumption, well. Secondly, we also see that the cellular network maintains the BER for data bits inside the TB at around $10^{-6}$ and that such a BER is slightly different across locations.

\subsubsection{Mobile user} 
In the mobile experiment, we move the UE along two indoor and two outdoor trajectories at a speed of two m/s. 
We repeat each trajectory 10 times, with varying offered loads from the server,
and record the RSRP range observed when moving along each trajectory. 
We calculate the TB error rate for each offered load and plot the results in Figure~\ref{fig:mobile_reTx_tb}. 
We see that the TB error rate of outdoor trajectory exhibits the same pattern as the static experiments: the higher the offered load, the higher the error rate. 
But for indoor trajectories, the TB rate is similar for all offered loads. 
Compared with outdoor trajectories, the channel along the indoor trajectories changes dramatically, due to indoor small-scale fading~\cite{LTE_UMTS}, resulting in more TB errors when LTE rate adaptation algorithms fail to cope with channel variations. 
We, however, observe that even in the challenging indoor mobile scenario, LTE still manages to keep the error rate below 12\%.


\subsection{Carrier Aggregation Monitoring using \systemname{}}
\label{s:eval_CA}

Accurate determination of congested cell status requires that the UE be aware of the instantaneous carrier aggregation configuration. 
In this section, we show that \systemname{} can track carrier aggregation in commercial LTE networks.

\begin{figure*}[htb]
     \begin{minipage}[htb]{0.23\linewidth}
        \centering
        \includegraphics[width=0.98\textwidth]{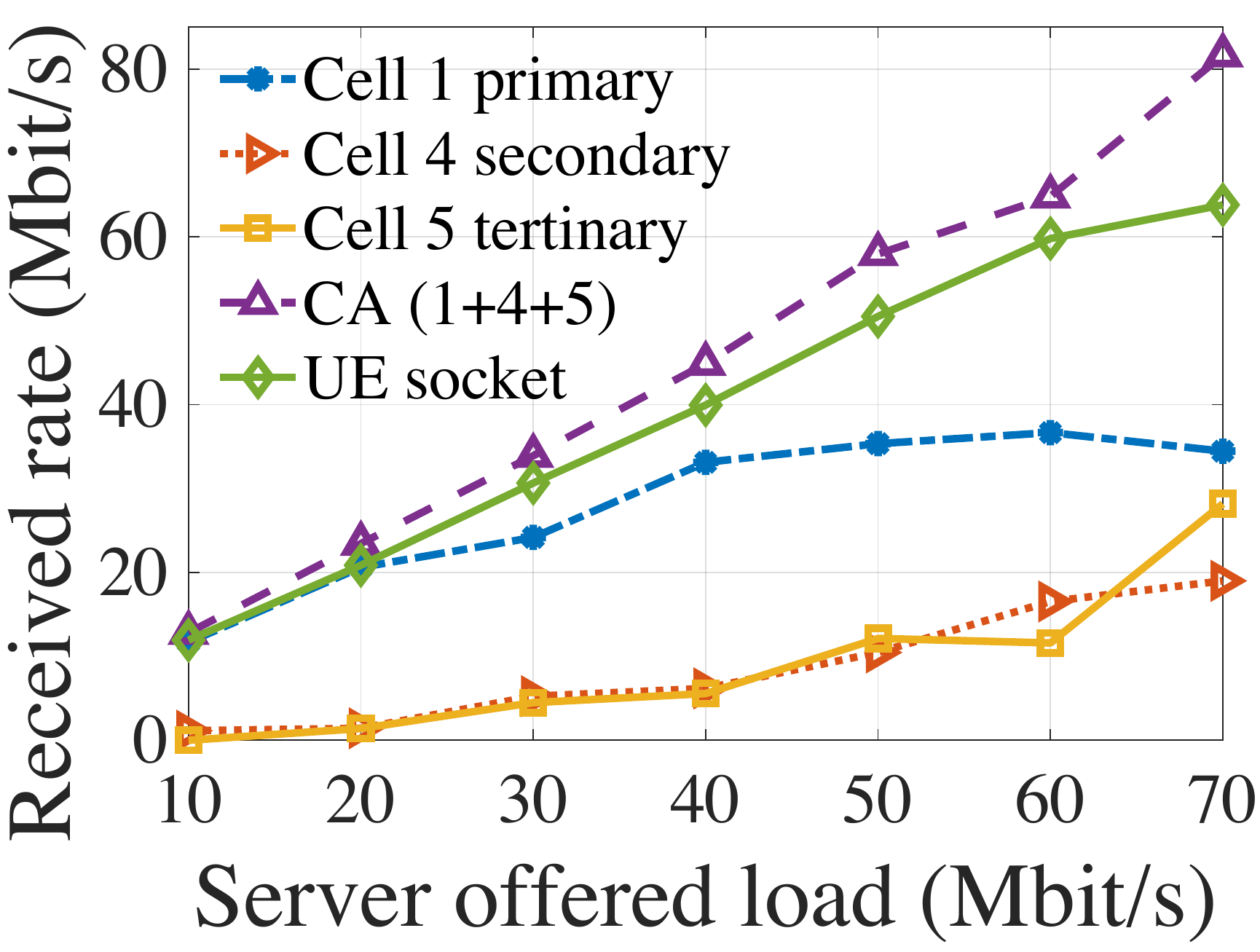}
        \caption{The received data rate from three aggregated cells and the UE socket.}
        \label{fig:CA_cell_rate}
    \end{minipage}
    \hfill
    \begin{minipage}[htb]{0.48\linewidth}
        \includegraphics[width=0.99\textwidth]{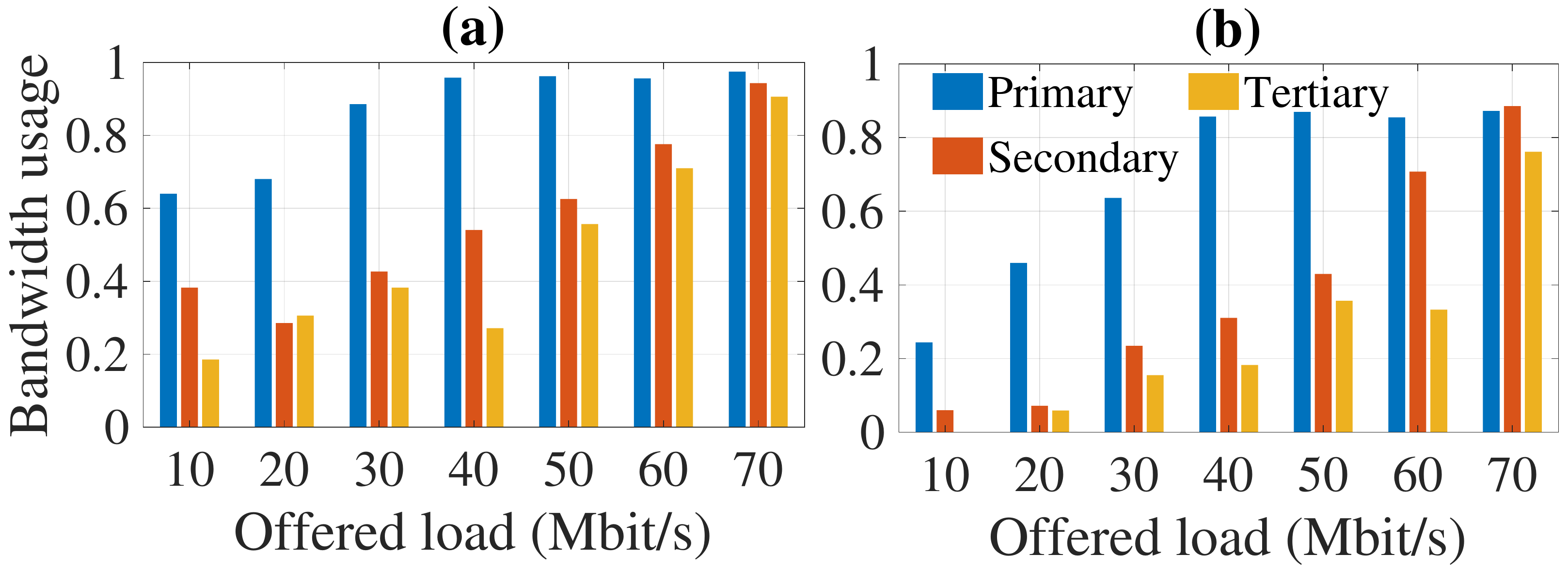}
      \caption{The ratio of utilized PRB to the total available PRBs (bandwidth usage) of three cells are given in \textbf{(a)}; and the bandwidth usages by the UE in three cells are plotted in \textbf{(b)}.}
      \label{fig:channel_usage}
    \end{minipage}
    \hfill
     \begin{minipage}[htb]{0.24\linewidth}
        \centering
        \includegraphics[width=0.98\textwidth]{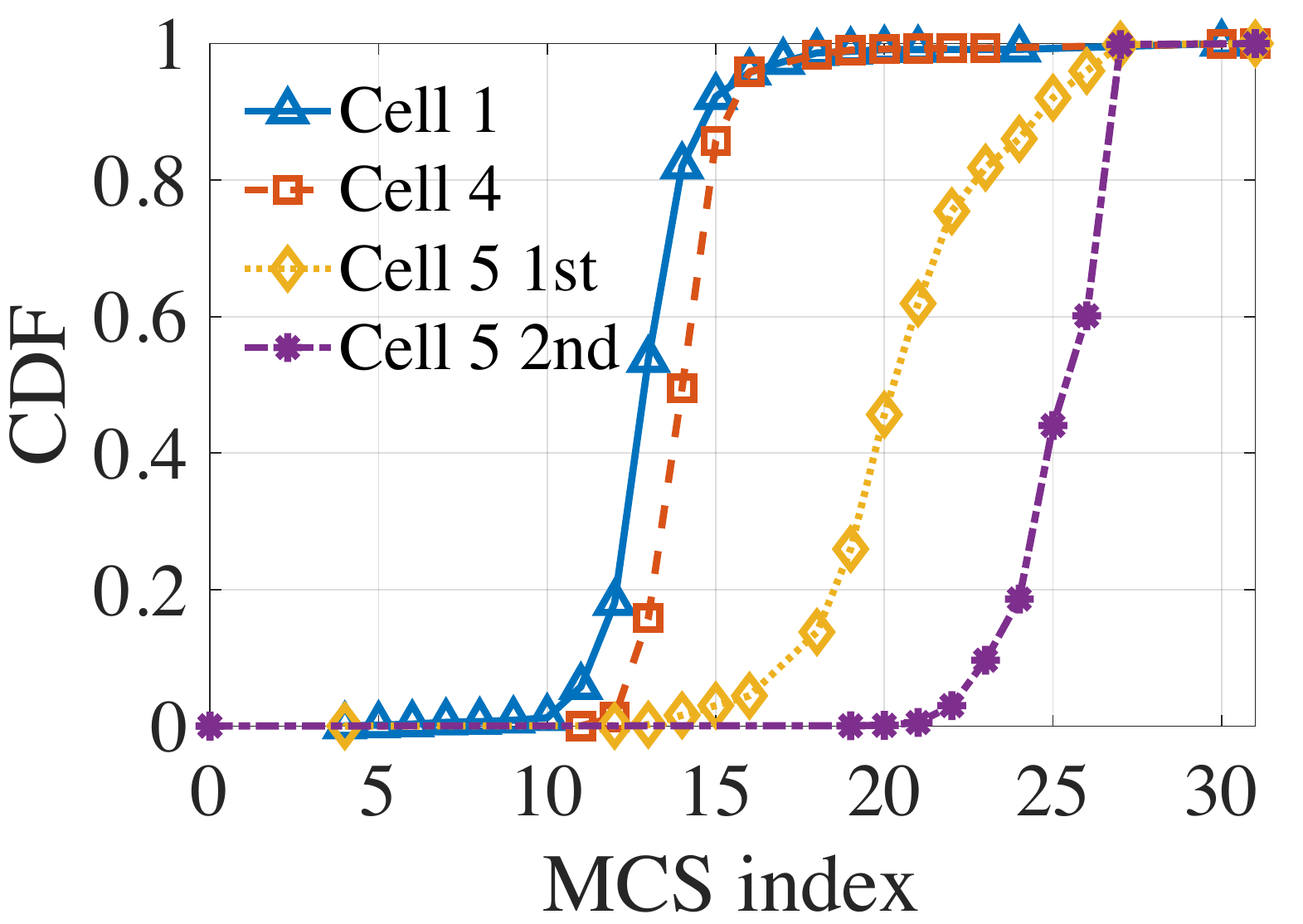}
        \caption{MCS index CDF for three cells (cell five uses two spatial streams).}
        \label{fig:CA_mcs_cdf}
    \end{minipage}
\end{figure*}

\subsubsection{Carrier Aggregation Load-balancing} \label{s:eval_CA_cells}
Understanding how carrier aggregation affects data delivery provides insights for designing an LTE-compatible congestion control algorithm.
Therefore, in this section, we investigate how data are load-balanced over different cells to the UE with carrier aggregation. 
In this experiment, the UE is static. LTE aggregates three cells (cell one as a primary cell, cell four as secondary, and five as tertiary) for this UE.
We let the remote server send UDP traffic at different speeds to the UE for five seconds, then count the number of data bits each cell tower delivers to the UE and calculate the average transmitted data rate of each cell. We also record the received data rate from the socket interface of the UE.

We plot the recorded data rates in Figure~\ref{fig:CA_cell_rate}. We have three observations from Figure~\ref{fig:CA_cell_rate}. 
Firstly, the aggregated rate (summation of three cells) is larger than the received data rate from the UE's socket because of protocol overhead and retransmission of erroneous TBs, which matches with the results in Figure~\ref{fig:cell_rate_inc}.
Secondly, the cellular network mainly uses the primary cell for data delivery until it is saturated. To confirm this observation, we also plot the bandwidth usage, \ie, the ratio of allocated PRBs to the total amount of PRBs the cell has, in Figure~\ref{fig:channel_usage}(a) and (b), depicting the bandwidth usage of the cell and the UE, respectively. We see that the primary cell gets saturated (bandwidth usage close to one) earlier (40~Mbit/s) than the other two aggregated cells (70~Mbit/s). 

Thirdly, we observe that when all cells are saturated at 70~Mbit/s, cell five with a bandwidth of 5~MHz supports a similar PHY data rate as cell one with a bandwidth of 20~MHz and a much higher data rate than cell four with a bandwidth of 10~MHz. To explain this, we plot the CDF of the MCS index each cell uses to transmit data to the UE in Figure~\ref{fig:CA_mcs_cdf}. We see that cell five uses two spatial streams and that the MCS used in each spatial stream are much larger than the other two cells, which means that cell five actually has better signal quality and higher spectral efficiency than cell one and five. Such a phenomenon indicates the inefficiency of the LTE load balancing algorithm with carrier aggregation. On the one hand, LTE prefers transmitting data to the UE via the primary cell. On the other hand, the primary cell may not be the one with the highest signal quality within the aggregated cells. Instead of always saturating the primary cell first, a load balancing algorithm that selects the cell for data transmission according to the channel quality of each aggregated cell, would significantly improve overall spectral efficiency.
\vspace{-0.26cm}

\begin{figure}
      \begin{subfigure}[htb]{\linewidth}
         \centering
         \includegraphics[width=0.83\linewidth]{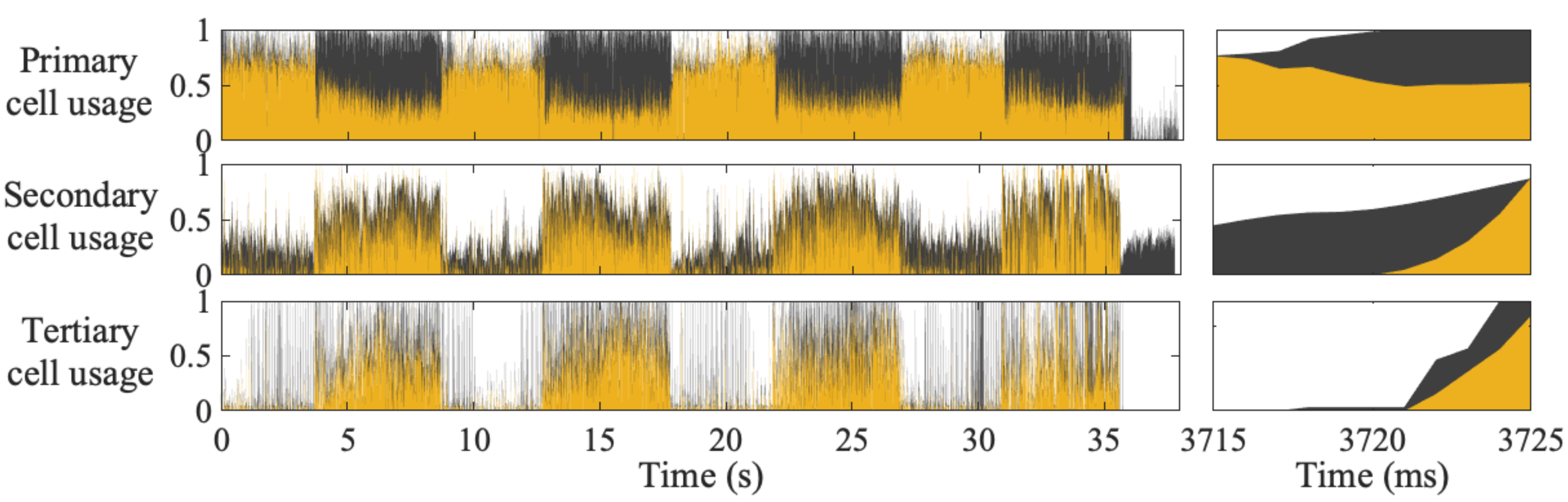}
          \caption{Channel usage of primary (cell 1), secondary (cell 4) , and tertiary  (cell 5) cells of Samsung S8.}
          \label{fig:CA_compete_s8}
      \end{subfigure}\hfill
       
      \begin{subfigure}[htb]{\linewidth}
         \centering
         \includegraphics[width=0.83\linewidth]{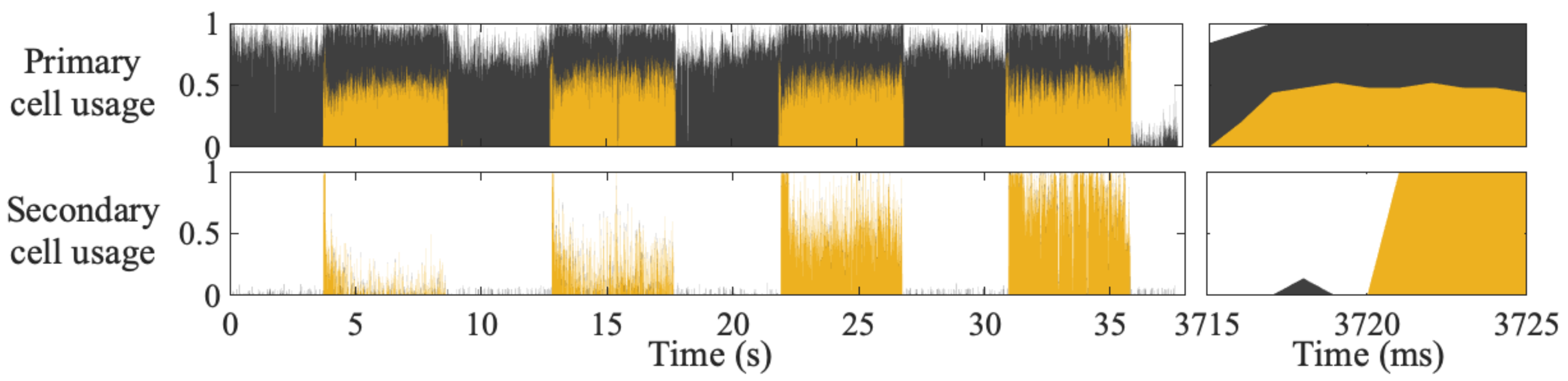}
          \caption{Channel usage of the primary (cell 1) and secondary (cell 3) cells of Xiaomi MIX3.}
          \label{fig:CA_compete_mix}
      \end{subfigure}
      \caption{The channel usage of the primary, secondary, and tertiary (if applicable) cells of two phones. 
      The black area represents the overall channel usage of the cell. 
      The yellow area represents the usage of mobile phones.}
      \label{fig:CA_2UE_compete}
\end{figure}

\subsubsection{Competing traffic}\label{s:eval_compete}
In this section, we investigate how competing traffic affects load balancing. 
In this experiment, we let a server transmit data to Samsung S8 with a constant offered load of 40~Mbit/s. 
We let another server transmit to a Xiaomi MIX3 with varying offered load from 30~Mbit/s to 60~Mbit/s (each for five seconds) 
and stop for four seconds between each transmission.  
The two phones connect to the same primary cell (20~MHz) but different secondary and tertiary cells (if applicable). 
Specifically, the network aggregates a 10~MHz secondary and a 5~MHz tertiary cell for Samsung S8, 
and a 10~MHz secondary cell Xiaomi MIX3. 

We plot the channel usage of the two phones in these four cells in Figure~\ref{fig:CA_2UE_compete}. 
From Figure~\ref{fig:CA_compete_s8}~(\textit{left}), we see that, when there is no competing traffic from the MIX3, 
most of the data for the S8 are delivered via the primary cell. When the competing traffic starts, 
the primary cell is saturated. The S8 and MIX3 share that bandwidth, 
so that a large portion of data for the S8 is offloaded to its secondary and tertiary cells. 
When the competing traffic is over, all the traffic for S8 is shifted back to the primary cell. 
We also zoom into the 10~ms period after the starting of competing traffic in Figure~\ref{fig:CA_compete_s8}~(\textit{right}). 
We see that the channel sharing in the primary cell and traffic offloading in secondary and tertiary cells are triggered within a few milliseconds, 
motivating the fine-grain physical layer information that \systemname{} provides. 
On the other hand, we see from Figure~\ref{fig:CA_compete_mix} that since the primary cell is shared with the S8, 
increasing the offered load for MIX3 only increases the channel usage in the secondary cell. 

From this experiment, we see that the impact of competing traffic in one cell could propagate to other cells via carrier aggregation. 
Such a cross-cell traffic correlation makes the resource allocation of each cell highly dynamic, resulting in significant link capacity variations. 
With the fine-grain information provided by \systemname{}, the UE can track these variations and make responsive actions.

\subsection{Capacity Tracking with CLAW}\label{s:CLAW}
In this section, we compare the accuracy of \systemnames{} capacity tracking with CLAW~\cite{CLAW}. 
CLAW implements a capacity tracker based on MobileInsight~\cite{MobileInsight}. 
Different from \systemname{}, MobileInsight reports the PRB allocation for a single user,
not the full PRB usage of the entire base station.
CLAW, therefore, proposes to estimate the PRB usage based on the power measurements reported by MobileInsight, 
with the intuition that higher utilization results in higher received power. 
We evaluate the performance of CLAW's PRB estimation and capacity estimation.

\parahead{Methodology}
We put two mobile users associated with the same base station (20~MHz and 100 PRBs) at one indoor and one outdoor location.
We implement CLAW on both mobile devices and log the estimation of the PRB utilization of every subframe. 
We put two \systemname{}s co-located with CLAW users for comparison.
We also deploy another three \systemname{}s at three different locations to estimate the PRB usage of the same base station
and extract the PRB usage of subframes where four \systemname{}s have exactly the same decoding results, as our ground truth.
\begin{figure}[htb]
    \centering
    \begin{subfigure}[htb]{0.495\linewidth}
        \centering
        \includegraphics[width=0.85\linewidth]{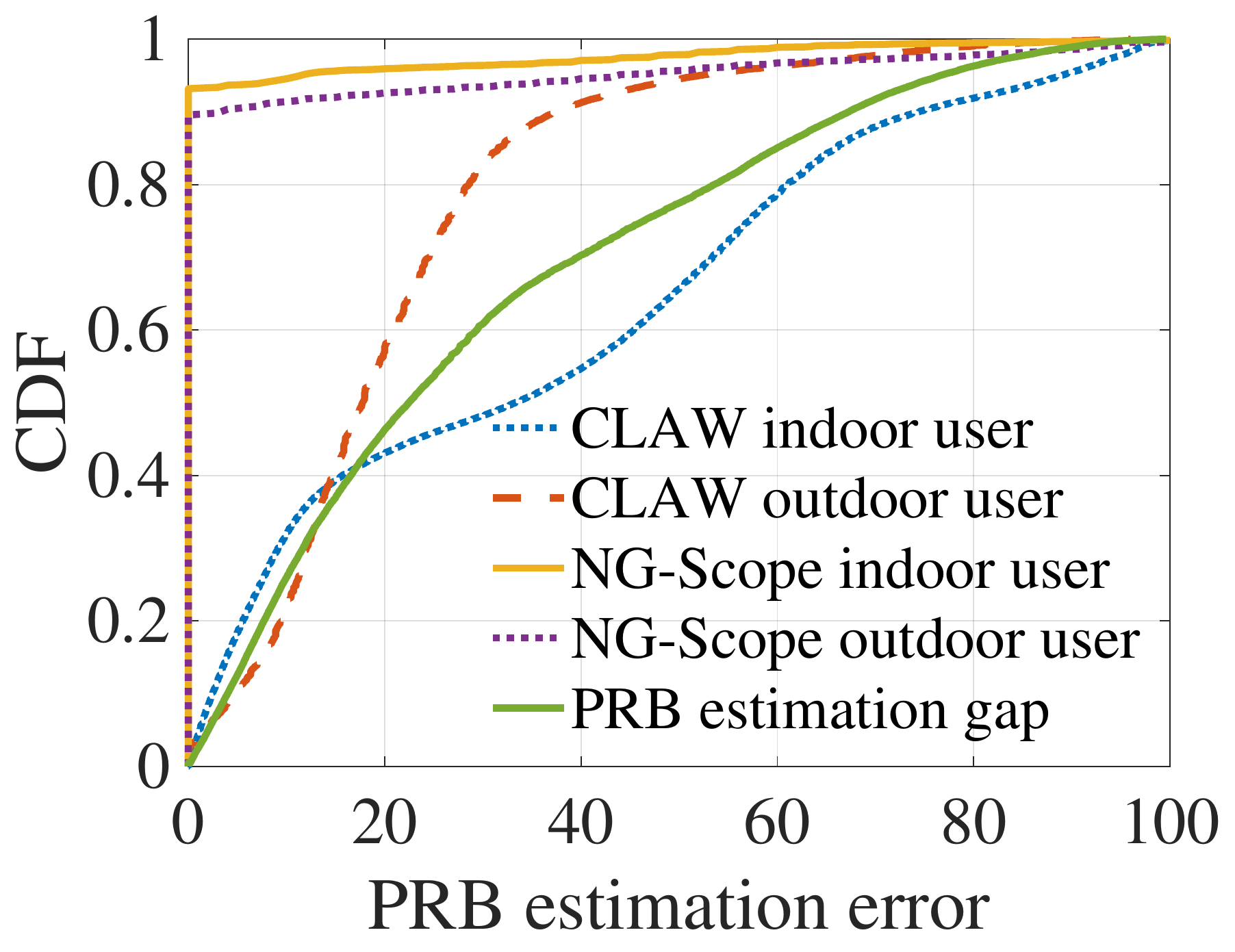}
        \caption{PRB estimation error.}
        \label{fig:claw_prb_err}
    \end{subfigure}
    \hfill
    \begin{subfigure}[htb]{0.495\linewidth}
        \centering
        \includegraphics[width=0.85\linewidth]{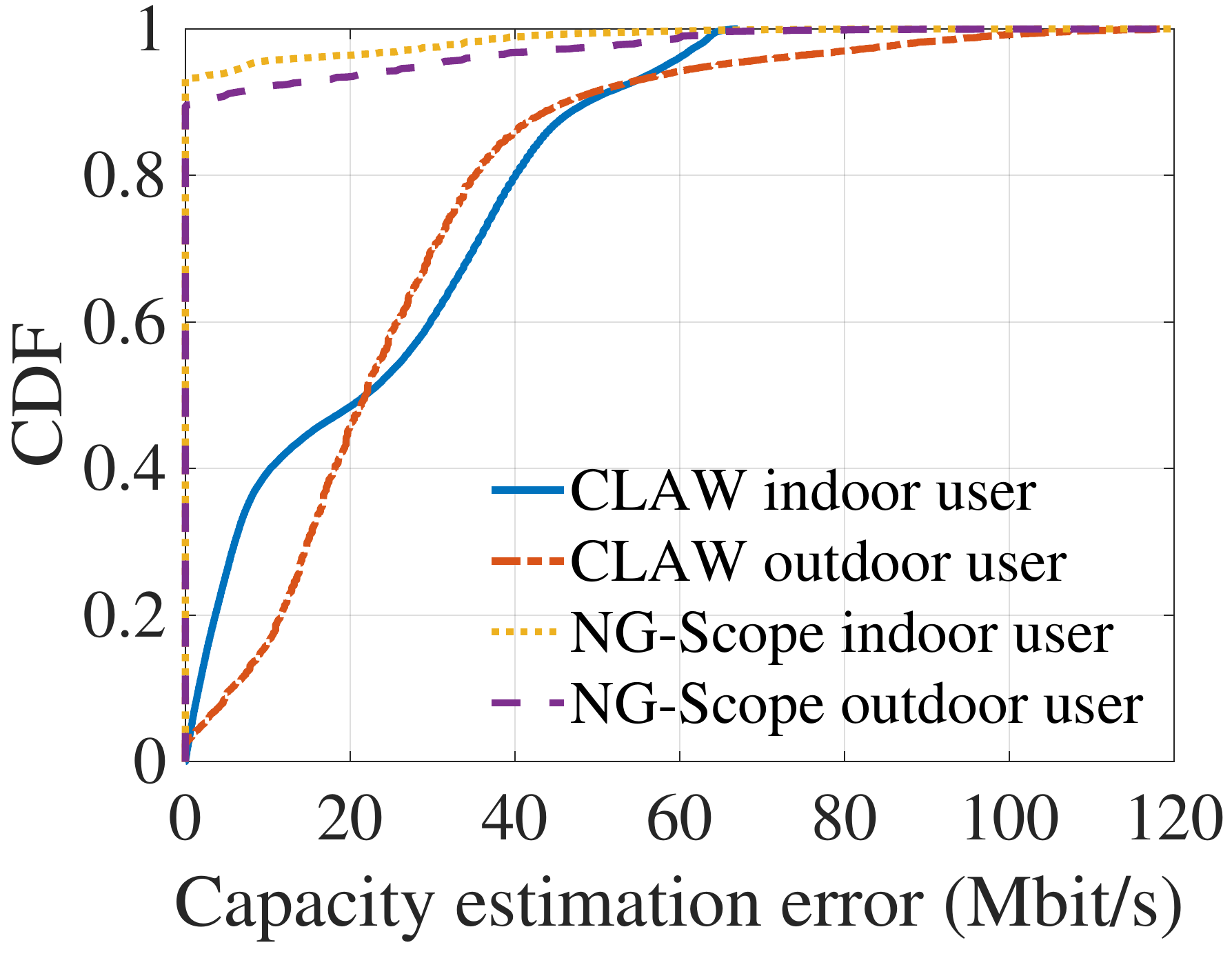}
        \caption{Capacity estimation error.}
        \label{fig:claw_cap_err}
    \end{subfigure}
    \caption{The PRB estimation error of a pair of indoor/outdoor CLAW users is plotted in (\textbf{a});
    the error of the capacity calculated using the inaccurate estimated PRB is given in (\textbf{b}).}
    \label{fig:claw}
\end{figure}
 
We plot the distribution of CLAW's PRB estimation error in Figure~\ref{fig:claw_prb_err}.
We see that CLAW has a much larger error in estimating utilized PRBs than \systemname{}. 
\systemname{} is accurate in estimation for 92\% and 88\% subframes for outdoor and indoor users respectively.
CLAW, however, has a median and 95-percentile error of 17 PRBs and 38 PRBs for the outdoor user, 
and 33 PRBs and 74 PRBs for the indoor user, respectively.
The underlying reason for such a large estimation error is that 
the power measurements are vulnerable to interference 
of neighboring base stations that operate at the same frequency band and thus become quite noisy.
Consequently, the PRB estimation based on such a noisy power report is unreliable. 
Furthermore, CLAW works better if all the PRBs have similar received power at the mobile user.
Such an assumption is invalid in an indoor scenario due to the frequency selective fading, 
so we observe a much higher estimation error for the indoor user. 
At last, we calculate the gap between two CLAW users' PRB estimations for the same subframe  
and plot the distribution of the gap across subframes in Figure~\ref{fig:claw_prb_err}.
We see that, due to the noise in the power measurements, 
two CLAW users almost never get the same utilized PRB estimation for the same subframe of the same base station.

We also translate the error of PRB estimation into the error of capacity estimation and plot the results in Figure~\ref{fig:claw_cap_err}. 
We see that, on average, CLAW has an RMS capacity error of 31~Mbit/s, which is 3.3$\times$ of \systemname{} (9.2~Mbit/s). 
The indoor CLAW user has a larger PRB error but a similar capacity error due to its lower signal strength.

\subsection{Bottleneck Detection with BurstTracker}\label{s:burstTracker}
In this section, we compare \systemnames{} performance with BurstTracker on one task: 
detecting whether the cellular link is the bottleneck of a connection.
The cellular link becomes the bottleneck when 
all the bandwidth of the base station is fully utilized.
BurstTracker is a dedicated tool built atop of MobileInsight for such a task.
MobileInsight reports only the PRB allocation for the mobile device it is implemented on, 
so it cannot be directly applied to determine the bottleneck.
To bypass such a constrain, BurstTracker makes a hypothesis about the base station's resource 
allocation algorithm-- cell tower allocates all the bandwidth to one user in one subframe, 
evoking the TDMA. 
We, however, observe different phenomena in our experimental result. 
Specifically, from Figure~\ref{fig:comp_ueFreq}, 
we see that a 20~MHz base station serves and thus allocates PRBs to more than one user
in 47.5\% of the active subframes.

\begin{figure*}[htb]
    \begin{minipage}[htb]{0.74\linewidth}
        \begin{subfigure}[htb]{0.49\linewidth}
            \centering
            \includegraphics[width=\linewidth]{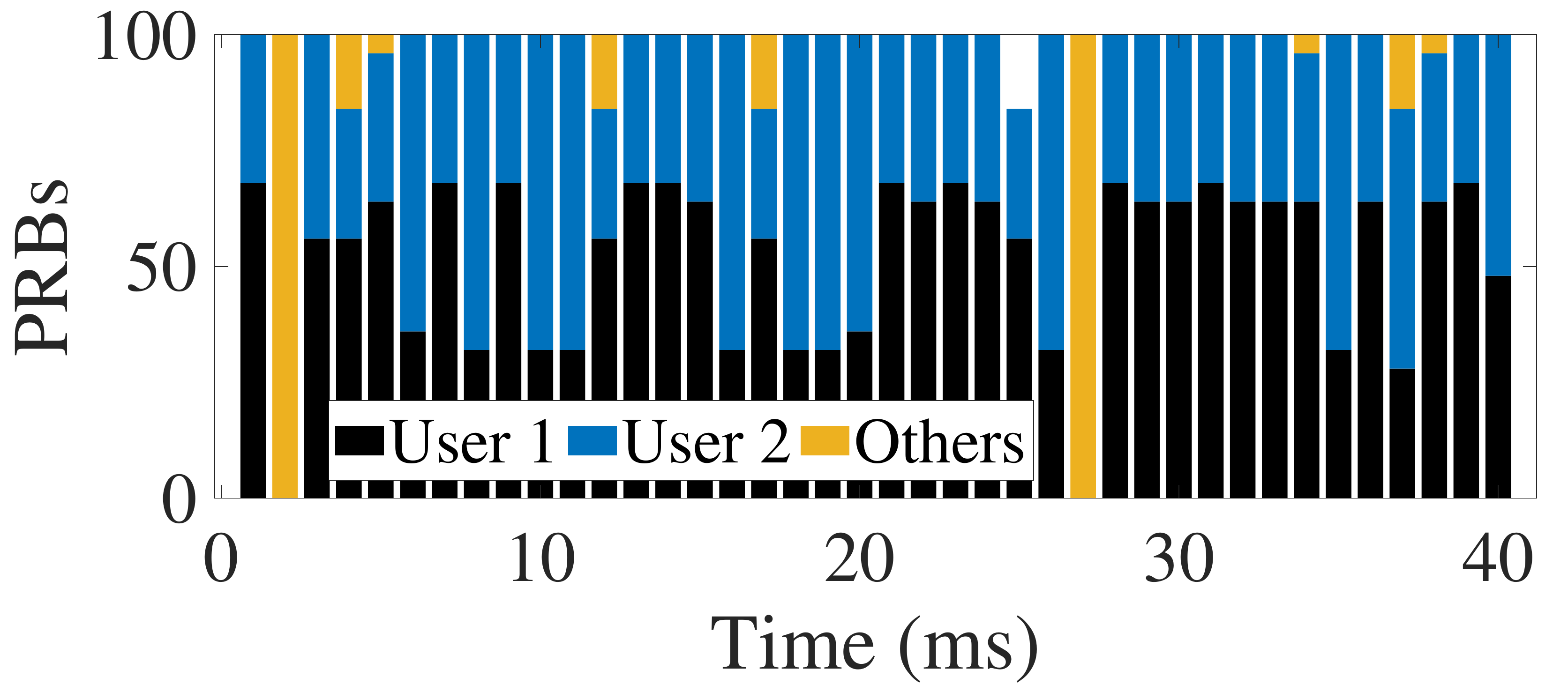}
            \caption{Two competing users.}
            \label{fig:burstTracker_compete}
        \end{subfigure}
        \hfill
        \begin{subfigure}[htb]{0.49\linewidth}
            \centering
            \includegraphics[width=\linewidth]{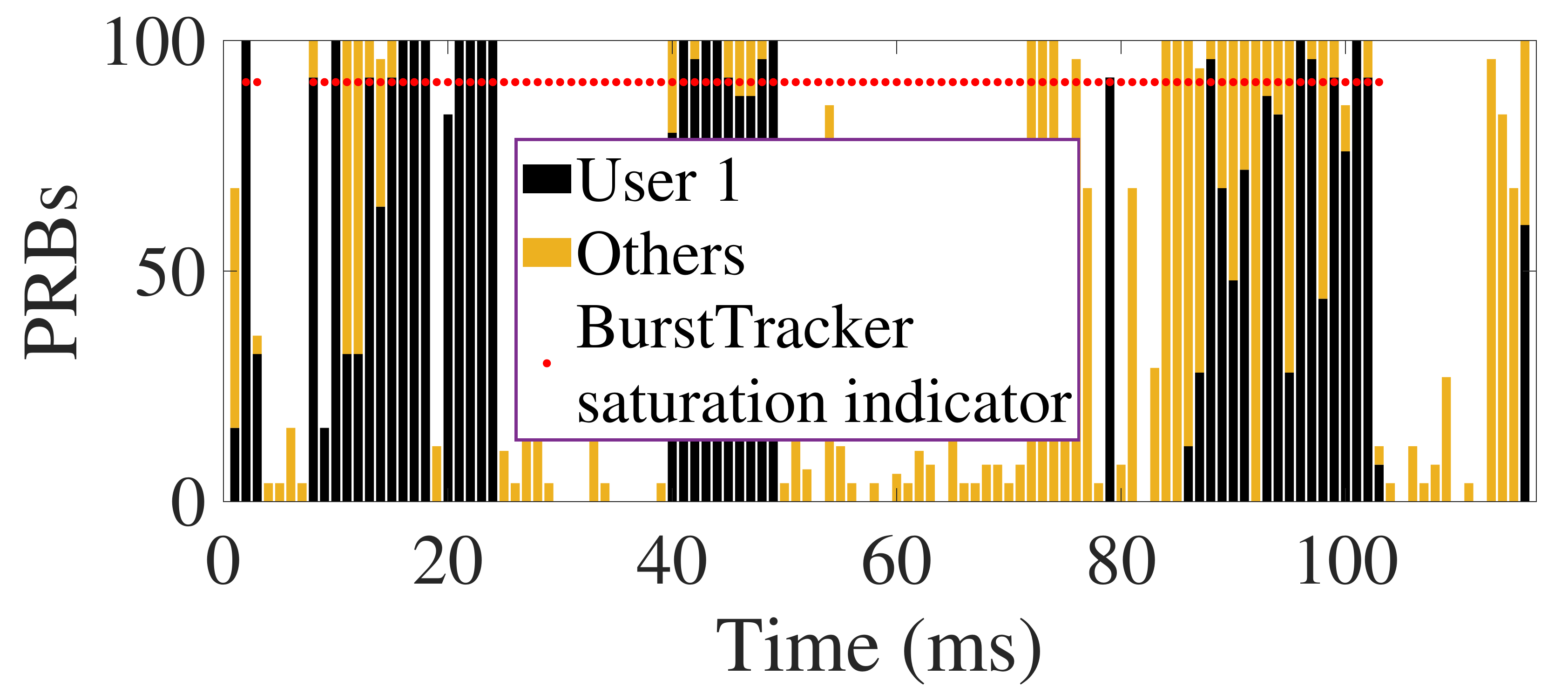}
            \caption{One user web-browsing.}
            \label{fig:burstTracker_empty}
        \end{subfigure}
        \caption{Resource allocation results measured using \systemname{}. 
        \textbf{(a)} The cell shares the PRBs inside each subframe (represented by a single bar) among multiple backlogged users, 
        so BurstTracker \cite{BurstTracker} detects no saturation when it in fact exists; 
        \textbf{(b)} when the user is web browsing, 
        BurstTracker indicates long periods of saturation even when the link is not fully saturated. }
        \label{fig:burst}
    \end{minipage}
    \hfill
    \begin{minipage}[htb]{0.24\linewidth}
    \centering
    \includegraphics[width=0.99\linewidth]{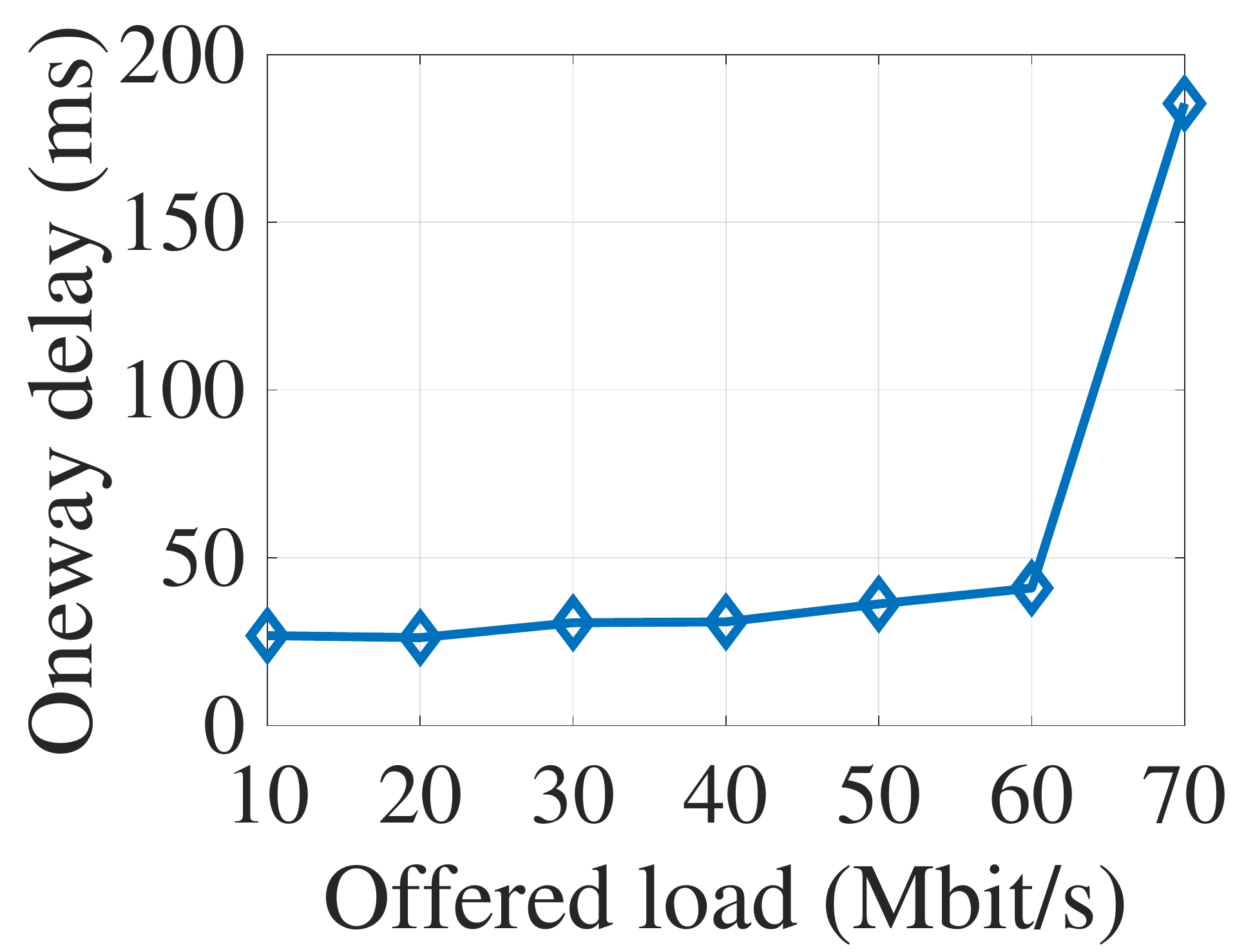}
    \caption{The achieved average oneway packet delay, with offered load varying from 10~Mbit/s to 70~Mbit/s.}
    \label{fig:CA_delay}
    \end{minipage}
\end{figure*}

Based on such a hypothesis, BurstTracker identifies 
the start of cellular wireless link becoming a bottleneck,
when MobileInsight reports more than 90\% of PRB occupation by this user. 
To investigate how this heuristic works in practice,
we let two users download bulky data simultaneously 
and plot the PRB allocation results of every subframe, 
measured by \systemname{}, 
in Figure~\ref{fig:burstTracker_compete}. 
We clearly see that these two users almost equally share 
the 100 PRBs of the base station, 
and there is no subframe in which one user occupies more than 90\% of the PRBs.
On the other hand, almost all the PRBs of this cell are fully utilized, 
so the cellular wireless link is indeed the bottleneck for both users. 
We feed such a trace to BurstTracker (we use the author's implementation)
and find that BurstTracker  detects no saturation 
because the utilization of each user stays around 50\%.

BurstTracker's method to identify the end of wireless saturation 
relies on a relatively slow decreasing traffic
load, \textit{i.e.}, a subframe with less than 40\% PRBs allocated for
the user, immediately followed by an empty subframe for that user. 
We, however, observe that 
the end of wireless saturation in significant amount of traces 
do not exhibit such a traffic pattern. 
For example, Figure~\ref{fig:burstTracker_empty} 
depicts the PRB allocation allocated for one web browsing 
user along time.
We see from this figure that
BurstTracker tells us that there exists a long
saturation period that lasts for around 100~ms. But if we check the
PRB allocation results, we see that the cell tower has multiple idle
intervals during the saturation period indicated by BurstTracker.
BurstTracker erroneously identifies those idle intervals as saturated
because the PRB allocation for the user does not end as required by 
BurstTracker. Specifically, the PRBs allocated for the user abruptly
become zero after a subframe with more than 90\% of PRBs allocated.

Detecting the wireless link saturation is an easy task for \systemname{} 
with the overall cell bandwidth usage decoded from the control channel.
For example, we see from Figure~\ref{fig:channel_usage}(a) 
that three aggregated cells are saturated with offered load 70~Mbit/s 
so that the LTE wireless link becomes the bottleneck.
We confirm this by checking the average oneway delay, 
which is plotted in Figure~\ref{fig:CA_delay}. 
We see that the average delay is constantly small with offered load from 10~Mbit/s to 60~Mbit/s. 
The average delay suddenly increases to 185~$ms$, 
which is caused by the packet buffering at the base station.

We run BurstTracker with 200 traces when the user is conducting different activities, including file downloading, video streaming, and web browsing, 
and record the length of the saturation period in each trace. We use the channel usage measured from \systemname{} as ground truth. 
According to our results, 42\% of the time BurstTracker identifies as saturated are truly positive and the rest 58\% are false positive. 
BurstTracker cannot identify the start and end of LTE link saturation correctly and robustly, due to its invalid hypothesis of all\hyp{}or\hyp{}nothing resource allocation, and its requirement of a
slowly decreasing traffic load. While these heuristics sometimes do function as intended, the fundamental reason that forces BurstTracker to rely on such heuristics is that they can only extract the PRB
allocation results for the current user, and not the resource usage of the whole cell.

\section{Related Work}
\label{s:related}
\parahead{Congestion control for cellular network}
The key challenge of designing congestion control for cellular network 
is to estimate and track the rapidly varying capacity. 
A large body of prior work rely on end\hyp{}to\hyp{}end packet statistics 
to infer link capacity~\cite{Copa,Machina,PCC-v,PCC,PROTEUS,ExLL}. 
End-to-end measurements cannot track the fast-varying cellular link capacity, 
so that these algorithms either under-utilize network bandwidth and/or introduce excess delays.
On the other hand, CLAW~\cite{CLAW} and piStream~\cite{piStream} 
adjust the transmission speed according to the base station's bandwidth usage, 
which is inferred from the received power measurements. 
Power measurements, however, are vulnerable to interference 
and thus result in significant estimation error~(\S\ref{s:CLAW}). 
CQIC~\cite{CQIC} adjusts its congestion window purely based on the wireless channel quality 
derived from the \textit{channel quality indicator} (CQI) reported by UE, 
and thus can only track the capacity variations caused by fluctuations in channel quality.
ABC~\cite{Accel-Brake-Control, ABC-NSDI} and throughput guidance~\cite{ThroughputG} proposes to redesign 
the cellular cell tower such that 
the cell tower can estimate the link capacity of each user and 
feed this information back to the sender, which requires modifications 
to the infrastructure and is not compatible with the deployed commercial LTE network.
PBE-CC~\cite{PBE} uses a similar method of capacity estimation as \systemname{}, 
but provides no evaluation on the computational cost and estimation accuracy. 
\systemname{}, however, fully describes the design and implementation details,
conducts extensive evaluation, and performs head\hyp{}to\hyp{}head comparison with the state-of-the-art,
\ie, OWL~\cite{OWL} and MobileInsight\hyp{}based CLAW~\cite{CLAW}. 
       
\parahead{Video streaming and video telephony}
The capacity of the end\hyp{}to\hyp{}end connection is required 
by the adaptive bitrate algorithms (ABR)~\cite{robustMPC, BOLA, Oboe, bufferABR} of video streaming applications 
and video codec~\cite{Vantage,Concerto,webRTC} of videotelephony applications, 
to perform real\hyp{}time video resolution selection.
These systems either rely on the capacity reported by the transport layer protocols~\cite{PCC-v, PCC, PROTEUS}
or directly derive the capacity based on coarse\hyp{}grained historical statistics about video delivery at the application layer, 
and thus cannot track the fast varying capacity in cellular networks.

\parahead{LTE monitoring tools}
LTEye~\cite{LTEye} and OWL~\cite{OWL} are two passive LTE sniffers. 
LTEye does not decode control information for MIMO transmission, 
while OWL cannot work with a cellular network that implements carrier aggregation. 
Furthermore, both LTEye and OWL use bit errors inside a control message as an indicator 
to conduct message validation, which is vulnerable to channel fading and interference, 
and thus introduces a huge amount of false positives (\S\ref{s:eval:accuracy}). 
QXDM~\cite{QXDM} and MobileInsight~\cite{MobileInsight} 
are tools that measure only the PRB allocation of a single user, 
but not the bandwidth usage of the whole cell, 
resulting in significant errors in capacity estimation.
Systems built atop of MobileInsight, 
including CLAW~\cite{CLAW}, PERCEIVE~\cite{PERCEIVE} 
and BurstTracker~\cite{BurstTracker} suffers from the same limitation.

\section{Discussion and Future Work}

\parahead{Telemetry at base station}
\systemname{} performs network telemetry at the mobile devices. 
An alternative solution could be collecting the millisecond-granular capacity information at the base station.
We note that, for any end-to-end applications, the base station is a third-party,
so the telemetry data collected by the base station cannot be trusted without proper authentication. 
Building an authentication system between the base station and every end-to-end application that runs atop of it 
involves significant overhead. 
Transmitting the telemetry data from the base station back to the mobile devices, however, incurs tremendous communication overhead.
For example, if we would like to update the capacity at the same frequency as \systemname{} provides, 
then the base station has to send 1000 messages to each mobile user every second. 
Even if we lower the frequency to every 20ms (which results in delayed reactions), 
the base station still needs to transmit 50 messages per second per user. 
Considering the number of UE the base station serves, 
the total number of messages the base station sends will be quite large. 
Base on the above analysis, we argue in this paper that performing telemetry at the mobile devices is a better choice,
at the cost of slightly increased computation overhead for each mobile user, as we have demonstrated in Section~\S\ref{s:comp}. 

\parahead{Time division duplexing (TDD)}
The current version of \systemname{} focuses on decoding the control channel of 
the cellular network that adopts FDD in its physical layer.
The main difference between TDD and FDD is the frame architecture.
Specifically, FDD uses separate frequencies for uplink and downlink channels,
while TDD leverages the same frequency band for both uplink and downlink.
We note that the structure of downlink subframes where the physical control channel resides is identical in TDD and FDD.
Therefore, \systemname{} is directly applicable to TDD for decoding control messages and then monitoring the capacity.
We leave the extension of \systemname{} to TDD as our near-term future work.

\parahead{Evolving cellular architecture}
The design of the physical control channel changes with evolving of the cellular network architecture, from 4G LTE to 5G and 5G beyond.
For example, the physical control channel of 5G NR and 4G LTE differs from the following three aspects:
firstly, the 5G NR encodes the physical control messages using polar codes instead of convolutional code;
secondly, the location of the control channel inside each subframe does not follow the configuration of 4G LTE as shown in Figure~\ref{fig:pdsch};
thirdly, the formats of the control message, \textit{i.e.,} the meaning of each bit inside a message, are different for 4G LTE and 5G.

We, however, note that two important design choices of the physical control channel still remains in the 5G NR 
and will keep staying in the 3GPP standards for the foreseeable future. 
First of all, the 5G NR does not encrypt the control messages they transmit so that each mobile user is capable of decoding all the control messages if they choose to. 
Secondly, the 5G NR carries the ID of the mobile devices, \ie, the C-RNTI, inside the control message via XOR-ing it with the CRC, as we have introduced in Section~\S\ref{s:dci_encode}.
According to the above analysis, with necessary customization to cope with the changes in the 5G control channel,
\systemname{} is able to decode all the control messages inside the control channel of the 5G network and associate each decoded message with its ID. 
Based on the decoded 5G control messages, \systemname{} is, therefore, capable of estimating the 5G network capacity, just as we have demonstrated for 4G LTE.
\section{Conclusion}
\label{s:concl}

\systemname{} sets new benchmarks for accuracy in mobile
cellular network monitoring hence enables the development of 
dramatically improved congestion control algorithms and applications that respond 
more effectively to fluctuations in the wireless channel.  
In the course of its design and implementation, we have 
additionally documented interesting and important experimental
phenomena surrounding carrier aggregation that are significant
to any network designer considering the traffic that flows through
a mobile cellular network.
\section{Acknowledgments}
We thank the anonymous reviewers and our shepherd
for their valuable feedback that has improved the quality of this
paper significantly. 
This material is based upon work supported by the National Science Foundation under Grant No.~CNS-2027647.  
We gratefully acknowledge a gift from the Amateur Radio Digital Communications Foundation.
\clearpage
\bibliographystyle{abbrv}
\bibliography{paper}

\begin{thebibliography}{10}

\bibitem{LTE-MAC}
3GPP.
\newblock {TS 36.321: Evolved Universal Terrestrial Radio Access (E-UTRA);
  Medium Access Control (MAC) protocol specification}.

\bibitem{TS211}
3GPP.
\newblock {TS36.211: Evolved Universal Terrestrial Radio Access (E-UTRA);
  Physical channels and modulation.}

\bibitem{TS213}
3GPP.
\newblock {TS36.213: Evolved Universal Terrestrial Radio Access (E-UTRA);
  Physical layer procedures.}

\bibitem{5G}
{5G specifications}.
\newblock \href{https://www.3gpp.org/release-15}{Available here}.

\bibitem{Oboe}
Z.~Akhtar, Y.~S. Nam, R.~Govindan, S.~Rao, J.~Chen, E.~Katz-Bassett,
  B.~Ribeiro, J.~Zhan, and H.~Zhang.
\newblock Oboe: Auto-tuning video abr algorithms to network conditions.
\newblock In {\em ACM SIGCOMM}, 2018.

\bibitem{Copa}
V.~Arun and H.~Balakrishnan.
\newblock Copa: Practical delay-based congestion control for the internet.
\newblock In {\em {NSDI}}, 2018.

\bibitem{BurstTracker}
A.~Balasingam, M.~Bansal, R.~Misra, K.~Nagaraj, R.~Tandra, S.~Katti, and
  A.~Schulman.
\newblock Detecting if lte is the bottleneck with bursttracker.
\newblock In {\em ACM MobiCom}, 2019.

\bibitem{Birthday}
{Birthday problem}.
\newblock \href{https://en.wikipedia.org/wiki/Birthday_problem}{Wiki}.

\bibitem{OWL}
N.~Bui and J.~Widmer.
\newblock {OWL}: A reliable online watcher for lte control channel
  measurements.
\newblock In {\em ACM AllThingsCellular}, 2016.

\bibitem{BBR}
N.~Cardwell, Y.~Cheng, C.~S. Gunn, S.~H. Yeganeh, and V.~Jacobson.
\newblock Bbr: Congestion-based congestion control.
\newblock {\em Queue}, 14(5), Oct. 2016.

\bibitem{CellurVideo}
J.~Chen, R.~Mahindra, M.~A. Khojastepour, S.~Rangarajan, and M.~Chiang.
\newblock A scheduling framework for adaptive video delivery over cellular
  networks.
\newblock In {\em ACM MobiCom}, 2013.

\bibitem{webRTC}
Webrtc.
\newblock \href{https://webrtc.org/}{Official website}.

\bibitem{CiscoReport}
Cisco: Global mobile data traffic forecast update.
\newblock
  \href{https://www.cisco.com/c/en/us/solutions/collateral/service-provider/visual-networking-index-vni/white-paper-c11-738429.html}{Available
  here}, 2019.

\bibitem{DASH}
Dynamic adaptive streaming over http.
\newblock
  \href{https://en.wikipedia.org/wiki/Dynamic_Adaptive_Streaming_over_HTTP}{Website}.

\bibitem{PCC}
M.~Dong, Q.~Li, D.~Zarchy, P.~B. Godfrey, and M.~Schapira.
\newblock {PCC}: Re-architecting congestion control for consistent high
  performance.
\newblock In {\em { NSDI}}, 2015.

\bibitem{PCC-v}
M.~Dong, T.~Meng, D.~Zarchy, E.~Arslan, Y.~Gilad, B.~Godfrey, and M.~Schapira.
\newblock {PCC Vivace}: Online-learning congestion control.
\newblock In {\em {NSDI}}, 2018.

\bibitem{Salsify}
S.~Fouladi, J.~Emmons, E.~Orbay, C.~Wu, R.~S. Wahby, and K.~Winstein.
\newblock Salsify: Low-latency network video through tighter integration
  between a video codec and a transport protocol.
\newblock In {\em {NSDI}}, 2018.

\bibitem{ABC-NSDI}
P.~Goyal, A.~Agarwal, R.~Netravali, M.~Alizadeh, and H.~Balakrishnan.
\newblock {ABC}: A simple explicit congestion controller for wireless networks.
\newblock In {\em {USENIX} {NSDI}}, 2020.

\bibitem{Accel-Brake-Control}
P.~Goyal, M.~Alizadeh, and H.~Balakrishnan.
\newblock Rethinking congestion control for cellular networks.
\newblock In {\em ACM HotNets}, 2017.

\bibitem{CUBIC}
S.~Ha, I.~Rhee, and L.~Xu.
\newblock {CUBIC}: A new {TCP}-friendly high-speed {TCP} variant.
\newblock {\em SIGOPS Oper. Syst. Rev.}, 42(5), July 2008.

\bibitem{bufferABR}
T.-Y. Huang, R.~Johari, N.~McKeown, M.~Trunnell, and M.~Watson.
\newblock A buffer-based approach to rate adaptation: Evidence from a large
  video streaming service.
\newblock In {\em ACM SIGCOMM}, 2014.

\bibitem{ThroughputG}
A.~Jain, A.~Terzis, H.~Flinck, N.~Sprecher, S.~Arunachalam, K.~Smith,
  V.~Devarapalli, and R.~Yanai.
\newblock Mobile throughput guidance inband signaling protocol.
\newblock {\em IETF, work in progress}, 2015.

\bibitem{LTEye}
S.~Kumar, E.~Hamed, D.~Katabi, and L.~Erran~Li.
\newblock {LTE} radio analytics made easy and accessible.
\newblock In {\em ACM SIGCOMM}, 2014.

\bibitem{PERCEIVE}
J.~Lee, S.~Lee, J.~Lee, S.~D. Sathyanarayana, H.~Lim, J.~Lee, X.~Zhu,
  S.~Ramakrishnan, D.~Grunwald, K.~Lee, and S.~Ha.
\newblock Perceive: Deep learning-based cellular uplink prediction using
  real-time scheduling patterns.
\newblock In {\em ACM MobiSys}, 2020.

\bibitem{MobileInsight}
Y.~Li, C.~Peng, Z.~Yuan, J.~Li, H.~Deng, and T.~Wang.
\newblock Mobileinsight: Extracting and analyzing cellular network information
  on smartphones.
\newblock In {\em ACM MobiCom}, 2016.

\bibitem{LTE-Rel-10}
{LTE Release 10}.
\newblock
  \href{https://www.3gpp.org/specifications/releases/70-release-10}{Available
  here}.

\bibitem{CQIC}
F.~Lu, H.~Du, A.~Jain, G.~M. Voelker, A.~C. Snoeren, and A.~Terzis.
\newblock {CQIC}: Revisiting cross-layer congestion control for cellular
  networks.
\newblock In {\em ACM HotMobile}, 2015.

\bibitem{Pensieve}
H.~Mao, R.~Netravali, and M.~Alizadeh.
\newblock Neural adaptive video streaming with pensieve.
\newblock In {\em ACM SIGCOMM}, 2017.

\bibitem{Mahimahi}
R.~Netravali, A.~Sivaraman, S.~Das, A.~Goyal, K.~Winstein, J.~Mickens, and
  H.~Balakrishnan.
\newblock Mahimahi: Accurate record-and-replay for http.
\newblock In {\em USENIX ATC}, 2015.

\bibitem{ExLL}
S.~Park, J.~Lee, J.~Kim, J.~Lee, S.~Ha, and K.~Lee.
\newblock Exll: An extremely low-latency congestion control for mobile cellular
  networks.
\newblock In {\em ACM CoNEXT}, 2018.

\bibitem{QXDM}
Qualcomm qxdm tool.
\newblock
  \href{https://www.qualcomm.com/documents/qxdm-professional-qualcomm-extensible-diagnostic-monitor}{Website}.

\bibitem{Vantage}
D.~Ray, J.~Kosaian, K.~V. Rashmi, and S.~Seshan.
\newblock Vantage: Optimizing video upload for time-shifted viewing of social
  live streams.
\newblock In {\em ACM SIGCOMM}, 2019.

\bibitem{LTE_UMTS}
S.~Sesia, I.~Toufik, and M.~Baker.
\newblock {\em LTE, The UMTS Long Term Evolution: From Theory to Practice}.
\newblock Wiley Publishing, 2009.

\bibitem{BOLA}
K.~{Spiteri}, R.~{Urgaonkar}, and R.~K. {Sitaraman}.
\newblock {BOLA}: Near-optimal bitrate adaptation for online videos.
\newblock In {\em IEEE INFOCOM}, 2016.

\bibitem{USRP}
{USRP}.
\newblock \href{https://www.ettus.com/}{Website}.

\bibitem{Machina}
K.~Winstein and H.~Balakrishnan.
\newblock {TCP Ex Machina}: Computer-generated congestion control.
\newblock In {\em ACM SIGCOMM}, 2013.

\bibitem{Sprout}
K.~Winstein, A.~Sivaraman, and H.~Balakrishnan.
\newblock Stochastic forecasts achieve high throughput and low delay over
  cellular networks.
\newblock In {\em USENIX NSDI}, 2013.

\bibitem{piStream}
X.~Xie, X.~Zhang, S.~Kumar, and L.~E. Li.
\newblock {piStream}: Physical layer informed adaptive video streaming over
  lte.
\newblock In {\em ACM MobiCom}, 2015.

\bibitem{CLAW}
X.~Xie, X.~Zhang, and S.~Zhu.
\newblock Accelerating mobile web loading using cellular link information.
\newblock In {\em ACM MobiSys}, 2017.

\bibitem{PBE}
Y.~Xie, F.~Yi, and K.~Jamieson.
\newblock {PBE-CC}: Congestion control via endpoint-centric, physical-layer
  bandwidth measurements.
\newblock In {\em ACM SIGCOMM}, 2020.

\bibitem{PROTEUS}
Q.~Xu, S.~Mehrotra, Z.~Mao, and J.~Li.
\newblock {PROTEUS}: Network performance forecast for real-time, interactive
  mobile applications.
\newblock In {\em ACM MobiSys}, 2013.

\bibitem{fugu}
F.~Y. Yan, H.~Ayers, C.~Zhu, S.~Fouladi, J.~Hong, K.~Zhang, P.~Levis, and
  K.~Winstein.
\newblock Learning in situ: a randomized experiment in video streaming.
\newblock In {\em {USENIX} {NSDI}}, 2020.

\bibitem{Pantheon}
F.~Y. Yan, J.~Ma, G.~D. Hill, D.~Raghavan, R.~S. Wahby, P.~Levis, and
  K.~Winstein.
\newblock Pantheon: The training ground for internet congestion-control
  research.
\newblock In {\em USENIX ATC}, 2018.

\bibitem{robustMPC}
X.~Yin, A.~Jindal, V.~Sekar, and B.~Sinopoli.
\newblock A control-theoretic approach for dynamic adaptive video streaming
  over {HTTP}.
\newblock In {\em ACM SIGCOMM}, 2015.

\bibitem{MuVi}
J.~Yoon, H.~Zhang, S.~Banerjee, and S.~Rangarajan.
\newblock {MuVi}: A multicast video delivery scheme for {4G} cellular networks.
\newblock In {\em ACM MobiCom}, 2012.

\bibitem{Verus}
Y.~Zaki, T.~P\"{o}tsch, J.~Chen, L.~Subramanian, and C.~G\"{o}rg.
\newblock Adaptive congestion control for unpredictable cellular networks.
\newblock In {\em ACM SIGCOMM}, 2015.

\bibitem{Concerto}
A.~Zhou, H.~Zhang, G.~Su, L.~Wu, R.~Ma, Z.~Meng, X.~Zhang, X.~Xie, H.~Ma, and
  X.~Chen.
\newblock Learning to coordinate video codec with transport protocol for mobile
  video telephony.
\newblock In {\em ACM MobiCom}, 2019.

\end{thebibliography}

\end{document}